\documentclass[fleqn,usenatbib]{mnras}

\usepackage{newtxtext,newtxmath}

\usepackage[T1]{fontenc}

\usepackage{graphicx}	
\usepackage{amsmath}	

\usepackage{graphicx}
\usepackage{amsmath}
\usepackage{subfigure}
\usepackage{multirow}
\usepackage{threeparttable}
\usepackage{array}
\usepackage{natbib}
\usepackage{xcolor}
\usepackage{hyperref}
\usepackage[version=4]{mhchem}
\usepackage{bm}
\usepackage{ulem}

\newcommand{\msun}{\mbox{${\rm M}_\odot$}}
\newcommand{\mstar}{\mbox{${M}_{\rm star}$}}
\newcommand{\cmjj}{\mbox{${\rm cm^{-2}}$}}

\newcommand{\kms}{\,{\rm{km\,s}^{-1}}}
\newcommand{\cc}{{\rm cm^{-3}}}

\defcitealias{CUBSV}{CUBS V}

\title[CUBS VI: Environmental investigation of CGM at $z\approx1$]{The Cosmic Ultraviolet Baryon Survey (CUBS) VI: Connecting Physical Properties of the Cool Circumgalactic Medium to Galaxies at ${\bm z\approx 1}$}

\author[Qu et al.]{
Zhijie Qu$^{1}$,\thanks{E-mail: quzhijie@uchicago.edu}
Hsiao-Wen Chen$^{1}$, 
Gwen C. Rudie$^{2}$,
Sean D. Johnson$^{3}$,
Fakhri S. Zahedy$^{2}$,
\newauthor
David DePalma$^{4}$,
Erin Boettcher$^{5,6,7}$,
Sebastiano Cantalupo$^{8}$,
Mandy C. Chen$^{1}$,
Kathy L. Cooksey$^{9}$,
\newauthor
Claude-Andr\'e Faucher-Gigu\`ere$^{10}$,
Jennifer I-Hsiu Li$^{3}$,
Sebastian Lopez$^{11}$,
Joop Schaye$^{12}$,
and 
\newauthor
Robert A. Simcoe$^{4}$
\\
$^{1}$ Department of Astronomy \& Astrophysics, The University of Chicago, 5640 S. Ellis Ave., Chicago, IL 60637, USA\\
$^{2}$ The Observatories of the Carnegie Institution for Science, 813 Santa Barbara Street, Pasadena, CA 91101, USA\\
$^{3}$ Department of Astronomy, University of Michigan, Ann Arbor, MI 48109, USA\\
$^{4}$ MIT-Kavli Institute for Astrophysics and Space Research, 77 Massachusetts Ave., Cambridge, MA 02139, USA\\
$^{5}$ Department of Astronomy, University of Maryland, College Park, MD 20742, USA\\
$^{6}$ X-ray Astrophysics Laboratory, NASA/GSFC, Greenbelt, MD 20771, USA\\
$^{7}$ Center for Research and Exploration in Space Science and Technology, NASA/GSFC, Greenbelt, MD 20771, USA\\
$^{8}$ Department of Physics, University of Milan Bicocca, Piazza della Scienza 3, I-20126 Milano, Italy \\
$^{9}$ Department of Physics and Astronomy, University of Hawai'i at Hilo, Hilo, HI 96720, USA\\
$^{10}$ Department of Physics \& Astronomy, Center for Interdisciplinary Exploration and Research in Astrophysics (CIERA), Northwestern University, \\
1800 Sherman Avenue, Evanston, IL 60201, USA \\
$^{11}$ Departamento de Astronomía, Universidad de Chile, Casilla 36-D, Santiago, Chile \\
$^{12}$ Leiden Observatory, Leiden University, PO Box 9513, NL-2300 RA Leiden, the Netherlands
}

\date{Accepted XXX. Received YYY; in original form ZZZ}

\pubyear{2023}

\begin{document}
\label{firstpage}
\pagerange{\pageref{firstpage}--\pageref{lastpage}}
\maketitle

\begin{abstract}
This paper presents a newly established sample of galaxies and galaxy groups at redshift $z\approx 1$ from the Cosmic Ultraviolet Baryon Survey (CUBS), for which sensitive constraints can be placed on the circumgalactic medium (CGM) using high-quality far-ultraviolet (FUV) and optical absorption spectra of background QSOs.
The CUBS program has uncovered 19 unique galaxies or galaxy groups at $z= 0.89$ to $1.21$ in six QSO fields, which is designated as the CUBSz1 sample.
In this CUBSz1 sample, nine galaxies or galaxy groups show absorption features, while ten systems do not have detectable absorption with 2-$\sigma$ upper limits of $\log N (\mbox{\ion{He}{I}})/\cmjj \lesssim 13.5$ and $\log N (\mbox{\ion{O}{V}})/\cmjj \lesssim 13.3$.
Environmental properties of the galaxies, including galaxy overdensities, the total stellar mass and gravitational potential summed over all nearby neighbors, and the presence of local ionizing sources, are found to have a significant impact on the observed CGM absorption properties.
Specifically, massive galaxies and galaxies in overdense regions exhibit a higher rate of incidence of absorption. 
At the same time, the observed CGM absorption properties in galaxy groups appear to be driven by the galaxy closest to the QSO sightline, rather than by the most massive galaxy or by mass-weighted properties.
We introduce a total projected gravitational potential $\psi$, defined as $-\psi/G = \sum M_{{\rm halo}}/d_{{\rm proj}}$ summed over all group members, to characterize the overall galaxy environment.
This projected gravitational potential correlates linearly with the maximum density detected in each sightline (i.e., a power law slope of $0.95_{-0.14}^{+0.15}$), consistent with higher-pressure gas being confined in deeper gravitational potential wells.
In addition, we find that the radial profile of cool gas density exhibits a general decline from the inner regions to the outskirts, and the amplitude is consistent with the cool gas being in pressure balance with the hot halo.
Finally, we note that the ionizing flux from nearby galaxies can generate an elevated $N$(\ion{H}{I})/$N$(\ion{He}{I}) ratio, which in turn provides a unique diagnostic of possible local sources contributing to the ionizing radiation field.
\end{abstract}




\section{Introduction}
The fate of a galaxy is determined by the competition between mass accretion and different feedback processes that also determine the physical properties of its extended gaseous halo, known as the circumgalactic medium (CGM; see \citealt{Donahue2022} for a recent review).
Gathering feedback and accreted materials, the CGM retains a sensitive record of these fundamental processes throughout the history of a galaxy's growth.

QSO spectroscopy provides a sensitive tool for characterizing the diffuse CGM and intergalactic medium (IGM).
The rest-frame UV ionic transitions cover a wide range of ionization states from neutral species (e.g., \ion{H}{I}) to highly ionized species (e.g., \ion{O}{VI} and \ion{Ne}{VIII}).
Previous surveys of different absorption features along random sightlines over a broad redshift range have yielded accurate measurements of the cosmic evolution of these atomic species (e.g., \citealt{Rahmati2016} and reference therein for a list of empirical studies).
Furthermore, detailed physical modelling of the ionization mechanism and line profile analyses have been developed to extract more physical properties, such as the gas density and temperature, for individual absorption components \citep[e.g.,][]{Zahedy:2019aa, Zahedy:2021aa, Sankar:2020aa, Sameer:2021aa, Cooper:2021aa}.

Independent of random sightline studies, absorption-line surveys conducted in the vicinity of known galaxies provide a direct avenue to connect absorption properties with the properties of the galaxies.
Previous studies primarily focus on luminous ($L^*$) galaxies in two epochs, $z\lesssim 0.5$ (e.g., \citealt{Chen:2017aa, Tumlinson:2017aa} for reviews) and $z\approx 2-3$ \citep[e.g.,][]{Steidel:2010aa, Turner:2014aa, Rudie:2012aa, Rudie:2019aa}.
Absorption properties of the CGM are found to depend strongly on the galactic environment, which can be characterized by the number densities of nearby galaxies and how massive the galaxies are (e.g., \citealt{Lopez:2008aa, Chen:2009aa, Johnson:2013aa, Johnson:2015aa, Burchett:2016aa, Chen:2020aa, Lan:2020aa, Beckett:2021aa, Huang:2021aa, Dutta:2021aa}).

At the same time, the connection between absorption properties and galaxy properties together with their associated environments for low-mass galaxies ($M_{\rm star} \approx 10^8 - 10^9\rm~ M_\odot$) remain uncertain \citep[e.g.,][]{Johnson:2017aa, Putman:2021aa}, particularly at moderate-to-high redshifts ($z\gtrsim 0.5$) due to survey incompleteness.
In addition, understanding the rapid decline of the cosmic star formation rate density from the {\it Cosmic Noon} at $z\approx2$ to the local Universe also requires systematic studies of the CGM at $z\approx 1$, which are still lacking.

Over the past few years, deep galaxy surveys have been carried out around QSO sightlines with high-quality QSO spectra to enable comprehensive CGM studies at $z\approx 1$ (e.g. the CASBaH survey, the CUBS survey, and the MUSEQuBES survey; \citealt{Burchett:2019aa, Chen:2020aa, Muzahid:2021aa}).
These survey data provide a deep look at the impact of galaxy environments on the CGM over a broader range of redshift, particularly for lower-mass galaxies (e.g., dwarf galaxies).
As part of the Cosmic Ultraviolet Baryon Survey (CUBS) program \citep[see details in ][]{Chen:2020aa}, we leverage the high signal-to-noise ratio (S/N) QSO spectra and deep galaxy surveys to construct a sample of galaxies at $z\approx 1$ for which the CGM properties can be constrained sensitively based on absorption features obtained in the spectra of background QSOs.
In \citet[][hereafter, \citetalias{CUBSV}]{CUBSV}, we investigated the thermodynamic properties of the CGM based on a kinematic line profile analysis.
Together with literature samples, we found that massive and passive galaxies exhibit higher non-thermal broadening and turbulent energy in their CGM than star-forming galaxies at $z \lesssim 1$.

In the current study, we focus on the impact of galaxy environment on the CGM at $z\approx 1$.
The paper is organized as follows.
Section \ref{data} presents a summary of available data for both galaxy surveys and QSO spectroscopy.
For each galaxy-CGM system, we infer a stellar mass and star formation rate (SFR) for each galaxy based on their observed spectral energy distribution (SED), and gas properties using photoionization modelling and line-profile analysis, which are presented in Section \ref{analysis}.
The impact of galaxy environments on the incidence rate of absorption is described in Section \ref{sec:obs_inc}.
In Section \ref{sec:obs_col} and Section \ref{sec:obs_density}, we examine how the ionization state and density of the gas also depend on the galaxy environments.
In Section \ref{sec:obs_he1}, we investigate possible local contributions to the ionizing flux from nearby galaxies based on anomalous $N$(\ion{H}{I})/$N$(\ion{He}{I}) ratios.
Finally, in Section \ref{discussion}, we discuss the implications of our findings.
Throughout the paper, we adopt a $\Lambda$ cosmology with $\Omega_{\rm m} = 0.3$, $\Omega_{\Lambda} = 0.7$, and a Hubble constant of $H_0 = 70\rm ~ km ~s^{-1}~ Mpc^{-1}$.

\section{Data}
\label{data}
We introduce a new sample of galaxies at $z\approx 1$, designated as CUBSz1, for which sensitive constraints are available for both absorption properties of their host halos and surrounding galaxy environments.
The galaxy survey data and quasar absorption spectra are drawn from the CUBS program, which covers 15 fields with NUV-bright (NUV$_{\rm AB} < 17$) background QSOs at $z_{\rm QSO}\gtrsim 0.8$ in the footprint of the Dark Energy Survey\footnote{https://www.darkenergysurvey.org/} \citep[DES; ][]{Drlica-Wagner:2018aa}.
See \citet{Chen:2020aa} for a full description of the CUBS program.
Seven of the CUBS QSOs occur at $z_{\rm QSO} \gtrsim 1$, and enable absorption line studies of diffuse gas at $z\approx 1$.
The CUBSz1 galaxies are found in six QSO fields based on their close proximity to the QSO sightline {\it without prior knowledge of any absorption features}.
It is a galaxy-centric sample to enable an unbiased study of the incidence of absorption features near galaxies in this redshift range.
In particular, we select galaxies at $z>0.883$ to ensure that \ion{He}{I} $\lambda$ 584 is covered by the COS spectra to provide a proxy for $N$(\ion{H}{I}).
In total, the sample contains 19 galaxies and galaxy groups at $0.89<z<1.21$ with constrained absorption properties.

\begin{figure*}
    \centering
    \includegraphics[width=0.31\textwidth]{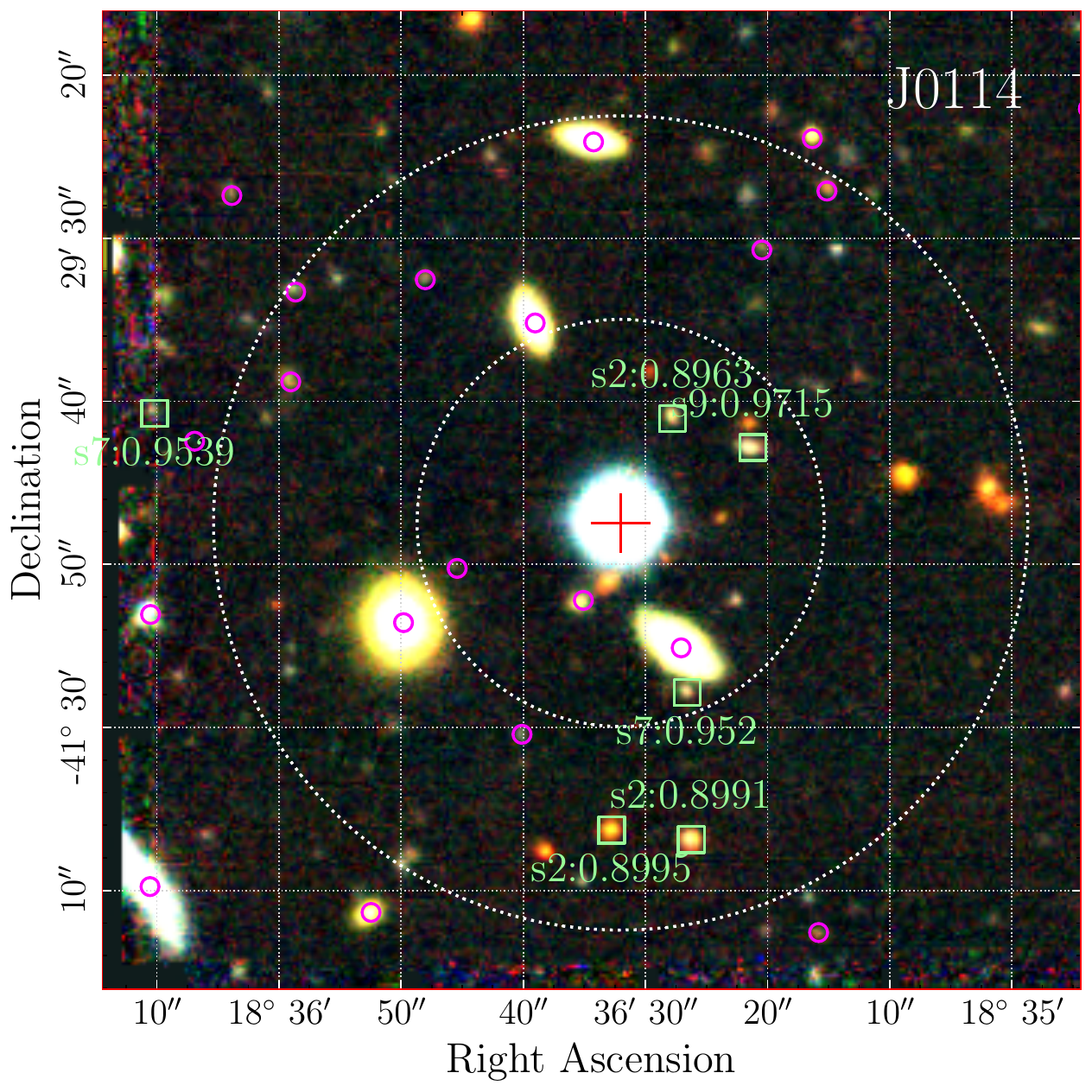}
    \includegraphics[width=0.31\textwidth]{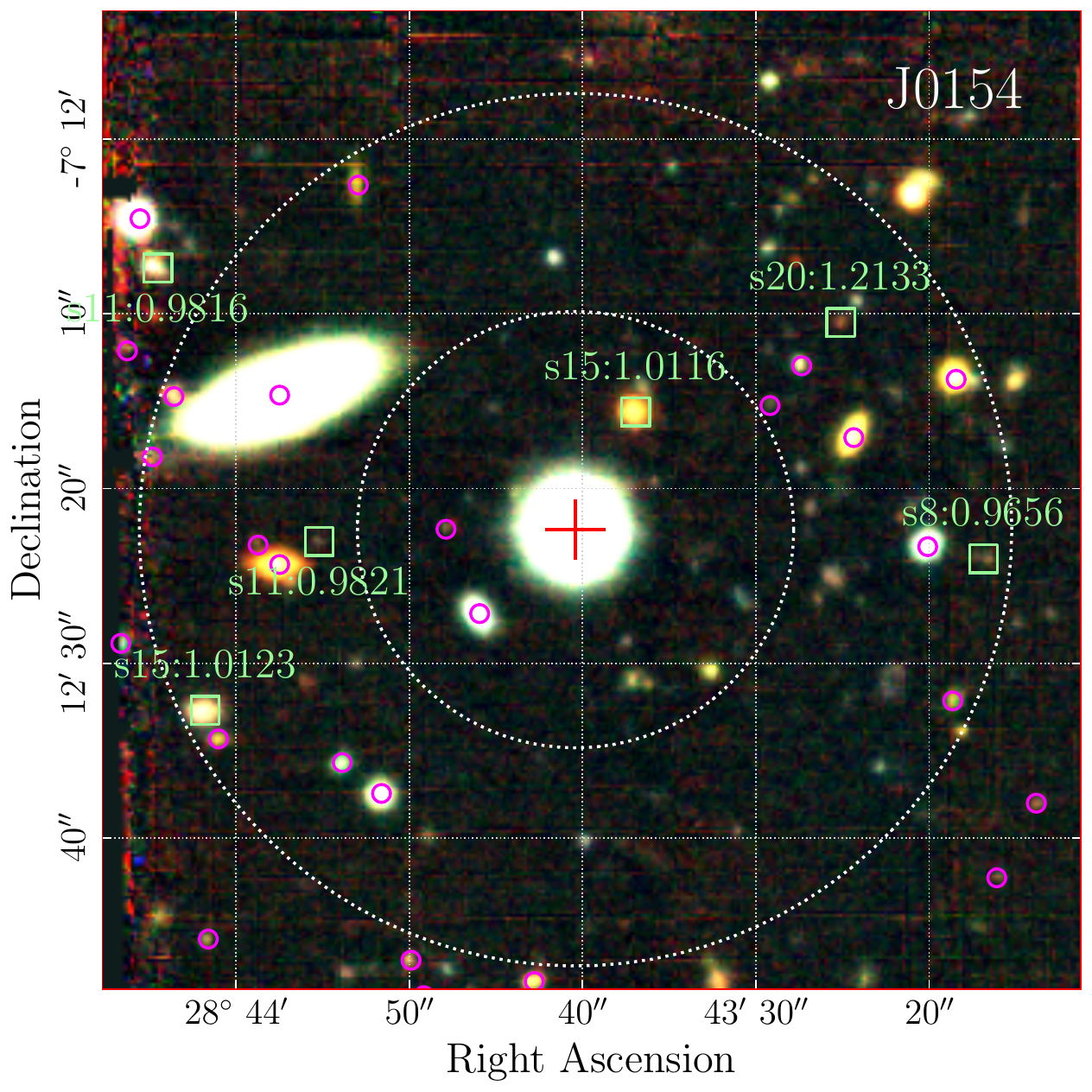}
    \includegraphics[width=0.31\textwidth]{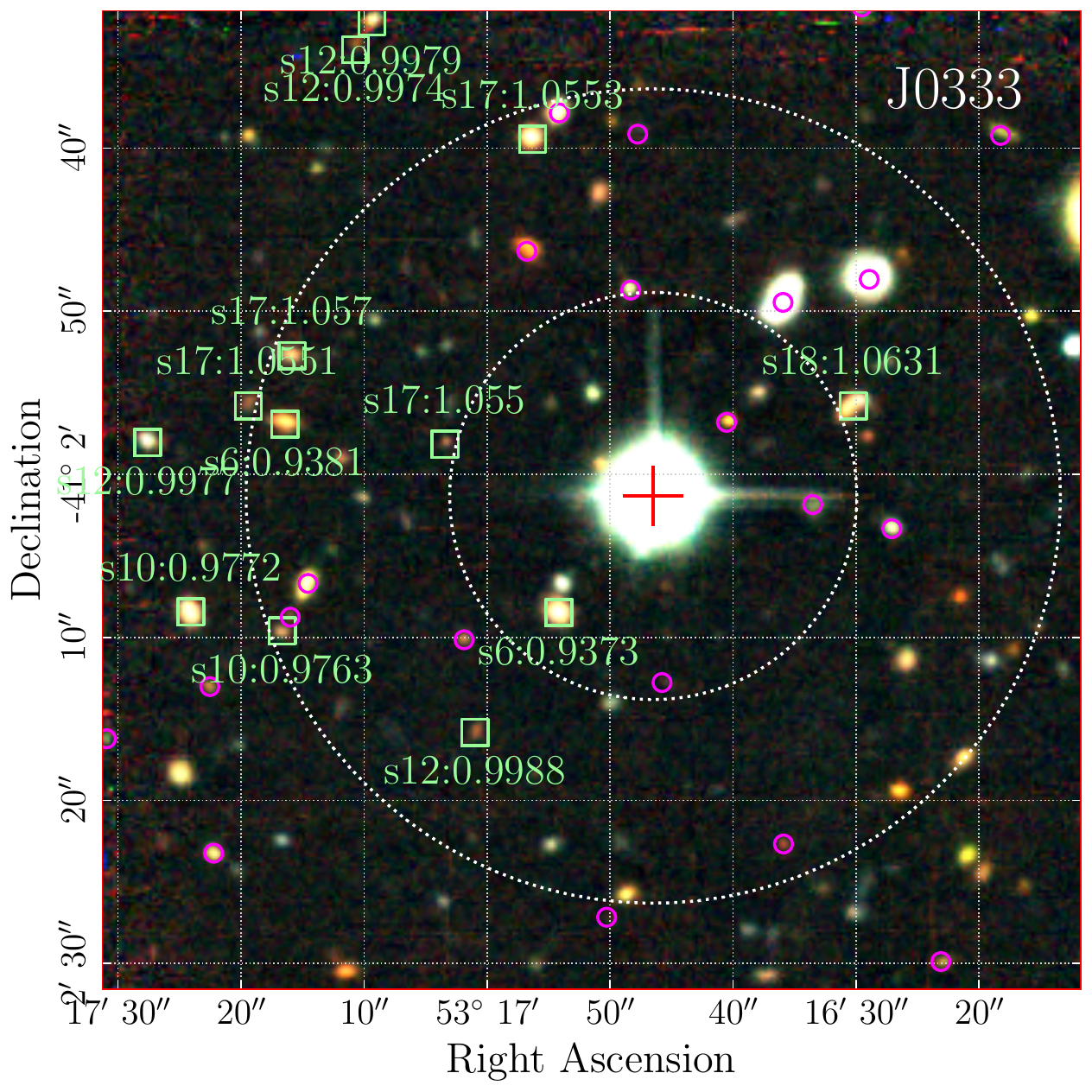}\\
    \hspace{0.02cm}
    \includegraphics[width=0.31\textwidth]{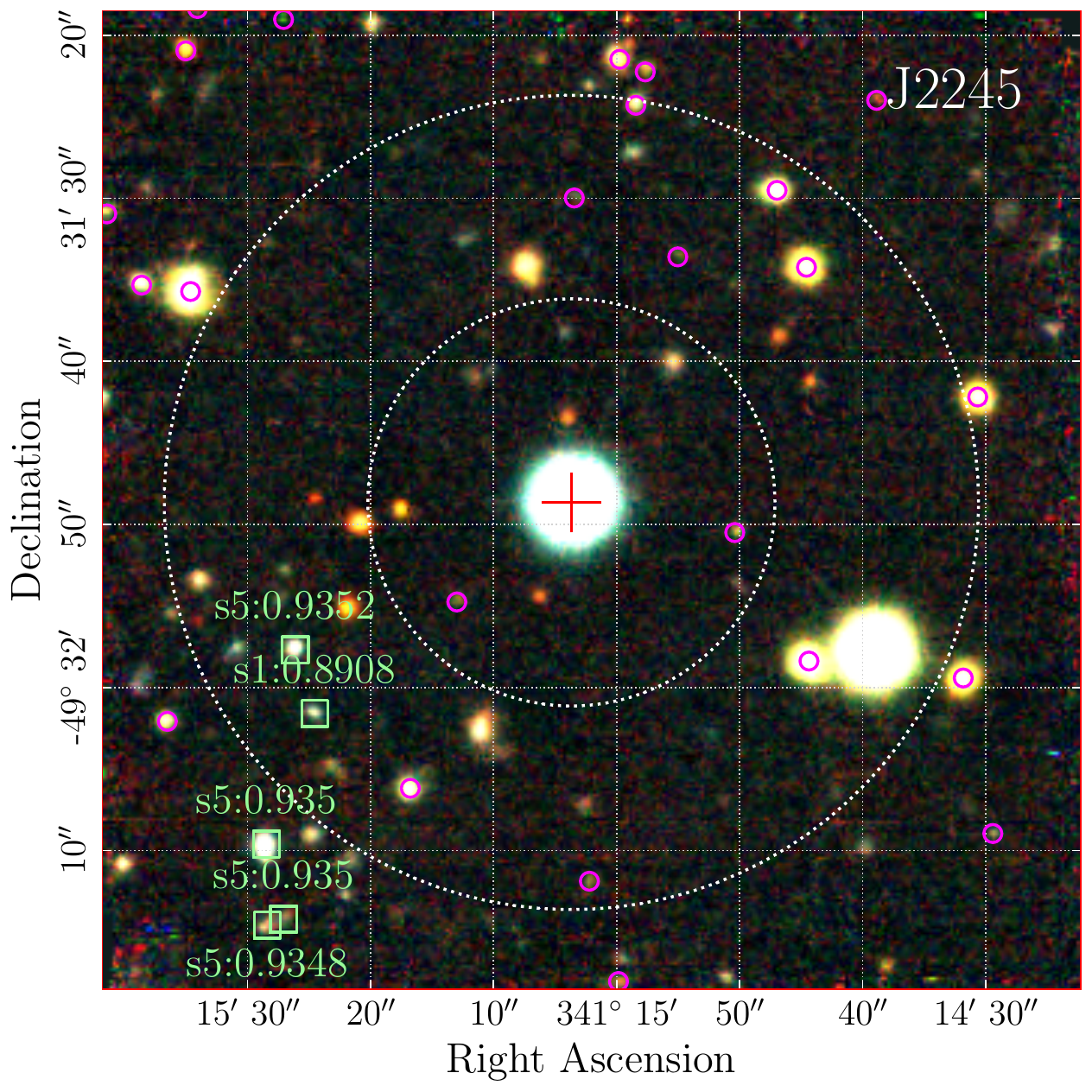}
    \includegraphics[width=0.31\textwidth]{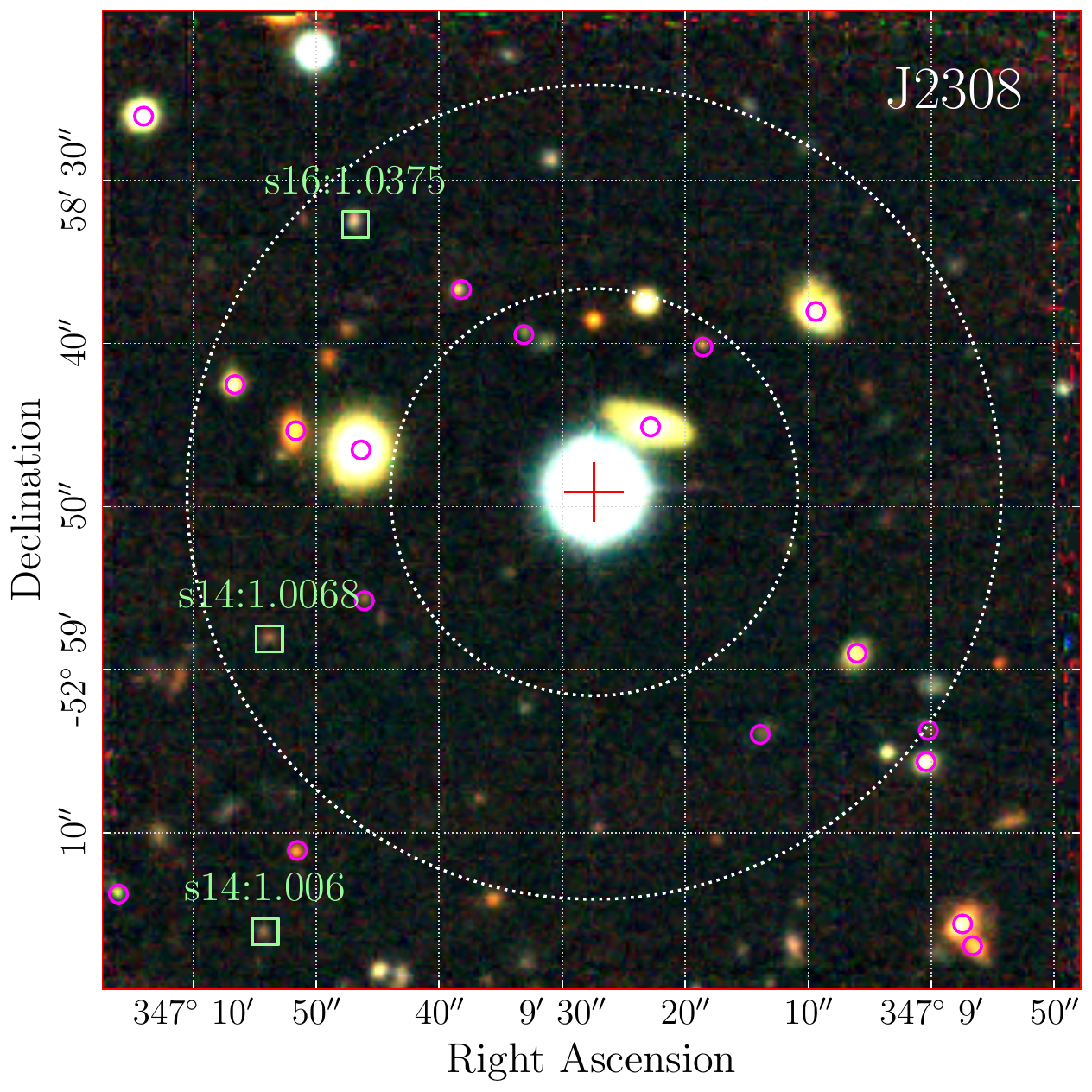}
    \includegraphics[width=0.31\textwidth]{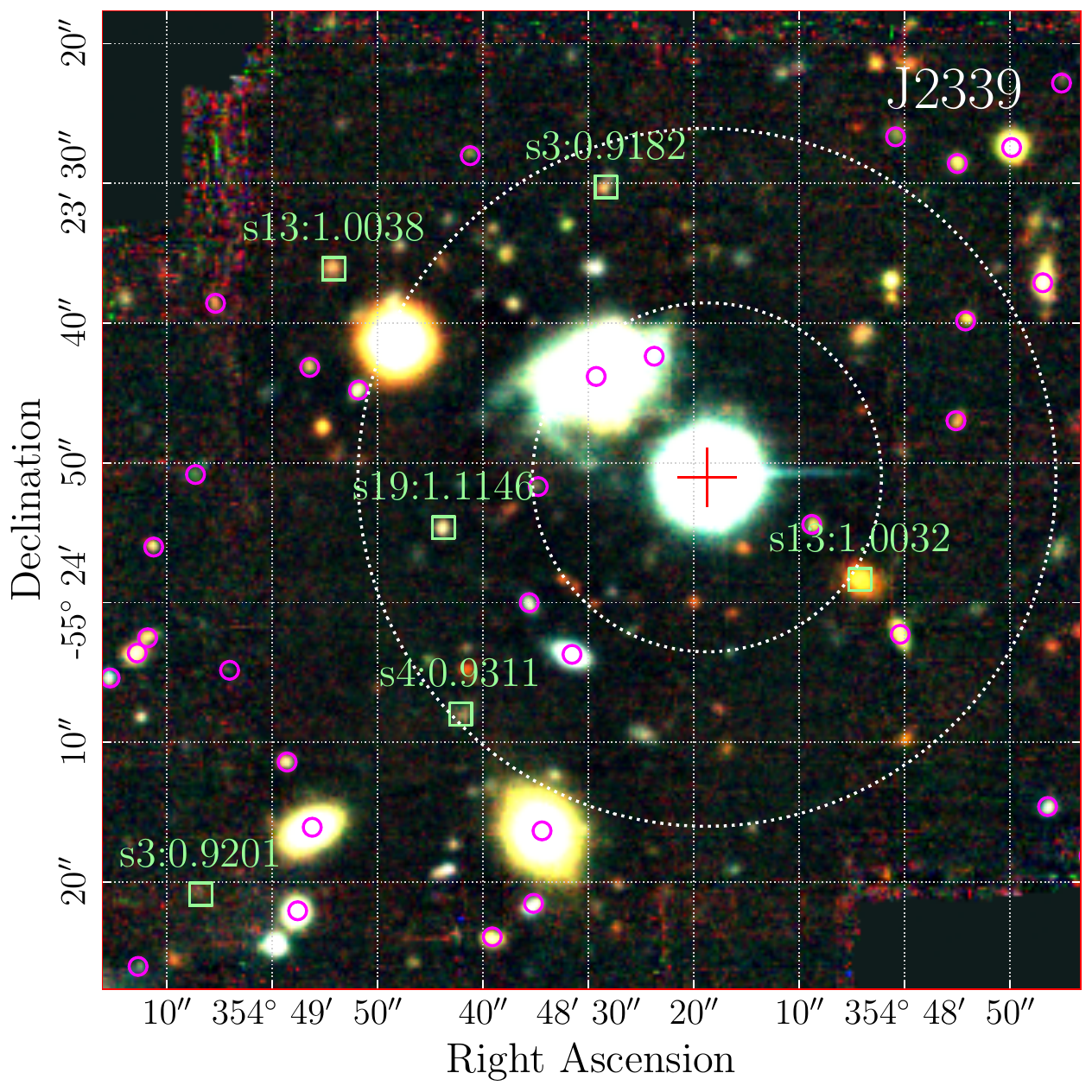}
    \caption{Colour images of six CUBS QSO fields are produced from integrating the MUSE cube over wavelength ranges of $4800\,-\,5800$ \r{A} (pseudo $g$ band; blue), $6000\,-\,7500$ \r{A} (pseudo $r$ band; green), and $7500\,-\,9000$ \r{A} (pseudo $i$ band; red).
    In each field, the QSO is located at the centre, and marked by a red "plus" symbol.
    The galaxies selected for the CUBSz1 sample are marked by a green box, and indicated by their system IDs and redshifts.
    Known foreground galaxies collected for the CUBS program are also marked with magenta open circles.
    Two dotted white circles indicate 100 kpc and 200 kpc at $z=1$.
    }
    \label{fig:muse_imgs}
\end{figure*}

\subsection{QSO absorption spectra and galaxy surveys}
\label{sec:cubs_data}
For each CUBS QSO, high-quality absorption spectra were obtained both in the FUV spectral window using the {\it Hubble Space Telescope} ({\it HST}) and the Cosmic Origins Spectrograph (COS; \citealt{Green2012}), and in the optical using the Magellan Inamori Kyocera Echelle Spectrograph (MIKE; \citealt{Bernstein2003}) on the Magellan Clay Telescope.
Medium-resolution (FWHM $\approx 20 \kms$) {\it HST}/COS spectra were obtained using the G130M and G160M gratings and multiple central wavelengths to yield contiguous spectral coverage from 1100 \r{A} to 1800 \r{A} (PID$=$15163; PI: Chen).
High-resolution MIKE spectra, covering the wavelength range from 3300\!~\! \r{A} to 9300\!~\! \r{A} with a spectral resolution of FWHM\,$\approx\!8-10$\,$\kms$, were processed using custom software to optimize the wavelength calibration and to achieve the best S/N in final coadded spectra \citep{Zahedy:2016aa, Chen:2018aa, Chen:2020aa}.
The coadded {\it HST}/COS FUV spectra have typical S/N of $12\,-\,31$ per resolution element, while the MIKE optical spectra have S/N of $22-63$ per resolution element.
At $z\!\approx\!1$, {\it HST}/COS FUV spectra cover a wide range of species, including \ion{H}{I}, \ion{He}{I}, \ion{C}{II}, \ion{N}{II} to \ion{N}{IV}, \ion{O}{II} to \ion{O}{V}, \ion{S}{II} to \ion{S}{V}, \ion{Ne}{IV} to \ion{Ne}{VI}, and \ion{Ne}{VIII}.
In addition to the FUV spectra, the MIKE spectra provide additional coverage for \ion{Mg}{I}, \ion{Mg}{II}, \ion{Fe}{II}, and \ion{C}{IV} at $z\!\approx\!1$.

The CUBS galaxy survey includes photometric and spectroscopic observations.
Both consist of three components defined by their field of views and survey depths.
For the photometric observations, the DES provides the initial multi-bandpass optical imaging data reaching limiting magnitudes of $\approx 24.5$, 24.0, and 23.3 in the $g$, $r$, and $i$ bands, respectively.
Deep optical $g'$, $r'$, and $z'$-band images using Magellan IMACS \citep{Dressler2011} and LDSS3 \citep{Osip:2008aa} are also available, together with near-infrared $H$-band images using FourStar \citep{Persson:2013aa}.
These imaging data reach the limiting magnitudes down to $\approx 26$, 25.5, 25.5, and 24.5 in the $g'$, $r'$, $z'$, and $H$ bands, respectively.
Optical magnitudes are calibrated using bright sources detected both in DES and Magellan observations, while $H$-band magnitudes are calibrated with 2MASS.
Integral field unit (IFU) spectroscopic data obtained using the Multi Unit Spectroscopic Explorer (MUSE) on the Very Large Telescope (VLT; \citealt{Bacon2010}) provide the deepest view of the inner $1'\times 1'$ region centred around the QSO (PID$=$0104.A-0147; PI: Chen).
The limiting magnitude is $m_{r, {\rm AB}}\approx 26$ in the pseudo-$r$ band ($6000-7500$ \r{A}).
Most galaxies in the CUBSz1 sample are too faint to be detected in the DES images, but are detected in the MUSE data (Figure \ref{fig:muse_imgs}).

For the redshift survey, the deepest component is completed using available IFU data from MUSE observation in the inner $1'\times 1'$ region. 
At $z\!\approx1$, faint galaxies down to stellar mass of $\mstar\!\approx\! 10^9~\msun$ ($m_{r, {\rm AB}}\!=\!25$ at $z\!\approx\!1$) can be identified at projected distance $d_{\rm proj} \!\lesssim\!250$ kpc from the background QSO.
Complementary multi-slit spectroscopic observations for larger field of views were obtained using LDSS3 and IMACS \citep{Osip:2008aa, Dressler2011} on the Magellan Telescopes to reach a limiting magnitude of $m_{r, {\rm AB}}\!=\!24$ and $22.5$, respectively.
These wide-field survey data enable investigations of large-scale structures beyond the MUSE footprints to $\approx 1-5$ Mpc.

Here, we briefly summarize the methodology for redshift measurements.
A complete galaxy catalogue of the ongoing CUBS program will be presented in a separate paper.
For each object spectrum, independent redshift measurements were obtained by two team members together with associated quality flags.
The most secure redshift measurement is determined by at least two spectral features (e.g., the [\ion{O}{II}] emission and/or the \ion{Ca}{II} H\&K absorption doublets).
There are also single-line redshift determinations, which were attributed to either H$\alpha$ or [\ion{O}{II}] $\lambda\lambda 3727, 3729$ emission lines if the [\ion{O}{II}] doublet is not resolved in the MUSE field.
Spectra with no apparent spectral features were not assigned a redshift.
A third observer compared the two independent redshift measurements together with their quality flags, and consolidated all available information to reach a final redshift assignment and a quality flag for each object.

\begin{table*}
    \caption{Summary of galaxies and absorption}
    \begin{center}
    \begin{tabular}{rrrrrrrrrrr}
    \hline
    \hline
ID & QSO & $z$ & $d_{\rm proj}$ & $(u-r)_{\rm rest}$ & $M_{r}$ & $\log (M_{\rm star}/M_\odot)$ & SED SFR & $d_{\rm proj}$ & $\log \psi$ & Type$^a$ \\
 & & & (kpc) & (mag) & (mag) & (dex) & ($M_\odot~{\rm yr^{-1}}$) & $/r_{\rm virial}$ \\
 (1) & (2) & (3) & (4) & (5) & (6) & (7) & (8) & (9) & (10) & (11)\\
\hline
s1 & J2245 & 0.8908 & 158 & $0.22_{-0.09}^{+0.07}$ & $-18.75\pm 0.30 $ & $8.14_{-0.09}^{+0.09}$ & $0.5_{-0.1}^{+0.2}$ & $2.60$ & $8.64$ &  E \\
s2 & J0114 & 0.8963 & 55 & $0.61_{-0.07}^{+0.07}$ & $-18.97\pm 0.19 $ & $8.89_{-0.12}^{+0.15}$ & $1.3_{-0.6}^{+1.1}$ & $0.71$ & $9.42$ & E \\
&  & 0.8991 & 155 & $1.24_{-0.04}^{+0.04}$ & $-20.51\pm 0.06 $ & $9.82_{-0.08}^{+0.10}$ & $7.4_{-2.7}^{+4.6}$ & $1.55$ & $9.31$ & E/A \\
&  & 0.8995 & 146 & $2.85_{-0.02}^{+0.02}$ & $-21.33\pm 0.03 $ & $10.87_{-0.02}^{+0.02}$ & $<0.3$ & $0.80$ & $10.11$ &  E/A \\
s3 & J2339 & 0.9182 & 172 & $0.79_{-0.08}^{+0.05}$ & $-19.16\pm 0.18 $ & $8.93_{-0.09}^{+0.11}$ & $3.5_{-0.8}^{+0.9}$ & $2.21$ & $8.95$ & E\\
&  & 0.9201 & 368 & $0.61_{-0.09}^{+0.08}$ & $>-17.8$ & $8.31_{-0.22}^{+0.21}$ & $0.6_{-0.3}^{+0.4}$ & $5.74$ & $8.36$ & E \\
s4 & J2339 & 0.9311 & 192 & $0.52_{-0.08}^{+0.06}$ & $-19.43\pm 0.21 $ & $8.77_{-0.06}^{+0.11}$ & $2.4_{-0.5}^{+0.7}$ & $2.59$ & $8.84$ & E/A \\
s5 & J2245 & 0.9348$^b$ & 251 & $0.70_{-0.13}^{+0.16}$ & $>-20.1$ & $9.11_{-0.20}^{+0.20}$ & $1.3_{-0.5}^{+1.3}$ & $3.08$ & $8.85$ & E \\
&  & 0.9350 & 221 & $0.67_{-0.12}^{+0.13}$ & $-21.12\pm 0.07 $ & $9.31_{-0.04}^{+0.11}$ & $7.5_{-0.9}^{+2.6}$ & $2.57$ & $8.97$ & E/A \\
&  & 0.9350$^b$ & 245 & $0.42_{-0.06}^{+0.06}$ & $>-20.1$ & $8.59_{-0.20}^{+0.20}$ & $0.4_{-0.2}^{+0.4}$ & $3.49$ & $8.67$ & E \\
&  & 0.9352 & 151 & $0.35_{-0.05}^{+0.04}$ & $-20.16\pm 0.15 $ & $8.90_{-0.07}^{+0.09}$ & $3.2_{-0.5}^{+0.9}$ & $1.96$ & $9.00$ & E \\
s6 & J0333 & 0.9373 & 72 & $0.79_{-0.04}^{+0.05}$ & $-21.04\pm 0.16 $ & $9.95_{-0.09}^{+0.08}$ & $5.6_{-1.3}^{+2.6}$ & $0.70$ & $9.70$ & E/A \\
&  & 0.9381 & 181 & $1.36_{-0.04}^{+0.04}$ & $-20.79\pm 0.09 $ & $10.02_{-0.08}^{+0.10}$ & $10.7_{-3.7}^{+7.6}$ & $1.71$ & $9.34$ & E/A \\
s7 & J0114 & 0.9520 & 88 & $0.73_{-0.13}^{+0.15}$ & $>-19.1$ & $8.84_{-0.20}^{+0.21}$ & $0.7_{-0.3}^{+0.6}$ & $1.17$ & $9.21$ & E \\
&  & 0.9539 & 232 & $0.35_{-0.13}^{+0.14}$ & $>-19.1$ & $8.50_{-0.18}^{+0.26}$ & $0.9_{-0.2}^{+0.4}$ & $3.42$ & $8.65$ & E/A \\
s8 & J0154 & 0.9656 & 185 & $1.08_{-0.08}^{+0.08}$ & $-18.36\pm 0.35 $ & $9.08_{-0.12}^{+0.12}$ & $0.9_{-0.3}^{+0.7}$ & $2.31$ & $8.98$ & E/A \\
s9 & J0114 & 0.9715 & 74 &  $0.60_{-0.06}^{+0.07}$ & $-19.87\pm 0.12 $ & $9.23_{-0.13}^{+0.17}$ & $3.0_{-1.3}^{+2.0}$ & $0.89$ & $9.43$ & E \\
s10 & J0333 & 0.9763 & 192 & $0.86_{-0.08}^{+0.08}$ & $-19.36\pm 0.27 $ & $9.02_{-0.15}^{+0.16}$ & $3.1_{-1.5}^{+1.1}$ & $2.44$ & $8.94$ & E \\
&  & 0.9772 & 232 & $0.83_{-0.03}^{+0.03}$ & $-20.83\pm 0.12 $ & $9.70_{-0.06}^{+0.09}$ & $18.6_{-4.4}^{+2.6}$ & $2.45$ & $9.10$ & E/A \\
&  & 0.9772 & 341 & $0.79_{-0.04}^{+0.04}$ & $-19.85\pm 0.25 $ & $9.59_{-0.09}^{+0.10}$ & $5.0_{-2.1}^{+3.6}$ & $3.71$ & $8.89$ & E \\
s11 & J0154 & 0.9813 & 597 & $0.75_{-0.03}^{+0.03}$ & $-21.18\pm 0.06 $ & $9.86_{-0.09}^{+0.10}$ & $11.5_{-5.2}^{+6.4}$ & $6.00$ & $8.76$ & E \\
&  & 0.9816 & 224 & $0.78_{-0.02}^{+0.03}$ & $-20.84\pm 0.07 $ & $9.70_{-0.09}^{+0.10}$ & $10.8_{-5.6}^{+4.7}$ & $2.36$ & $9.12$ & E/A \\
&  & 0.9821 & 117 & $1.11_{-0.16}^{+0.16}$ & $>-19.1$ & $8.72_{-0.24}^{+0.19}$ & $0.5_{-0.2}^{+0.5}$ & $1.62$ & $9.04$ & E \\
s12 & J0333 & 0.9974 & 263 & $2.02_{-0.14}^{+0.13}$ & $>-19.1$ & $9.75_{-0.19}^{+0.19}$ & $4.5_{-1.7}^{+1.9}$ & $2.75$ & $9.07$ & E \\
&  & 0.9977 & 249 & $0.49_{-0.05}^{+0.05}$ & $>-19.1$ & $9.01_{-0.10}^{+0.11}$ & $3.5_{-1.0}^{+0.9}$ & $3.19$ & $8.83$ & E \\
&  & 0.9979 & 270 & $0.85_{-0.07}^{+0.06}$ & $-19.85\pm 0.19 $ & $9.17_{-0.09}^{+0.11}$ & $6.3_{-1.4}^{+1.8}$ & $3.31$ & $8.85$ & E \\
&  & 0.9988 & 145 & $1.28_{-0.15}^{+0.16}$ & $>-19.1$ & $9.28_{-0.18}^{+0.22}$ & $2.0_{-0.9}^{+2.1}$ & $1.73$ & $9.16$ & E \\
s13 & J2339 & 1.0032 & 105 & $1.94_{-0.01}^{+0.01}$ & $-23.54\pm 0.01 $ & $11.47_{-0.01}^{+0.01}$ & $<0.3$ & $0.19$ & $11.79$ & E/A \\
&  & 1.0038 & 245 & $0.92_{-0.04}^{+0.04}$ & $-20.30\pm 0.11 $ & $9.48_{-0.16}^{+0.17}$ & $9.0_{-3.9}^{+3.0}$ & $2.77$ & $9.00$ & E \\
s14 & J2308 & 1.0060 & 270 & $1.41_{-0.12}^{+0.12}$ & $-19.60\pm 0.37 $ & $9.61_{-0.18}^{+0.17}$ & $2.8_{-1.2}^{+1.9}$ & $2.94$ & $9.01$ & E \\
&  & 1.0068 & 175 & $0.93_{-0.08}^{+0.09}$ & $>-18.8$ & $9.02_{-0.12}^{+0.14}$ & $0.9_{-0.3}^{+0.8}$ & $2.24$ & $8.98$ & E \\
s15 & J0154 & 1.0116 & 60 & $1.53_{-0.03}^{+0.02}$ & $-23.08\pm 0.03 $ & $10.87_{-0.02}^{+0.02}$ & $4.0_{-0.5}^{+0.4}$ & $0.35$ & $10.49$ & E/A \\
&  & 1.0123 & 189 & $0.88_{-0.03}^{+0.02}$ & $-21.67\pm 0.04 $ & $9.90_{-0.06}^{+0.12}$ & $34.0_{-5.3}^{+6.7}$ & $1.89$ & $9.28$ & E/A \\
s16 & J2308 & 1.0375 & 177 & $0.49_{-0.09}^{+0.10}$ & $-19.13\pm 0.47 $ & $9.07_{-0.19}^{+0.18}$ & $1.1_{-0.1}^{+0.2}$ & $2.26$ & $9.00$ & E\\
s17 & J0333 & 1.0550 & 106 & $1.04_{-0.16}^{+0.17}$ & $-19.25\pm 0.38 $ & $9.00_{-0.21}^{+0.21}$ & $0.9_{-0.4}^{+0.9}$ & $1.38$ & $9.20$ & E/A \\
&  & 1.0551 & 206 & $0.91_{-0.10}^{+0.10}$ & $-18.83\pm 0.35 $ & $8.72_{-0.13}^{+0.15}$ & $2.1_{-0.6}^{+0.8}$ & $2.92$ & $8.81$ & E \\
&  & 1.0553 & 187 & $1.23_{-0.03}^{+0.02}$ & $-21.77\pm 0.05 $ & $10.17_{-0.06}^{+0.10}$ & $58.7_{-6.7}^{+7.2}$ & $1.71$ & $9.42$ & E/A \\
&  & 1.0570 & 192 & $0.66_{-0.05}^{+0.05}$ & $>-19.5$ & $9.30_{-0.11}^{+0.13}$ & $4.6_{-2.2}^{+1.9}$ & $2.31$ & $9.05$ & E \\
s18 & J0333 & 1.0631 & 109 & $0.93_{-0.07}^{+0.04}$ & $-21.46\pm 0.09 $ & $9.86_{-0.07}^{+0.08}$ & $31.2_{-5.4}^{+6.3}$ & $1.12$ & $9.51$ & E/A \\
s19 & J2339 & 1.1146 & 157 & $0.51_{-0.05}^{+0.07}$ & $-20.12\pm 0.15 $ & $9.07_{-0.06}^{+0.06}$ & $4.6_{-0.6}^{+0.8}$ & $2.04$ & $9.06$ & E \\
s20 & J0154 & 1.2133 & 160 & $1.63_{-0.07}^{+0.07}$ & $-19.89\pm 0.14 $ & $9.83_{-0.11}^{+0.11}$ & $7.7_{-2.8}^{+5.6}$ & $1.72$ & $9.36$ & E\\
\hline
    \end{tabular}
    \end{center}
    \begin{flushleft}
    $^a$ Galaxy spectral types were classified based on the incidence of emission (E) and absorption (A) features, and E/A indicates galaxies displaying both features. 
    \label{tab:gal_sample}
    $^b$ These two galaxies are separated by $1.06''$, which cannot be resolved in the Magellan H band image. The reported lower limits on $M_r$ were derived from the total $H$ band magnitude. 
    \end{flushleft}

\end{table*}

\subsection{Galaxies at $\bm{z\approx1}$}

In the CUBSz1 sample, the galaxies were selected based on their close proximity projected around the QSO sightline ($d_{\rm proj} < 200$ kpc), without reference to the presence or absence of absorption features in the QSO spectroscopy.
A radius of 200 kpc corresponds to $\approx 1.5$ $r_{\rm virial}$ of an $L_*$ galaxy at $z\approx1$.
In this sample construction, we only consider galaxies with the most secure redshift determined by at least two spectral features (see Section \ref{sec:cubs_data}).
A lower limit of galaxy redshift at $z=0.883$ was imposed to ensure the coverage of \ion{He}{I} in the {\it HST}/COS FUV spectrum.
Considering the roughly constant helium abundance from $z=2-3$ to $z=0$ \citep{Cooke:2018aa}, \ion{He}{I} is an alternative tracer of the total gas column density (after an ionization correction; see Section \ref{sec:obs_he1} and \ref{sec:dis_h1_he1}), when Ly$\alpha$ and Ly$\beta$ are not covered by the {\it HST}/COS FUV spectra.

\begin{figure*}
\begin{center}
\includegraphics[width=0.98\textwidth]{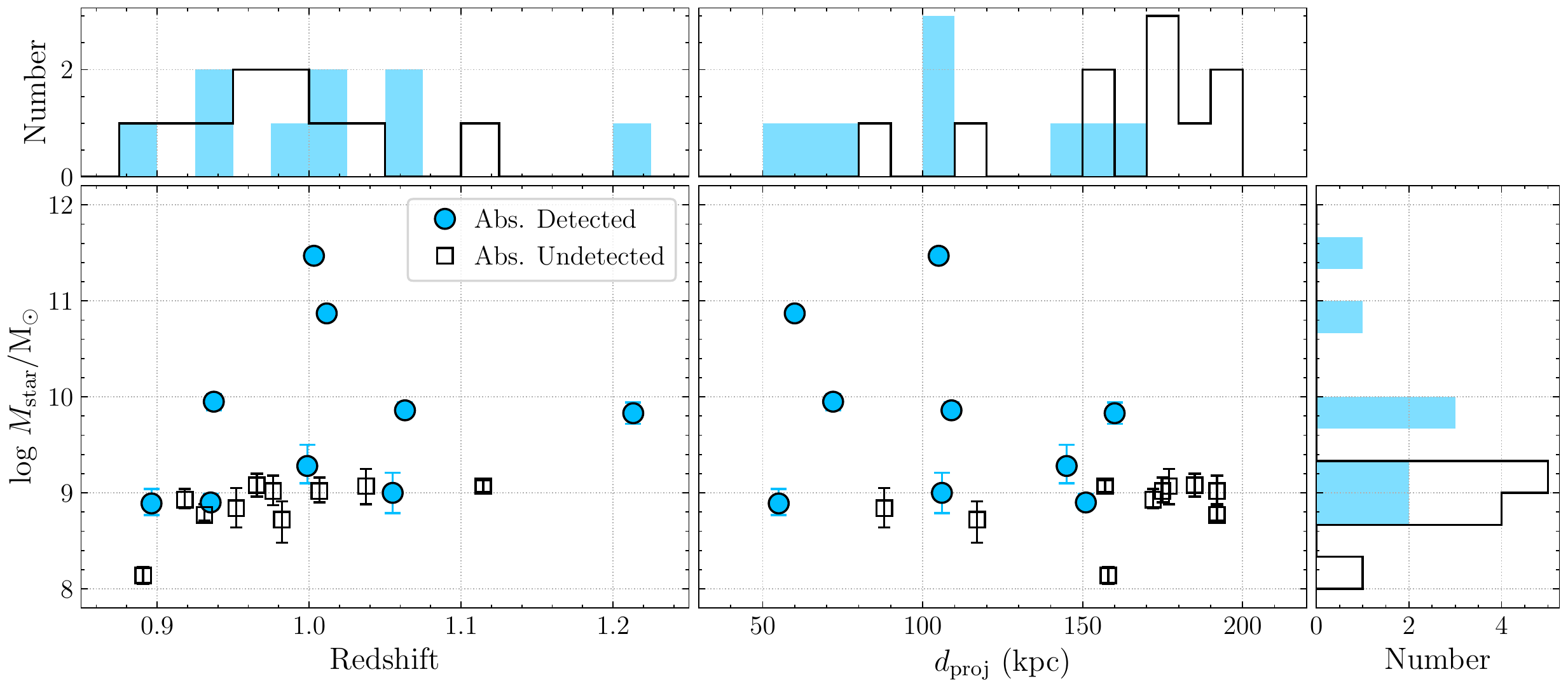}
\end{center}
\caption{Summary of the general characteristics of the CUBSz1 sample based on the closest galaxy in individual systems, including stellar mass ($M_{\rm star}$), redshift, and projected distance ($d_{\rm proj}$). 
Galaxy halos with and without absorption detected are highlighted by closed and open symbols, respectively.
The \textit{left panel} shows the distribution of stellar mass versus redshift for the closest galaxy in the CUBSz1 sample, while
the \textit{right panel} shows the distribution of stellar mass versus projected distance from the same galaxies.
Histograms at the top and right summarize the distributions in redshift, projected distance, and stellar mass, respectively.
Blue-filled histograms are for systems with absorption features, while black-open histograms are for non-detections.
}
\label{fig:gal_pro}
\end{figure*}

Within the six CUBS fields (i.e., J0114, J0154, J0333, J2245, J2308, and J2339), 26 galaxies at $d_{\rm proj} < 200$ kpc and $z = 0.89-1.21$ were securely identified.
In Figure \ref{fig:muse_imgs}, we show the three-color MUSE images using pseudo-$g$ ($4800\,-\,5800$ \r{A}), pseudo-$r$ ($6000\,-\,7500$ \r{A}), and pseudo-$i$ ($7500\,-\,9000$ \r{A}).
For these galaxies, we also searched for neighbouring galaxies with line-of-sight velocity $|\Delta v| \lesssim 500 \kms$ and projected distance of $\lesssim 1$ Mpc, and identified them as part of a galaxy group.
This exercise led to 20 unique galaxy systems (i.e., either an isolated galaxy or a galaxy group) with a total of 41 member galaxies at $z=0.89-1.21$.
The typical number of galaxy members in a galaxy system is between 1 and 3.

In Table \ref{tab:gal_sample}, we summarize all galaxies in the CUBSz1 sample.
All 20 galaxy systems are sorted by their redshifts, and system IDs are assigned accordingly.
For each galaxy, we report system ID, CUBS QSO field, redshift, projected distance, rest-frame $u-r$ colour,  rest-frame $r$ band absolute magnitude, stellar mass ($M_{\rm star}$) predicted by the SED modelling (see Section 3.1), predicted average star formation rate (SFR) over the past 300 Myr, virial-normalized distance, projected gravitational potential (details in \ref{sec:obs_inc}), and spectral type in columns (1) through (11).
The rest-frame colour $(u-r)_{\rm rest}$ was calculated from a SED fitting exercise (see details in Section \ref{sec:med_gal}). 
The absolute rest-frame $r$ band magnitude was calculated using the Magellan $H$ band magnitude, with non-detections represented by lower limits.
Galaxy spectra are presented in the Appendix (Fig. \ref{fig:g1} to \ref{fig:g20}), and the galaxy spectral types are assigned as absorption-dominated (A), emission-dominated (E), or composite (A/E) using detected spectral features.

In Figure \ref{fig:gal_pro}, we show the sample distributions between redshift, projected distance, and stellar mass for isolated galaxies and, in cases of a group, the closest member galaxy.
Particularly, in the CUBSz1 sample, the galaxies with the smallest $d_{\rm proj}$ also have the smallest $d_{\rm proj}/r_{\rm virial}$ (see Table \ref{tab:gal_sample}).
The limiting stellar mass of the CUBSz1 sample is $\log (M_{\rm star}/{\rm M_\odot}) \approx 9$.
For five of the 20 galaxy systems, the closest galaxies have stellar mass $\log (M_{\rm star}/{\rm M_\odot}) \gtrsim 10$.
In Figure \ref{fig:sm_sfr}, we compare the stellar mass-SFR correlation for the CUBSz1 sample with the star-forming main sequence.
The CUBSz1 galaxies span a range in SFR from $\lesssim 0.3$ to 60 $\rm M_\odot~yr^{-1}$.
As shown in Figure \ref{fig:sm_sfr}, the stellar mass-SFR relationship of the CUBSz1 sample is consistent with the star-forming main sequence at $z \approx 1$, which has SFR about one order of magnitude higher than the local Universe.
Therefore, most CUBSz1 galaxies are star-forming $L^*$ and sub-$L^*$ galaxies at $z\approx 1$.
If only considering the closest galaxies, only two galaxies are qualified as passive galaxies (i.e., sSFR $< 10^{-10}~\rm yr^{-1}$ at $z\approx 1$; Systems s13 and s15).
If considering all galaxies in each group, another System s2 contain one passive galaxy.

\subsection{Absorption properties in the CUBSz1 sample}
\label{sec:sample_abs}
We search for associated absorption features within $|\Delta v| \lesssim 500 \kms$ (i.e., about twice the escape velocity of $L^*$ galaxies) of all individual galaxies in the CUBSz1 sample.
Nine galaxies or galaxy groups host confirmed absorption components with multiple kinematically aligned transitions, while ten galaxy systems do not have detectable transitions (with $2\sigma$ limiting column densities of $\log N(\ion{He}{I})/{\rm cm^{-2}} \lesssim 13.5$ or $\log N(\ion{O}{V})/{\rm cm^{-2}} \lesssim 13.3$).
Assuming cool gas has a typical density of $\log n_{\rm H}/{\rm cm^{-3}} \approx -3$ and is photoionized by the HM05 UVB, we infer a corresponding limiting total hydrogen column density of $\log N/{\rm cm^{-2}} \approx 18$ based on the limiting column densities of both \ion{H}{I} and \ion{O}{V}. Accordingly, we divide the galaxy sample into two groups, systems with detected absorption and those without for the subsequent analysis (see \S\ \ref{sec:obs_inc}).

One special system is System s9 at $z=0.9715$ in the J0114 field, which may have metal transitions due to \ion{O}{III}, \ion{O}{IV}, and \ion{O}{V}.
However, all of the transitions are contaminated by other features (Figure \ref{fig:a9}).
As a result, no robust constraints can be obtained for this system, which is excluded from subsequent analyses.
Available COS and MIKE spectra cover a wide range of ionic transitions from \ion{S}{II} and \ion{O}{II} to \ion{Ne}{VIII}.

In Figures \ref{fig:gal_pro} and \ref{fig:sm_sfr}, we divide the CUBSz1 sample into two sub-samples, galaxies displaying absorption features versus those with no absorption found.
It is clear that absorption systems are preferentially detected in sightlines projected within 160 kpc.
Specifically, nine of 13 galaxies at $d_{\rm proj}< 160$ kpc exhibit associated absorption features, whereas zero out of six at larger distances show any.
At the same time, absorption features are more frequently found around more massive galaxies.
In particular, five of 13 galaxy systems at $d_{\rm proj}< 160$ kpc are more massive than $\log M_{\rm star}/{\rm M_\odot} \approx 9.5$, and all five exhibit absorption features.
The influence of the galaxy environments on the absorption features will be discussed in Section \ref{results}.

\section{Analysis}
\label{analysis}
Investigating the impact of the galaxy environment on the CGM requires knowledge of both gas properties (e.g., column density, kinematics, and temperature) and galaxy properties (e.g., stellar mass and star formation history).
The detailed absorption analysis has been presented in \citetalias{CUBSV}, while details of the galaxy SED fitting will be presented in a separate paper.
Here, we briefly summarize the absorption analysis to obtain the gas properties and the galaxy SED modelling.

\subsection{Galaxy SED modelling and halo properties}
\label{sec:med_gal}

\begin{figure}
\begin{center}
\includegraphics[width=0.47\textwidth]{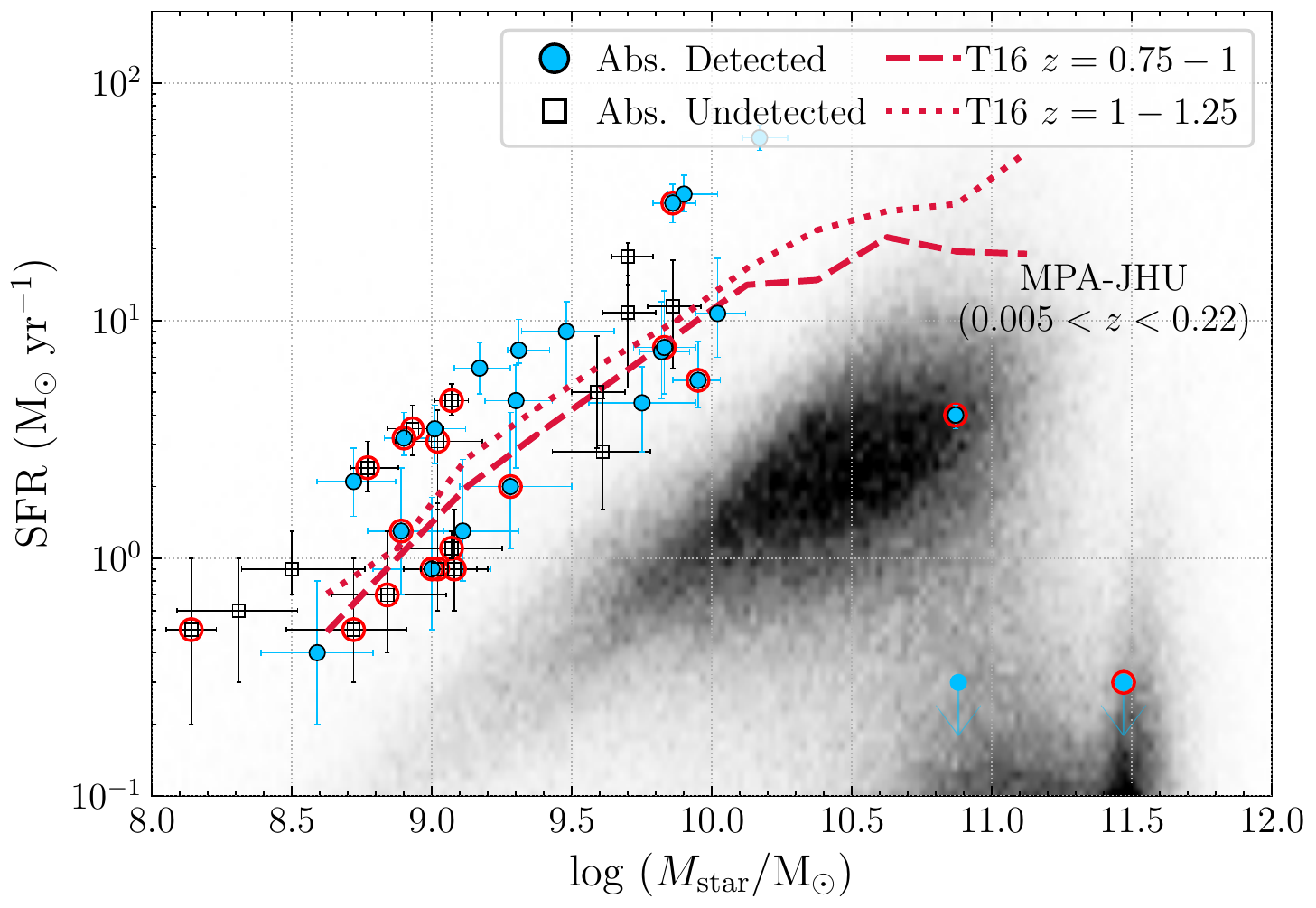}
\end{center}
\caption{
Comparisons of the CUBSz1 sample with the general galaxy populations in different epochs in the star formation rate versus stellar mass parameter space.
In this plot, all CUBSz1 galaxies are shown, with the closest galaxies marked by open red circles.
The star-forming main sequence at $z\approx 1$, shown as dashed and dotted lines, is adopted from \citet[][ designated as T16]{Tomczak:2016aa}.
The SFR in the T16 sample is derived based on FUV and IR continuum, which traces the star formation activities on time scales of $\approx 100$ Myr.
Therefore, it can be directly compared with our SED-based SFR estimates.
It is clear that the CUBSz1 sample consists primarily of star-forming galaxies at $z\approx 1$.
As a comparison, the grey-shaded region represents the star-forming main sequence using galaxies in the MPA-JHU catalogue at $0.005 < z < 0.22$ \citep[][adopting the emission line-based SFR]{Brinchmann:2004aa}.
Galaxies at $z\approx1$ exhibit on average higher star formation rates than galaxies in the local Universe.}
\label{fig:sm_sfr}
\end{figure}

To characterize the galaxy properties, we first infer stellar masses and SFRs for individual galaxies by fitting the observed broad-band SEDs using the Bayesian Analysis of Galaxies for Physical Inference and Parameter EStimation package (BAGPIPES; \citealt{Carnall:2018aa}).
BAGPIPES generates model SEDs by synthesizing a series of simple stellar populations (SSPs), which are formed following an assumed star formation history.
The SSP evolution tracks come from the latest 2016 release of \citet{Bruzual:2003aa}, which sample models across a wide range in stellar age and metallicity.
We adopt the \citet{Kroupa:2001aa} initial mass function for this exercise.
The model spectrum is further attenuated by interstellar dust \citep{Calzetti:2000aa} and the IGM along the line of sight \citep{Inoue:2004aa}.

In BAGPIPES, we adopt a star formation history following ${\rm SFR}\propto t\times e^{-(t/\tau)}$, where $\tau$ is the characteristic $e$-folding timescale of star formation.
This function is a modification of the more commonly adopted closed-box model that follows a simple exponential decline star formation rate history.
The inclusion of another factor of $t$ is designed to accommodate the expectation of an initial rise in star formation activities in early times.

The posterior distribution of the SED modelling is obtained by matching the BAGPIPES models with the measured optical and infrared photometry (i.e., $g$, $g'$, $r$, $r'$, $i$, $z'$, and $H$; see Section \ref{data}), from which we infer a best-fit stellar mass and average SFR within the last 300 Myr with their 16th and 84th percentiles.
These best-fit quantities are presented in Table \ref{tab:gal_sample}.

The halo mass of each individual galaxy is calculated based on the stellar mass using a stellar mass-halo mass (SMHM) relation at $z\approx 1$, adopted from \citet{Behroozi:2019aa} after accounting for missing light as described in  \citet{Kravtsov:2018aa}. 
The halo radius ($r_{\rm virial}$) is approximated using $r_{200}$, within which the mean dark matter density is 200 times the critical density.
A detailed study of the galactic environment is presented in Section \ref{sec:obs_inc}.

\begin{figure*}
\begin{center}
\includegraphics[width=0.98\textwidth]{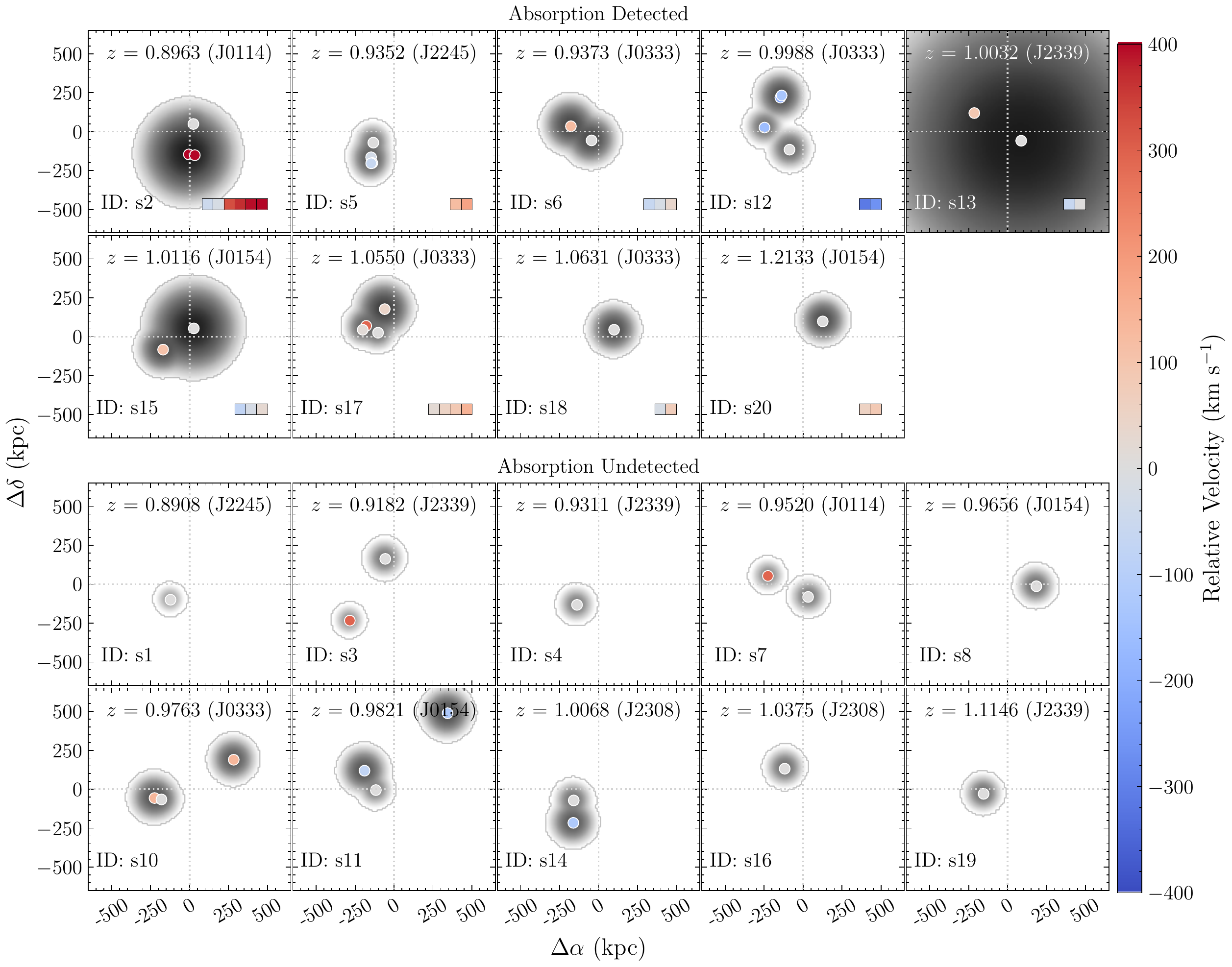}
\end{center}
\caption{Overview of the galactic environment around galaxy systems with absorption detected (top nine panels) and without (lower ten panels).
Velocity offsets of individual galaxy members relative to the redshift of the galaxy closest to the sightline are shown in colour-coded circles.
To visualize the halo extent of each galaxy, we overlay a 2D Gaussian with FWHM $= r_{\rm virial}$, and amplitude is scaled according to the inferred halo mass.
The resulting grayscale provides a qualitative assessment of the surface mass density of the dark matter, which is proportional to the total gravitational potential summed for all dark matter halos.
The presence of resolved absorption components is indicated by squares in the bottom-right corner of each panel.
These squares are colour-coded according to the component line-of-sight velocity offset from the galaxy closest to the QSO sightline as well.
}
\label{fig:gal_environment}
\end{figure*}

\subsection{Absorption analysis}
\label{sec:med_abs}
For each galaxy system, all absorption features in QSO spectra are decomposed into individual components, which is enabled by the available medium-resolution FUV COS spectrum and high-resolution optical MIKE spectrum (\citealt{Zahedy:2021aa, Cooper:2021aa}, \citetalias{CUBSV}).
The decomposition is guided by strong transitions such as \ion{H}{I}, \ion{Mg}{II}, and \ion{O}{IV}.
Kinematically aligned ionic transitions with consistent line centroids are grouped together for the purpose of constraining the ionization state of each component.
There are a total of 26 absorption components with the simultaneous presence of low-to-intermediate ionization species in nine galaxy systems, and additionally, three components were identified based on the presence of highly ionized \ion{Ne}{VIII} in three galaxy systems (see Figure \ref{fig:a1} to Figure \ref{fig:a20}).

For each resolved ionic transition, we perform a Voigt profile analysis to determine the column density ($N$), line centroid ($v$), and Doppler width ($b$).
These model parameters are optimized under a Bayesian framework implemented using \texttt{emcee} (\citealt{Foreman-Mackey:2013aa}; the likelihood function and details are described in \citetalias{CUBSV} Section 3.1).
The results are summarized in Table \ref{tab:ions} for all detected ions.
In this table, ions belonging to the same absorption component are shown together and sorted according to the atomic number and ionization state of each ion.

In \citetalias{CUBSV} and this study, we focus on the low-to-intermediate ionization phases (i.e., cool gas; $\log T/{\rm K} \approx 4$), which have multiple ions for constraining the gas properties.
Among the 26 low-to-intermediate ionization state absorption components, 20 have multiple ions detected, which enables photoionization modelling.
A three-dimensional grid of photoionization models (i.e., the density, hydrogen column density, and metallicity) is constructed using Cloudy (v17; \citealt{Ferland:2017RMxAA}), assuming photoionization equilibrium (PIE) with the HM05 ultraviolet background (UVB; an updated version of \citealt{Haardt:2001aa} in Cloudy; see \citealt{Zahedy:2021aa} for comparison with recent UVB in \citealt{Khaire:2019aa} and \citealt{Faucher-Giguere:2020aa}).
For each absorption component, the photoionization model grid is searched to find gas parameters that match the relative column density ratio between multiple ions (see details in \citetalias{CUBSV} Section 3.2).
The derived gas densities are reported in Table \ref{tab:ions}.
A single-phase model is a good representation of all but four absorption components.
For these four systems, where a single-phase model failed, a two-phase model is found to be sufficient to produce observed ion ratios \citepalias{CUBSV}.

\begin{figure*}
\begin{center}
\includegraphics[width=0.48\textwidth]{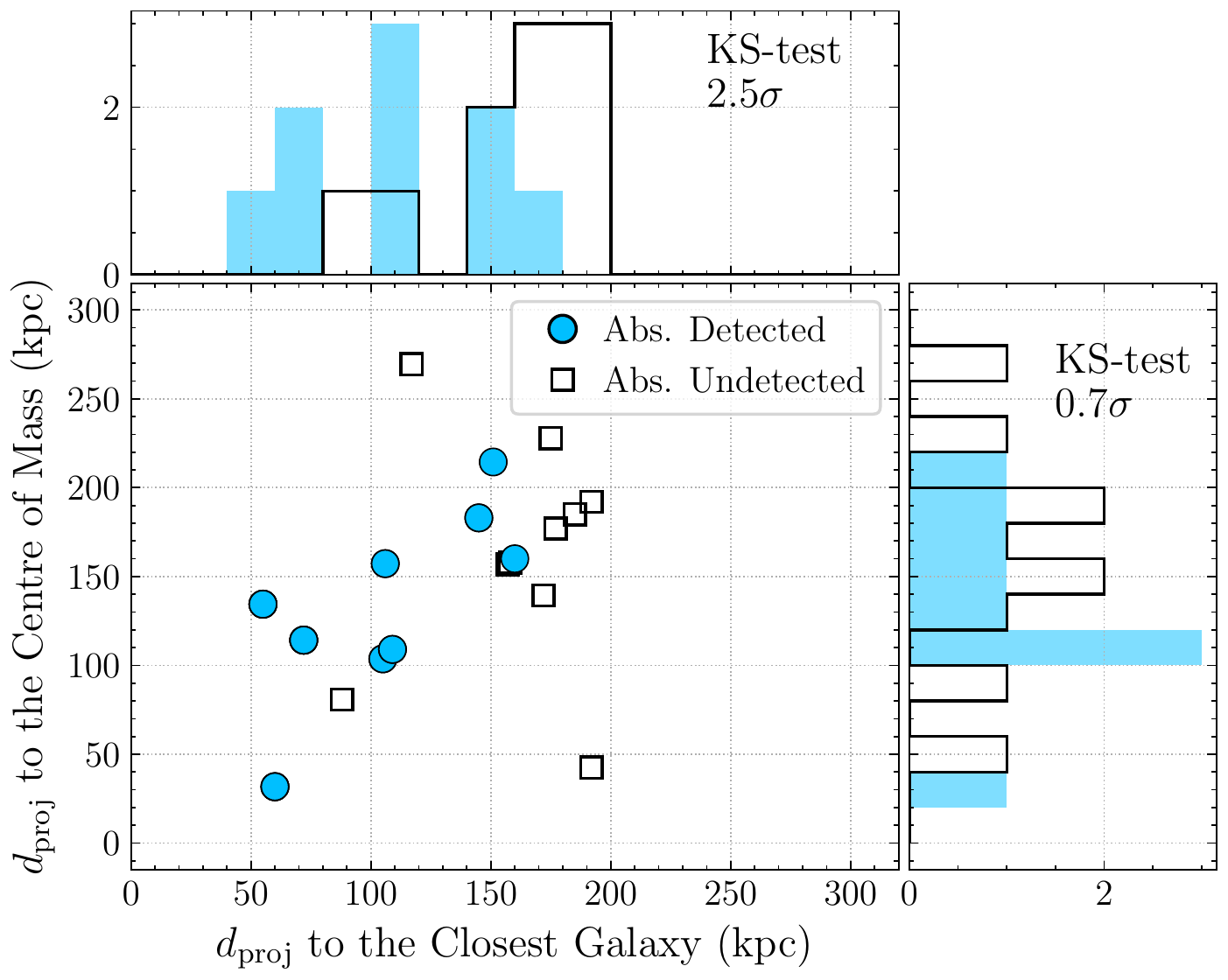}
\includegraphics[width=0.48\textwidth]{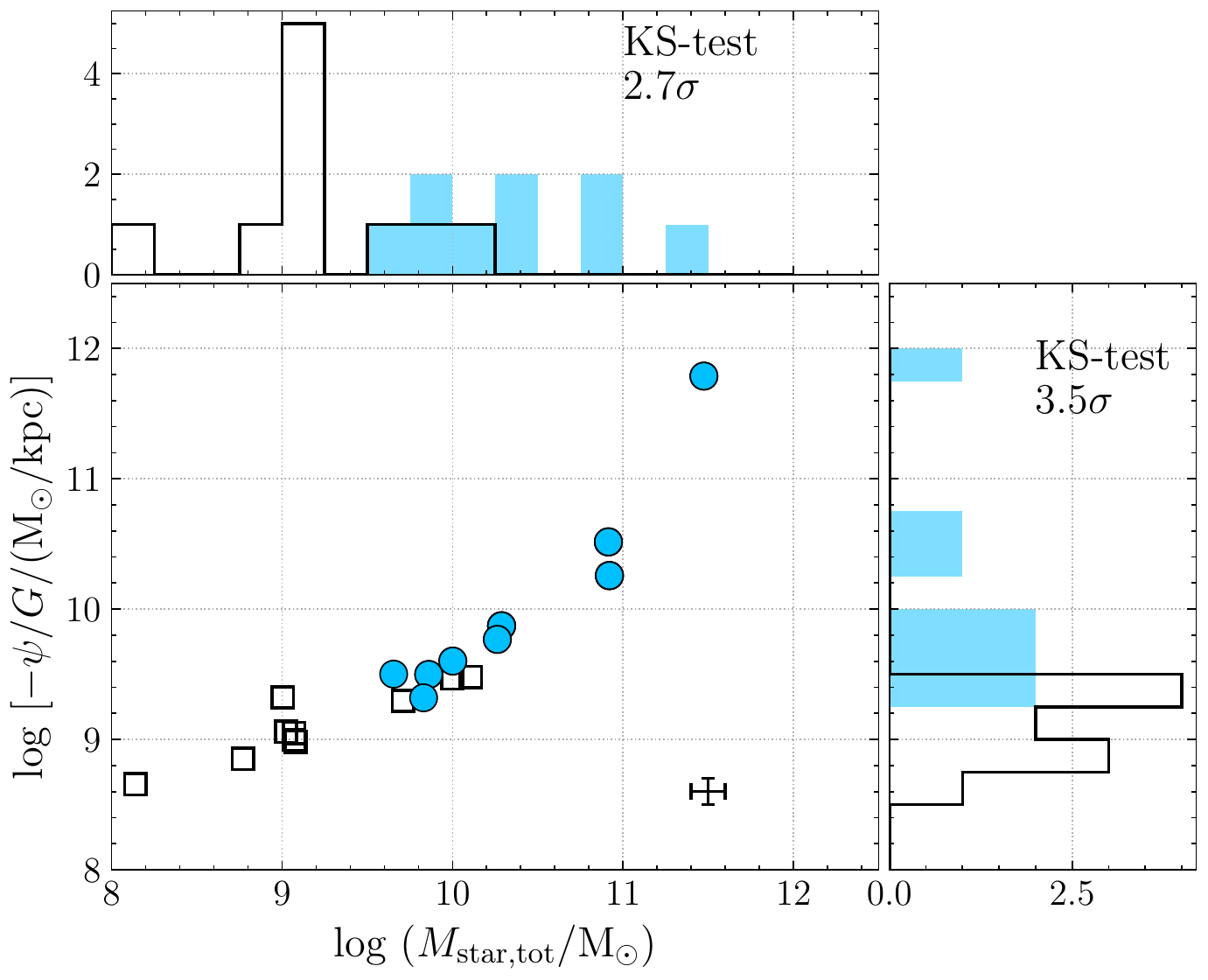}
\end{center}
\caption{Summary of absorbing gas presence and absence versus galaxy properties.
Symbols are the same as Figure \ref{fig:gal_pro}.
{\it Left panel:} the presence of absorbing gas dependence on the projected distance to the closet galaxy and the projected distance to the centre of mass of each galaxy system.
The absorbing gas in the CGM is more correlated with the closest galaxy ($2.5 \sigma$) than with the system centre of mass ($0.7 \sigma$), which suggests the closest galaxy has a larger impact than the other galaxies.
Specifically for the CUBSz1 sample, the physically closest galaxies are also galaxies with the smallest $r_{\rm virial}$-normalized distance.
{\it Right panel:} absorbing gas dependence on the total stellar mass and the projected gravitational potential (defined in Equation 2).
The uncertainty introduced by the stellar mass estimation and the SMHM relationship is shown as a cross in the low right corner.
Among these four galaxy characteristics, the projected gravitational potential is the most distinguishable between absorption and non-absorption systems with a significance of $3.5\sigma$.
}
\label{fig:comp_env}
\end{figure*}

Comparing the observed line widths of different elements allows us to infer the gas temperature and the underlying non-thermal motions.
Light elements have larger thermal velocities than heavy ones, while the non-thermal motion is assumed to be the same for all elements sharing a common line centroid.
Considering the resolutions of COS and MIKE, constraints on the temperature and non-thermal motion are most robust when comparing hydrogen or helium with metal ions (e.g., oxygen and magnesium, \citealt{Zahedy:2021aa}).
Again, a Bayesian framework is used to obtain the best-fit constraints (see details in \citetalias{CUBSV} Section 3.3).
As shown in Table 1 of \citetalias{CUBSV}, about half of the components with available density constraints also have constraints on temperatures and non-thermal motions.

\section{Results}
\label{results}
In this section, we examine how the properties of the CGM depend on the galaxy environment, including the incidence of metal-line absorbers, spatial distributions of ionic column densities, and the thermodynamics of the gas.
We also examine whether $N$(\ion{He}{I}) is indeed a good tracer of $N$(\ion{H}{I}).

\subsection{Incidence of absorption and galaxy environment}
\label{sec:obs_inc}

We first examine the incidence of metal-line absorbers for galaxies in different environments.
The high S/N ($\gtrsim 20$) COS and MIKE spectra allow us to place sensitive constraints on the diffuse gas in the foreground of CUBS QSOs.
Specifically, typical $2\sigma$ limiting column densities for two strong transitions \ion{He}{I} $\lambda 584$ and \ion{O}{V} $\lambda 629$ are $\log (N/{\rm cm^{-2}}) \approx 13.5$ and $13.3$, respectively, which is converted into a limiting total hydrogen column density of $\log N/({\rm H}){\rm cm^{-2}} \approx 18$ (Section \ref{sec:sample_abs}).

\begin{figure*}
\begin{center}
\includegraphics[width=0.98\textwidth]{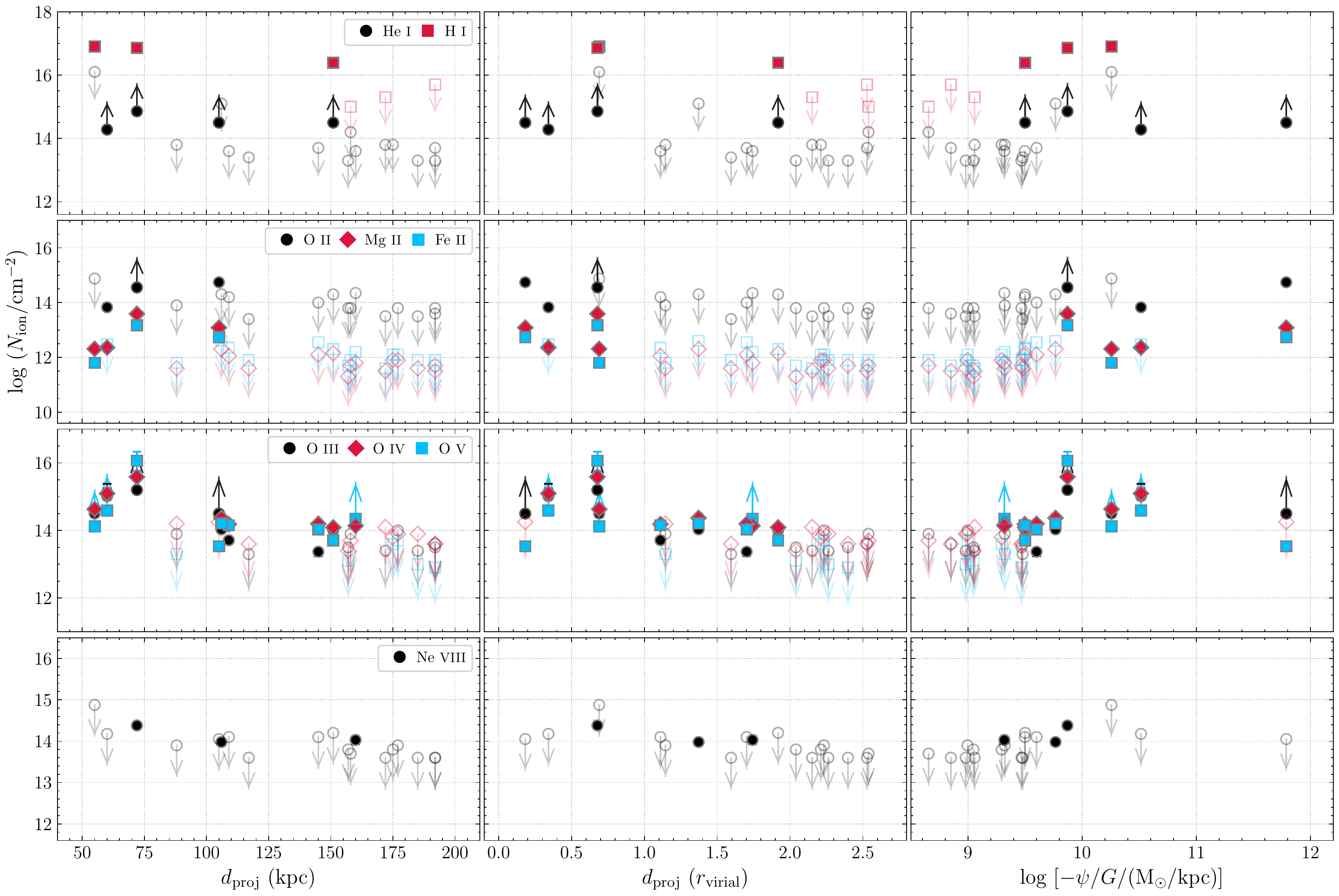}
\end{center}
\caption{The observed column density profiles versus the projected distance to the closest galaxy in units of physical distance (kpc) and $r_{\rm virial}$ in the left and middle panels, respectively.
Their dependences on the projected gravitational potential are displayed in the right panels.
The adopted elements include the neutral species (\ion{H}{I} and \ion{He}{I}), low ionization state (\ion{O}{II}, \ion{Mg}{II}, and \ion{Fe}{II}), intermediate ionization state (\ion{O}{III}, \ion{O}{IV}, and \ion{O}{V}), and \ion{Ne}{VIII} from top to bottom.
In all panels, the $2 \sigma$ upper limits for non-detections are shown as open symbols with downward arrows, and $2\sigma$ lower limits for saturated measurements are marked as open symbols with  upward arrows.
Low ionization species are detected primarily at small $d_{\rm proj}$ and high gravitational potentials $\psi$, while intermediate and high ions are more extended to the halo outskirts and where $\psi$ is low.}
\label{fig:ion_dist}
\end{figure*}

With these detection limits, the CUBSz1 sample is divided into two groups, the systems with detected absorption features and those without. 
In Figure \ref{fig:gal_environment}, we show the galaxy environments for these two groups.
The extent of the dark matter halo around each galaxy is represented by a 2D Gaussian profile, with the full width at half maximum corresponding to the virial radius ($r_{\rm virial}$).
For clarity, the 2D profiles are truncated at twice the virial radius.
The amplitude of the Gaussian profile is scaled according to the inferred halo mass.
In the overlapped regions, direct summations are calculated for all dark matter halos.
Then, the plotted grayscale indicates the surface mass density of all halos, which is proportional to the gravitational potential.
Figure \ref{fig:gal_environment} demonstrates that absorption features are preferentially found in overdense environments  with a larger number of neighbours and a larger number of more massive galaxies.
Nine systems with detected absorption features contain 23 member galaxies, while ten systems containing 17 galaxies show no detectable absorption features.
More quantitatively, we perform a Kolmogorov–Smirnov (KS) test to determine whether the number of members in each system is statistically different between systems with and without absorption features, and find an insignificant difference less than 0.5 $\sigma$ with a $p$-value of 0.58.
The difference of the stellar masses between absorption or non-absorption systems is more significant by $2.2~\sigma$ with a $p$-value of 0.03.

In Figure \ref{fig:comp_env}, we explore a combination of parameters to identify the galaxy properties that are most distinguishable between the absorption and non-absorpiton systems.
These include the projected distance to the closest galaxy, the projected distance to the centre of mass in each system, the total stellar mass, and the projected gravitational potential (defined below).
In this exercise, we search for the properties that best capture the distinction between absorbers and non-absorbers.
In the left panel, we compare two different ways of measuring the proximity of galaxies to the absorbers.
We find that the projected distance to the closest galaxy is more distinguishable with the incidence of absorbing gas than the distance to the mass centre of the galaxy system.
A KS test returns a $p$-value of $1.2\times 10^{-2}$ ($\approx 2.5 \sigma$) and 0.49 ($\approx 0.7 \sigma$) for the closest galaxy and the centre of mass, respectively.
Considering the smallest $d_{\rm proj}/r_{\rm virial}$, we obtain a $p$-value of $1.3\times 10^{-3}$ ($\approx 3.2\sigma$).
Similarly, we also examine the impact of the most massive galaxy in the system using the distance to the most massive galaxies, and found that the incidence of absorption is insensitive to the presence of massive galaxies in the group ($p$-value of 0.54 and $\approx 0.6 \sigma$).
This exercise suggests that the proximity to a galaxy is more deterministic in predicting the incidence of absorbing gas along a QSO sightline than the collective effect of all group members.

However, the strong overlap in $d_{\rm proj}$ between absorbing and non-absorbing galaxy systems in Figure \ref{fig:comp_env} also suggests that the closest galaxy may not be the only factor that drives the observed absorption properties of the CGM.
To explore additional driving factors, we introduce a projected gravitational potential to approximate the maximum gravitational potential along a sightline for each galaxy system.
At a given position along a sightline, the maximum gravitational potential available is the sum of the gravitational potential of each member galaxy assuming point masses,
\begin{eqnarray}
    -\frac{\phi}{G} = \sum_i \frac{M_{{\rm halo}, i}}{r_{{\rm 3d}, i}},
\end{eqnarray}
where $G$ is the gravitational constant, and $r_{{\rm 3d}, i}$ is the three-dimensional distance to each galaxy $i$.
However, because the three-dimensional distance is unavailable in observations, the projected distance is adopted instead.
We can therefore define a projected gravitational potential ($\psi$), following
\begin{eqnarray}
    -\frac{\psi}{G} = \sum_i \frac{M_{{\rm halo}, i}}{d_{{\rm proj}, i}}.
    \label{eqn:psi}
\end{eqnarray}
Equation (2) can be interpreted as representing the maximum possible gravitational potential experienced by an absorber along a QSO sightline.
In practice, the maximum potential can be achieved when all galaxies are aligned in a plane perpendicular to the sightline. Outside of this plane, there is no point that can have $r_{\rm 3d} = d_{\rm proj}$ for all galaxies.

In the right panel of Figure \ref{fig:comp_env}, we show that this projected gravitational potential ($\psi$) is tightly correlated with the total stellar mass, which is a tracer for the total dark matter mass \citep[e.g.][]{Yang:2007aa}.
However, this behaviour is a selection bias due to the maximum projected distance cutoff at 200 kpc for the closest galaxies.
The presence or absence of absorbing ions in the CGM is more cleanly decoupled when considering either the total stellar mass (KS $p$-value of $7.0\times 10^{-3}$; $2.7 \sigma$) or $\psi$ ($4.1\times 10^{-4}$; $3.5\sigma$) than what is shown in the left panel.

Among all the tested characteristics of galaxies along QSO sightlines, the projected gravitational potential $\psi$ appears to be the best indicator of the absorbing properties of the CGM.
In the following discussion, we will focus on $\psi$ for quantifying the impact of galaxy environments.

\subsection{Spatial distribution of ionic column densities}
\label{sec:obs_col}

Next, we examine how the observed spatial distributions of the integrated column densities depend on galaxy properties for different ions.
We consider species over a wide range of ionization states from the ground state species probed by \ion{H}{I} and \ion{He}{I}, low ionization species probed by \ion{O}{II}, \ion{Mg}{II}, and \ion{Fe}{II}, intermediate ionic transitions such as \ion{O}{III}, \ion{O}{IV}, and \ion{O}{V}, to \ion{Ne}{VIII}, the highest ionization species detected.
These are presented in Figure \ref{fig:ion_dist} from top to bottom rows, respectively.
For each species, we consider three independent variables to characterize the galaxy properties, the projected physical distance to the closest galaxy ($d_{\rm proj}$; left panels), $r_{\rm virial}$ normalized projected distance to the closest galaxy ($d_{\rm proj}/r_{\rm virial}$; middle panels), and projected gravitational potential ($-\psi/G$; right panels).

The detection rates of all species decline with increasing projected distance in both physical distances and $r_{\rm virial}$-normalized distances.
This trend supports that the UV-selected cool(-warm) absorbing clouds are physically associated with the closest galaxy and its environments, consistent with previous studies (e.g., \citealt{Chen:1998aa, Chen:2001aa, Chen:2010aa, Stocke:2013aa, Liang:2014aa, Werk:2014aa, Johnson:2015aa, Huang:2016aa, Zahedy:2021aa, Huang:2021aa}).
Specifically for the CUBSz1 sample, no absorption is detected beyond 160 kpc from the closest galaxy for six galaxy systems.

\begin{figure*}
\begin{center}
\includegraphics[width=0.98\textwidth]{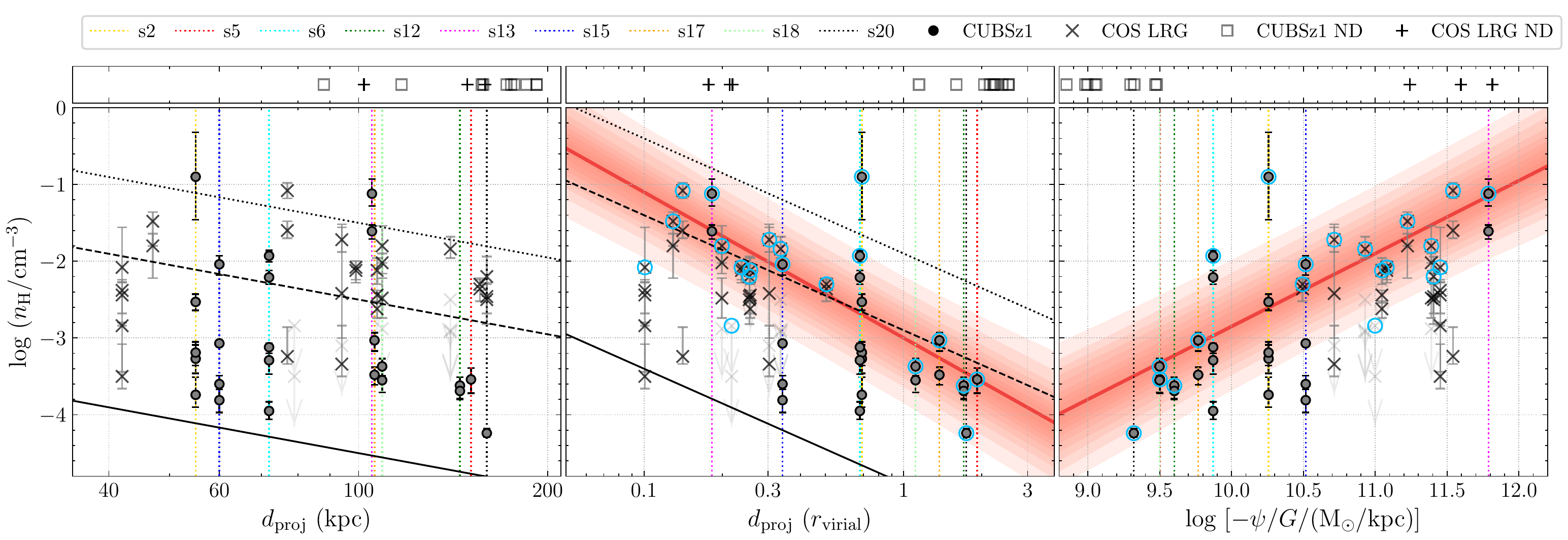}
\end{center}
\caption{The $\log T/{\rm K} \approx 4$ gas density of individual absorption components versus $d_{\rm proj}$ (left panel), $r_{\rm virial}$-normalized $d_{\rm proj}$ (middle panel), and projected gravitational potential ($\psi$; right panel).
Individual absorption components associated with galaxies in the CUBSz1 and COS LRG samples are marked by filled circles and crosses, respectively.
For visual guidance, absorption components sharing the same halo in the CUBSz1 sample are connected by vertical dotted lines.
The locations of galaxies without associated absorption are shown in the top panels as open boxes (pluses) for the CUBSz1 (COS LRG) sample.
For comparison, the density profile of the hot gas is shown as the black solid lines in the left and middle panels, while dashed and dotted lines are predicted cool gas density, which is 100 or 1000 times higher than the hot gas (see text for details).
The highest density (circled in blue) in each absorption system depends strongly on $r_{\rm virial}$-normalized $d_{\rm proj}$ and $\psi$.
The best-fit power-law relations are $\log (n_{\rm H}/\cc) = (-1.9\pm 0.3) \times \log (d_{\rm proj}/r_{\rm virial})] + (-3.0\pm0.1)$ and $\log (n_{\rm H}/\cc) = (0.95_{-0.14}^{+0.15}) \times \log[-\psi/G/({\rm M_\odot/kpc})] + (-12.4_{-1.6}^{+1.5})$ shown as the red shaded regions in middle and right panels.
}
\label{fig:n_dist}
\end{figure*}

Although the detection rate declines for all ions with increasing distance, ions with different ionization potentials exhibit different behaviours.
Because of the high abundance of hydrogen and helium, \ion{H}{I} and \ion{He}{I} can be detected in various phases, and are not robust indicators of the ionization state of the gas.
Therefore, we focus on metal ions (mainly oxygen ions due to their high abundance) to gain insight into the ionization structure of the CGM.
Figure \ref{fig:ion_dist} shows that low ionization state ions (e.g., \ion{Mg}{II} and \ion{O}{II}) are only detected within $\lesssim 100$ kpc or $\lesssim 0.8~ r_{\rm virial}$.
Within $r_{\rm virial}$, these low ions are detected in four out of four galaxies and galaxy groups, leading to a covering fraction of $\kappa = 100\%$ with a $1\sigma$ lower limit of $70\%$ (also see \citealt{Huang:2021aa}).
Beyond $r_{\rm virial}$, no detections of any absorption features are found across 15 sightlines.

In contrast, intermediate ionization state ions (e.g., \ion{O}{III}-\ion{O}{V}) are detected out to $\approx 160$ kpc or $\approx 2~ r_{\rm virial}$.
Five of seven sightlines at $1-2~r_{\rm virial}$ show associated \ion{O}{III}-\ion{O}{V} leading to a covering fraction of $\kappa \approx 70_{-30}^{+20}\%$.
Beyond $2~r_{\rm virial}$, none of the eight sightlines exhibits detectable ions, resulting in $\kappa = 0 \%$ with a $1\sigma$ upper limit of $<20\%$.
Therefore, the intermediate ionization phase is more extended than the low ionization phase.
Such a difference may be explained by a declining density profile from the inner halo to the outskirts, under a photoionization scenario (Section \ref{sec:obs_density}).

For still more highly ionized gas traced by \ion{Ne}{VIII}, no clear trend is found with radius.
Within $r_{\rm virial}$, only one \ion{Ne}{VIII} component is detected among four sightlines, while two of seven sightlines between $1-2~r_{\rm virial}$ show \ion{Ne}{VIII}.
In addition, all three observed \ion{Ne}{VIII} absorbers display a similar column density of $\log N/{\rm cm^{-2}}\approx 14$, which is comparable to the detection limits afforded by the data.
It remains inconclusive whether the spatial distribution of \ion{Ne}{VIII} changes with radius.

We have shown in Figure \ref{fig:comp_env} that the projected gravitational potential may serve as a discriminator between absorbing and non-absorbing sightlines.
In Figure \ref{fig:ion_dist}, we further show that the projected gravitational potential can also be a tracer for different ionization phases.
The low ionization species (e.g., \ion{O}{II}) are only detected in regions with $-\psi/G \gtrsim 10^{10}~ {\rm M_\odot~kpc^{-1}}$. 
In the relatively low potential regions ($-\psi/G \approx  10^9 - 10^{10}~ {\rm M_\odot~kpc^{-1}}$), intermediate and high ionization state ions are detected.
Although all three detected \ion{Ne}{VIII} systems are in the low projected gravitation potential system, the rate of incidence is overall significantly lower than lower-ionization species and as a result, we cannot rule out the possibility that it can also be detected in high potential systems ($-\psi/G \approx  10^{10} - 10^{12}~ {\rm M_\odot~kpc^{-1}}$).

\subsection{Density profile of cool gas}
\label{sec:obs_density}

Figure \ref{fig:ion_dist} clearly shows that the intermediate ionization phase is more extended than the low ionization phase, suggesting a declining gas density profile in the cool CGM.
As described in Section \ref{sec:med_abs}, we determine gas densities of individual absorption components by reproducing the relative column density ratios of kinematically-matched transitions under the assumption of PIE.
In \citetalias{CUBSV}, we have shown that  inferred gas densities of individual absorption components at $z\approx1$ span over three decades from $\log (n_{\rm H}/\cc)\approx -4$ to $-1$, and these are included in Table \ref{tab:ions} together with transitions used in the photoionization modelling for completeness.

In the following analysis, we also include results from the COS LRG survey, which focuses on luminous red galaxies with stellar masses of $\log M_{\rm star}/\msun \approx 11.2$ at $z\approx 0.4$ \citep{Zahedy:2019aa}.
Sightlines in the COS LRG sample probe the inner regions of these massive halos ($d_{\rm proj}/r_{\rm virial}\lesssim 0.5$), which also fall into the high gravitational potential regime ($-\psi/G \approx 10^{11}~{\rm M_\odot~kpc^{-1}}$).

Figure \ref{fig:n_dist} shows the gas density dependence on $d_{\rm proj}$ and $r_{\rm virial}$-normalized $d_{\rm proj}$.
For the CUBSz1 sample, the inferred cool gas densities exhibit an intriguing difference at small and large $d_{\rm proj}/r_{\rm virial}$.
At $d_{\rm proj}/r_{\rm virial} \lesssim 1$, three of four systems show a large minimum-to-maximum scatter in gas density from $\log (n_{\rm H}/\cc)\approx -4$ to $-1$.
In contrast, the scatter of gas densities is $\lesssim 0.5$ dex in all six sightlines at $d_{\rm proj}/r_{\rm virial} > 1$, where all absorption components have low gas densities of $\log (n_{\rm H}/\cc) \lesssim -3$.
The only exception is System s13, which is associated with a massive galaxy group.
These observations indicate that high-density gas is preferentially present in inner halos.

Despite a large scatter, the radial profile of gas density shows a general decline from the inner regions to the outskirts, which is most clearly displayed in $d_{\rm proj}/r_{\rm virial}$ (the middle panel of Figure \ref{fig:n_dist}).
Furthermore, this density decline is more significant when focusing on the highest-density component along each sightline.
We fit a power law to the highest gas density as a function of $d_{\rm proj}/r_{\rm virial}$, shown as red lines and shaded regions in Figure \ref{fig:n_dist}.
The fitting procedure follows what is described in \citetalias{{CUBSV}}, and is implemented by \texttt{emcee} \citep{Foreman-Mackey:2013aa}.
This exercise yields a best fit relationship of $\log (n_{\rm H}/\cc) = -(1.9\pm 0.3) \times \log (d_{\rm proj}/r_{\rm virial}) + (-3.0\pm0.1)$ for the joint sample of both CUBSz1 and COS LRG.
This empirical relation has an intrinsic scatter of $\sigma_{\rm p}\approx 0.4$ after excluding the contributions due to measurement uncertainties.

\begin{figure*}
\begin{center}
\includegraphics[width=0.98\textwidth]{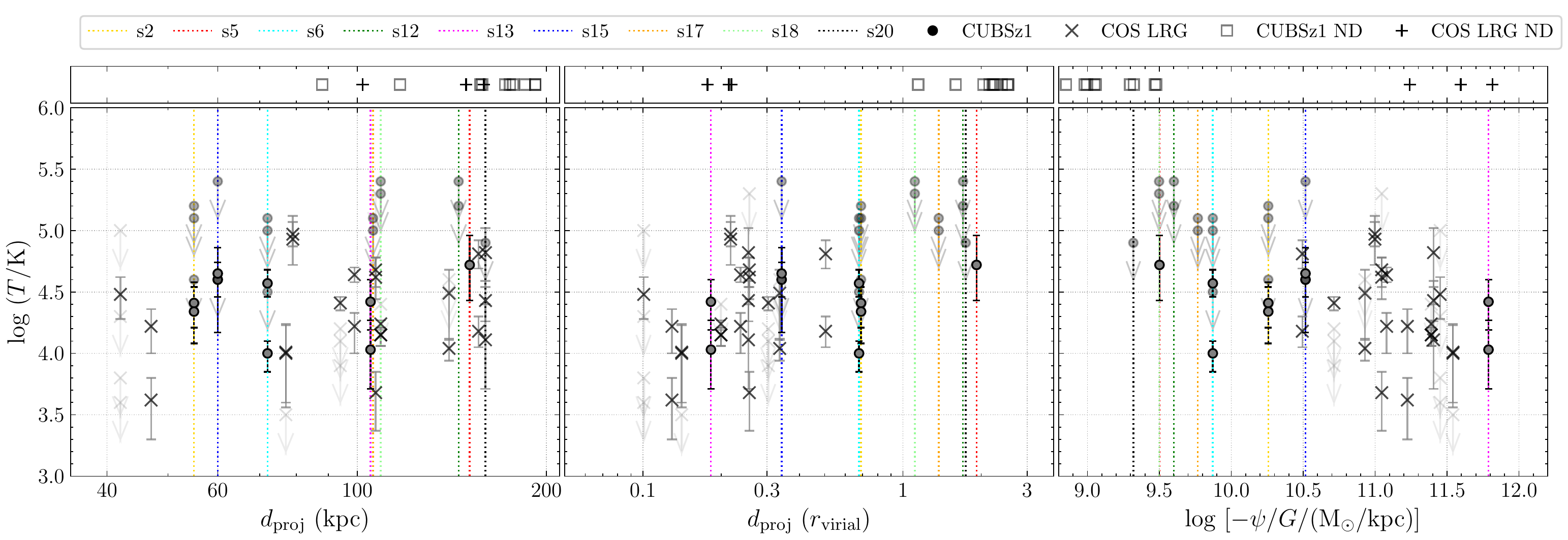}
\end{center}
\caption{Similar plots as Figure \ref{fig:n_dist} but for the gas temperature.
Being confined to a narrow range of $\log (T/ {\rm K}) \approx 4.3\pm 0.3$, the temperature only has a weak dependence on the projected distance or the gravitational potential.
}
\label{fig:t_dist}
\end{figure*}

We compare the radial decline of the cool gas density with the hot CGM profile to place the inferred cool CGM properties in a broad context.
Assuming hydrostatic equilibrium, the density distribution of virialized hot CGM can be approximated by a power-law profile $r^{-\alpha}$ with a slope of $\alpha \approx 1-2$ at $0.1 \lesssim d_{\rm proj}/r_{\rm virial} \lesssim 2$ \citep[e.g.,][]{Voit:2019aa, Stern:2019aa, Faerman:2020aa}.
The inferred power law slope of the cool CGM  is consistent with the expected slope of the hot CGM, which further motivates the question of whether the cool CGM is in pressure balance with the hot halo.

In Figure \ref{fig:n_dist}, we plot the hot CGM density profile of $L_*$ galaxies as black solid lines, which is approximated by a power law with a slope of 1.5 (\citealt{Li:2017aa}; also see \citealt{Singh:2018aa}).
In the CUBSz1 and COS LRG samples, the inferred cool gas densities are roughly a factor of $\approx 100$ above the density profile of the hot CGM.
Such a difference is expected because of the temperature difference between the cool and hot phases.
The temperature of the cool gas is $\approx 10^4$ K (\citealt{Zahedy:2021aa}; \citetalias{CUBSV}), which is $\approx 100$ times lower than the virial temperature of $L_*$ galaxies (i.e., $\approx 10^6$ K).
Therefore, if the cool gas is in pressure balance with the hot CGM, the cool gas density is expected to be $\approx 100$ times higher than the hot gas density.
A density model scaled up by 100 is shown in the dashed lines in Figure \ref{fig:n_dist}.
A second model scaled up by 1000 is also plotted as tilted dotted lines to allow for possibilities of a higher temperature (i.e., $T\approx 7\times 10^6$ K) for the hot CGM in more massive halos expected for the luminous red galaxies.
These hot CGM-based gas density models are consistent with the radial density profile of the cool CGM determined by the highest-density component along each sightline, supporting the hypothesis that the cool CGM is in pressure balance with the hot CGM, over a wide range of mass and environment (also see \citealt{Zahedy:2019aa}).

Next, we investigate the correlation between the cool gas density and the projected potential $\psi$, which is also plotted in Figure \ref{fig:n_dist} (right panel).
There is a significant increase in the highest density along each sightline with increasing $\psi$.
We characterize this relationship by a power law of $\log (n_{\rm H}/\cc) = (0.95_{-0.14}^{+0.15}) \times \log[-\psi/G/({\rm M_\odot/kpc})] + (-12.4_{-1.6}^{+1.5})$.
We infer a power law index consistent with the unity as would be the case if the high-density gas is confined in the high-pressure region, where the internal energy of the ambient gas is driven by the gravitational potential.
In virial equilibrium, then the virial temperature is $\log (T/{\rm K}) \approx -3.6+\log[-\psi/G/({\rm M_\odot/kpc})]$.
This temperature is $\approx 2\times 10^6$ K for $\log[-\psi/G/({\rm M_\odot/kpc})]=10$ (expected at 100 kpc in the halo of an $L_*$ galaxy), which is comparable with the virial temperature of an $L_*$ galaxy.
Therefore, a plausible scenario is that the hot ambient CGM, which also dominates the total gas mass, contains most of the internal energy from gravitational collapse, and the cool CGM is in pressure balance with the hot CGM.

Similar to the density, we also consider the impact of the galaxy environment on the temperature and non-thermal motion of the gas.
In \citetalias{CUBSV}, we found that the temperature-density profile can be approximated by a power law of $\log\,(T/{\rm K})\!=\!(3.92\pm0.17)\!-\!(0.16\pm 0.07)\,\log\,(n_{\rm H}/\cmjj)$.
Also, the non-thermal velocity depends on the gas density following a power law of $\log\,(b_{\rm NT}/{\rm km~s^{-1}})\!=\!(0.52\pm0.11)\!-\!(0.21\pm0.04)\,\log\,(n_{\rm H}/\cmjj)$.
Considering these correlations together, a weak radial dependence of temperature or non-thermal motions is expected (Figure \ref{fig:t_dist}).
However, adopting the generalized Kendall $\tau$ test, the radial trend and the dependence on $\psi$ are both observationally insignificant, showing 1.5 $\sigma$ and 1.2 $\sigma$, respectively.
There are two reasons for these insignificant trends.
First, the large scatter together with a small dynamic range at all radii makes it challenging to determine the radial dependence (Figure \ref{fig:n_dist}).
Second, the line width analysis requires higher-quality data than the photoionization modelling.
In the CUBSz1 sample, most absorption systems do not have \ion{H}{I} covered in COS/FUV spectra, which limits the ability to precisely decompose the contributions from thermal and non-thermal broadening.
As a result, only about half of the systems have a measured temperature or non-thermal broadening \citepalias{CUBSV}.
As shown in Figure \ref{fig:t_dist}, numerous upper limits on the temperature further contribute to the scatter, making it impossible to determine any definitive trend.
The situation for the non-thermal motion is similar, and therefore not shown here.

\begin{figure*}
\begin{center}
\includegraphics[width=0.98\textwidth]{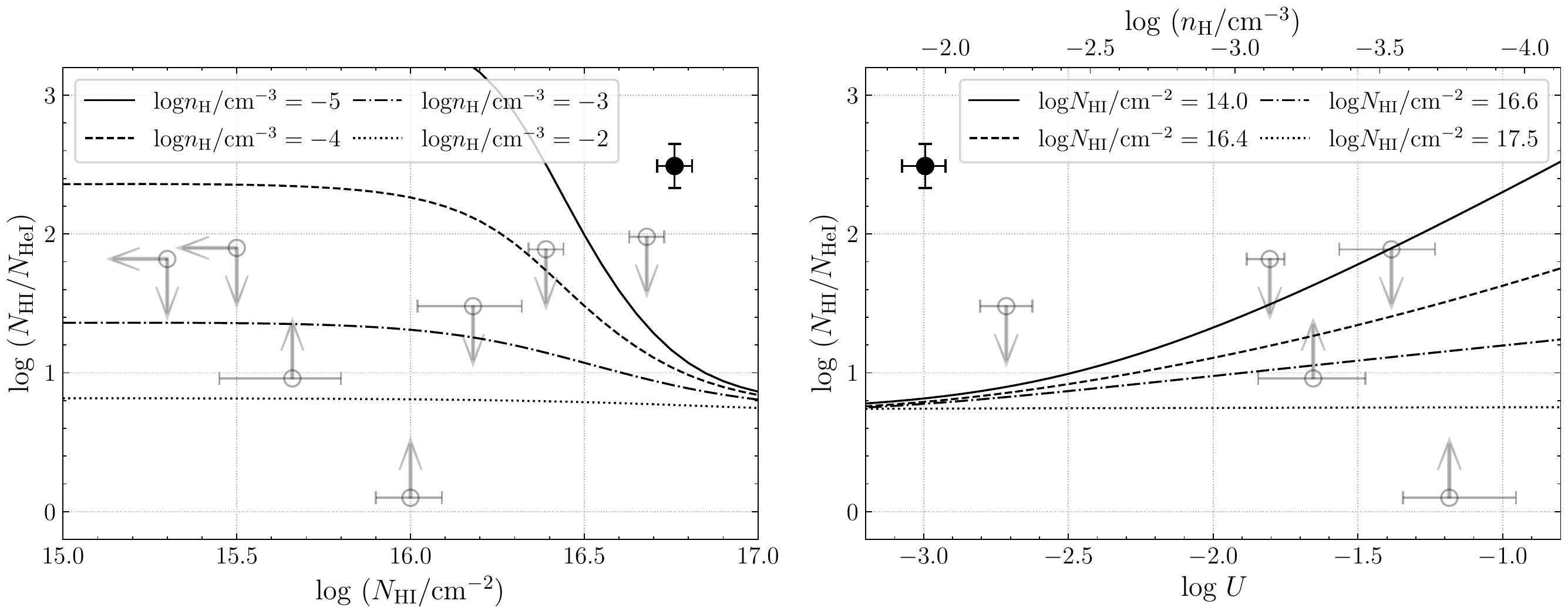}
\end{center}
\caption{Comparisons between the observed and predicted $N$(\ion{H}{I})/$N$(\ion{He}{I}) ratios for gas being photoionized by the HM05 UVB at $z\approx 1$.
In the left panel, we show model predictions of the $N$(\ion{H}{I})/$N$(\ion{He}{I}) ratio as a function of $N$(\ion{H}{I}).
At $\log N$(\ion{H}{I})$/\cmjj \lesssim 16$, the predicted $N$(\ion{H}{I})/$N$(\ion{He}{I}) ratio declines with increasing $n_{\rm H}$, but does not depend on $N$(\ion{H}{I}).
At higher $N$(\ion{H}{I}), the models quickly converge to $\log N$(\ion{H}{I})/$N$(\ion{He}{I}) $\approx 0.8$ by $\log N$(\ion{H}{I})$\approx 17$. 
In the right panel, model-predicted $N$(\ion{H}{I})/$N$(\ion{He}{I}) ratios are shown as a function of the ionization parameter $U$.
For observed values, six components are included, which have multiple metal transitions to determine $n_{\rm H}$ and $U$.
Although the predicted ratio remains flat with $\log N$(\ion{H}{I})/$N$(\ion{He}{I}) $\approx0.8$ over the tested range of $U$ for gas with high $\log N$(\ion{H}{I})$/\cmjj = 17.5$, the ratio increases with $U$ in the lower $N$(\ion{H}{I}) regime.
The model predictions broadly cover the observed $N(\mbox{\ion{H}{I}})/N(\mbox{\ion{He}{I}})$ ratios, with one exception at $\log N$(\ion{H}{I}) $/\cmjj=16.76$ and $\log U = -3$ (i.e., the filled black circle).
This is the component at $v=-20\kms$ in System s6.
The discrepancy can be reduced by including additional local ionization sources (see Figure \ref{fig:HI_HeI_bb}).
}
\label{fig:HI_HeI_hm05}
\end{figure*}

\subsection{$\bm{N}$(\ion{H}{I})/$\bm{N}$(\ion{He}{I}) ratio and incident radiation field}
\label{sec:obs_he1}
As described in Section \ref{data}, the CUBSz1 sample was selected to have $z>0.883$ to cover \ion{He}{I} $\lambda 584$ as a proxy for $N$(\ion{H}{I}) to constrain the total gas column density.
Strong \ion{H}{I} transitions, such as Ly$\alpha$ and Ly$\beta$, are only covered at $z<0.74$ in the COS/FUV spectra, so most systems only have \ion{He}{I} covered.
Because hydrogen and helium are both light elements that follow a nearly fixed cosmic abundance ratio, constraints on $N$(\ion{He}{I}) can in principle be translated to constraints of $N$(\ion{H}{I}).
However, the robustness of this method depends on the conversion from $N$(\ion{He}{I}) to $N$(\ion{H}{I}) for the appropriate ionization mechanism.

Here, we examine whether $N$(\ion{He}{I}) is indeed a good tracer of $N$(\ion{H}{I}).
In our analysis, the cool gas is assumed to be photoionized by the UVB under PIE, which is consistent with the density-temperature relationship presented in \citetalias{CUBSV}.
We compare the measured \ion{H}{I} to \ion{He}{I} column density ratios $N$(\ion{H}{I})/$N$(\ion{He}{I}) with predictions from a suite of photoionization models. 
In the CUBSz1 sample, high-order Lyman series lines together with the \ion{He}{I} $\lambda 584$ transition are covered along sightlines for six galaxies or galaxy groups at $z < 0.94$, but only three sightlines (Systems s2, s5, and s6) exhibit detectable absorption features (Table \ref{tab:gal_sample}).
These three absorption systems have 11 kinematically-matched components.
However, the \ion{He}{I} $\lambda$ 584 transitions of s2 occurs at the edge of the COS/FUV spectrum with S/N $\approx 4$ per resolution element, and three components spanning a velocity range of $v\approx 10-70\kms$ are blended with each other (Figure \ref{fig:a2}).
As a result, the \ion{He}{I} $\lambda$ 584 feature cannot be decomposed to constrain individual components, and these three components are omitted from the following consideration.

Among the eight components with both \ion{H}{I} and \ion{He}{I} coverage, six have measured \ion{H}{I} column densities, while the remaining two show no trace of \ion{H}{I} with only upper limits on $N$(\ion{H}{I}) available. 
Four components show strong \ion{He}{I} $\lambda$ 584 transitions, two of which are saturated with only lower limits available.
Combining these together leaves one component with measured $N$(\ion{H}{I})/$N$(\ion{He}{I}) ratio, while the rest have only single-sided limits.

In Figure \ref{fig:HI_HeI_hm05}, the observed $N$(\ion{H}{I})/$N$(\ion{He}{I}) ratios of these eight components are presented along with predictions from photoionization models, computed under the HM05 UVB ionizing field.
Among these eight components with \ion{H}{I} coverage, six components have multiple metal lines to measure the gas density.
The $N$(\ion{H}{I})/$N$(\ion{He}{I}) ratios are predicted to be high at low $N$(\ion{H}{I}) and low $n_{\rm H}$ (or high ionization parameter; $\log U = \log n_{\rm photon}/n_{\rm H}$).
At $\log N(\mbox{\ion{H}{I}})/{\rm cm^{-2}} \gtrsim 17$, all photoionization models converge to $\log N$ (\ion{H}{I})/$N$(\ion{He}{I}) $\approx 0.8$.
These models predict an upper limit of $\log N$(\ion{H}{I})/$N$(\ion{He}{I}) $\approx 2$ for gas with densities of $\log (n_{\rm H}/\cc) \gtrsim -4$, which are the densities that reproduce the column density ratios of different metal transitions.
Therefore, the range of observed $N$(\ion{H}{I})/$N$(\ion{He}{I}) ratios confirms the validity of the photoionization model ionized by the HM05 UVB for most absorption components in the CUBSz1 sample. 

The only exception is the absorption component at $\approx -20 \kms$ in System s6 (i.e., component s6c2 in Table \ref{tab:ions}).
This component has the highest component \ion{H}{I} column density of $\log N(\mbox{\ion{H}{I}})/{\rm cm^{-2}} = 16.76\pm 0.05$ in the CUBSz1 sample, and a relatively high density of $\log (n_{\rm H}/\cc) = -1.93_{-0.08}^{+0.07}$ (Table \ref{tab:ions}).
Therefore, a low $N(\mbox{\ion{H}{I}})/N(\mbox{\ion{He}{I}})$ ratio is predicted by photoionization models with an incident radiation field of the HM05 UVB.
However, the associated \ion{He}{I} column density is only $\log N(\mbox{\ion{He}{I}})/{\rm cm^{-2}} = 14.27\pm 0.15$, leading to a high ratio of $\log N(\mbox{\ion{H}{I}})/N(\mbox{\ion{He}{I}}) = 2.5 \pm 0.2$, which cannot be explained by the fiducial photoionization models.
This deviation may be in part due to the helium abundance variation, but it is unlikely to have a helium abundance 30 times lower than the cosmic abundance \citep[e.g.][]{Cooke:2018aa}.

\begin{figure*}
\begin{center}
\includegraphics[width=0.98\textwidth]{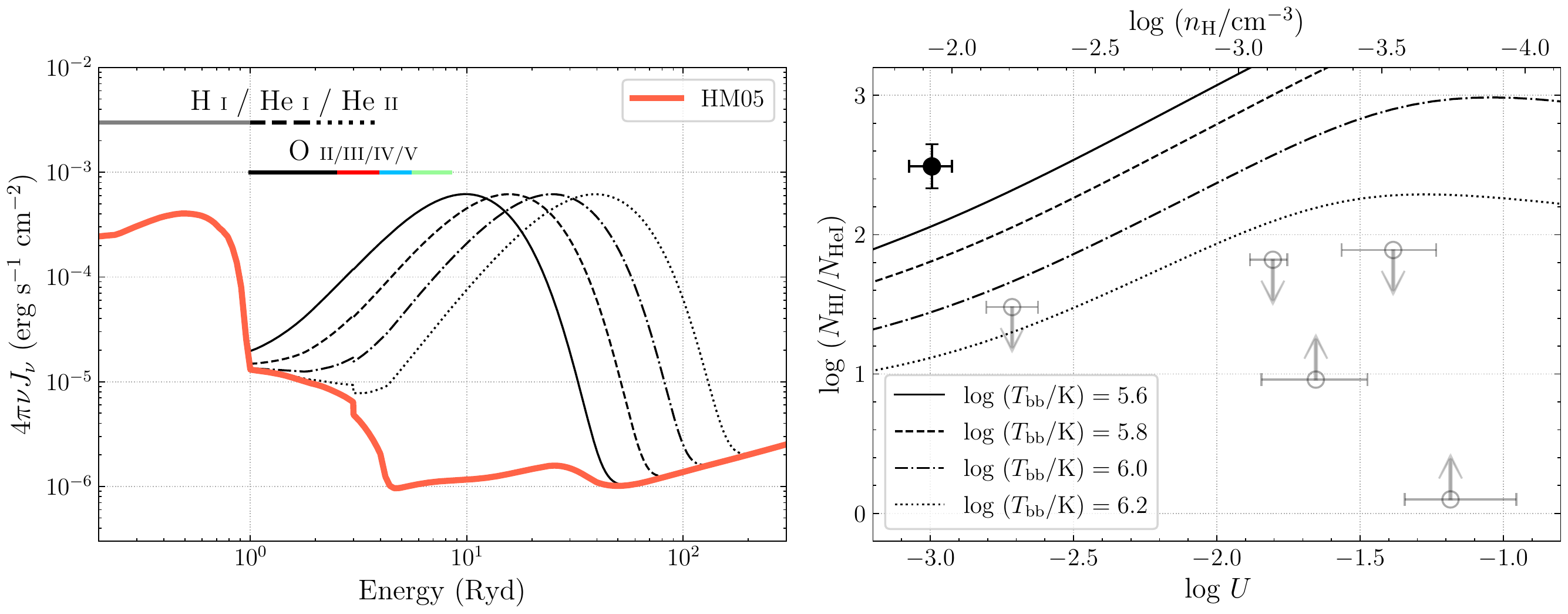}
\end{center}
\caption{Effects of local fluctuations in the ionizing radiation field.
Left panel: the fiducial $z\approx1$ HM05 UVB and different blackbody spectra are shown in red and black lines, respectively.
The ionization potentials of different ions are labelled on top of the panel to show the spectral window affecting the photoionization of different ions.
The additional blackbody mainly affects the intermediate ionization phase, which is traced by \ion{He}{II} and \ion{O}{III}-{\scriptsize V}.
Right panel: Dependence of predicted $N$(\ion{H}{I})/$N$(\ion{He}{I}) ratio on the ionization parameter $U$.
By photoionizing \ion{He}{II} to \ion{He}{III}, the additional blackbody can also reduce the \ion{He}{I} population, leading to a high $N$(\ion{H}{I})/$N$(\ion{He}{I}) value.
See Section \ref{sec:obs_he1} and \ref{sec:dis_h1_he1} for more discussion.
}
\label{fig:HI_HeI_bb}
\end{figure*}

Consequently, deviations from the HM05 UVB are needed to raise the $N$(\ion{H}{I})/$N$(\ion{He}{I}) ratio through photoionization in two different ways. 
First, considering the ionization potential of \ion{He}{I} (1.8 Ryd), a direct means to raise the $N$(\ion{H}{I})/$N$(\ion{He}{I}) ratio is to vary the slope of the ionizing spectrum at $1-2$ Ryd, where a harder field would lead to a higher ratio.
Alternatively, enhanced ionizing radiation beyond 4 Ryd may also increase the ionization of \ion{He}{II} and produce more \ion{He}{III}, thereby reducing the recombination rate of \ion{He}{II} to form \ion{He}{I}. 
To explore the range of modifications needed in the ionizing spectrum, we consider an additional blackbody spectrum with a range of temperature from $T_{\rm bb}=10^5$ K to $T_{\rm bb}=10^7$ K to generate different spectral slopes at $1-2$ Ryd (Figure \ref{fig:HI_HeI_bb}).
In this test, we assume that this additional radiative source is located 100 kpc away from the absorber, considering the projected distance of 72 kpc between the QSO sightline and the closest galaxy.
The luminosity of this blackbody is fixed at $10^{45} \rm~erg~s^{-1}$, which dominates the incident field between $1-100$ Ryd over the UVB.  

This exercise leads to a best-fit blackbody temperature of $\log (T_{\rm bb}/{\rm K}) = 5.6$, below which the additional blackbody radiation begins to have non-negligible effects on the relative abundances of low-ionization species such as \ion{O}{II} and \ion{O}{III}.
In Figure \ref{fig:HI_HeI_bb}, we present the results for four different blackbody temperatures of $\log (T_{\rm bb}/{\rm K}) = 5.6$, $5.8$, $6.0$, and $6.2$.
As expected, a hard ionizing radiation field at 1-2 Ryd and an additional ionizing source at 10 Ryd significantly increases the $N(\mbox{\ion{H}{I}})/N(\mbox{\ion{He}{I}})$ ratio, with $\log (T_{\rm bb}/{\rm K}) = 5.6$ predicting the highest $N(\mbox{\ion{H}{I}})/N(\mbox{\ion{He}{I}})$ ratio.
However, this model is still $2 \sigma$ away from the observed $N$(\ion{H}{I})/$N$(\ion{He}{I}) ratio of $2.5 \pm 0.2$, predicting a higher \ion{He}{I} column density than observed.
Although this model cannot completely reproduce the observed ratio, it shows that local fluctuations in the ionizing radiation field may be detectable based on the observed $N$(\ion{H}{I})/$N$(\ion{He}{I}) ratio.
The origin of this additional soft x-ray incident field is discussed in Section \ref{sec:dis_h1_he1}.

The need of an additional soft x-ray spectrum as dictated by the odd $N$(\ion{H}{I})/$N$(\ion{He}{I}) ratio in component s6c2 also raises a question regarding the accuracy of the best-fit ionization parameters for the remaining components with only $N$(\ion{He}{I}) available.  To directly address such concern requires measurements of $N$(\ion{H}{I}), which can only be obtained with additional NUV spectroscopy of these QSOs.  Here we assess the impact on the estimates of $n_{\rm H}$ without $N$(\ion{H}{I}) by repeating the photoionization analysis for component s6c2 without including $N$(\ion{H}{I}).   We find that while no robust constraints can be placed for the gas metallicity, we are able to recover a consistent $n_{\rm H}$ to within the uncertainties.

To better understand the $N$(\ion{H}{I})/$N$(\ion{He}{I}) ratio (e.g., the occurrence rate of the extremely high ratio), more data are needed to cover both \ion{H}{I} (Ly$\alpha$ or Ly$\beta$ to obtain a lower limit on the column density) and \ion{He}{I}.
For the remaining CUBSz1 sample with typical $N$(\ion{H}{I})/$N$(\ion{He}{I}) ratios, we adopt the standard HM05 UVB for all systems to determine the gas density.

\section{Discussion}
\label{discussion}
In Section \ref{results}, we have examined the impact of the galaxy environment on these properties on the spatial profile of the cool CGM, including the incidence of absorption, ionic column densities, and gas densities.
Specifically, we have introduced a projected gravitational potential ($\psi$), which is the sum of the projected gravitational potential of each group member.  Among all defining characteristics of galaxies, including $d_{\rm proj}$ and $d_{\rm proj}/r_{\rm virial}$ to either the closest galaxy or the centre of mass of a galaxy group, total stellar mass, and $\psi$, 
we find that $\psi$ is the best indicator of the absorbing properties of the CGM.
In addition, the $N$(\ion{H}{I})/$N$(\ion{He}{I}) ratio is found to be sensitive to local fluctuations in the ionizing radiation field.
In this section, we discuss the implications of these findings. 

\subsection{Contributions to a non-uniform ionizing radiation field due to the presence of neighbouring galaxies}
\label{sec:dis_h1_he1}

As described in Section \ref{sec:obs_he1}, the photoionization scenario under a PIE assumption is a self-consistent solution for most absorption components in the cool CGM.
However, uncertainties still exist in both the ionizing radiation intensity and spectral shape. 
In practice, the adopted incident field is usually the UVB, synthesizing emissions from both galaxies and QSOs over a cosmological volume, but two caveats remain in adopting the UVB as the primary source of photoionization. 

First, the current UVB model is still missing some components, including the emission from the warm-hot CGM or IGM, which has a low-energy tail into the soft x-ray to UV band at $\approx 10-100$ Ryd (e.g., \citealt{Upton-Sanderbeck:2018aa} and \citealt{Stofanova:2022aa}).
In addition, the emission properties of some components included in the synthesized model are poorly constrained. 
For example, the unknown population of x-ray binaries or clustering of young stars and associated emission spectra lead to significant uncertainties of the UV to x-ray SED from star-forming regions.

Second, while the synthesized UVB model provides a good estimate of the mean background radiation field at different redshifts, local fluctuations due to nearby galaxies or QSOs can play a significant role in offsetting the ionization balance.
For example, ionizing flux from star-forming $L^*$ galaxies may overcome the UVB within $\approx 50$ kpc \citep[e.g.,][]{Werk:2014aa, Chen:2017cc}, while a luminous QSO could ionize surrounding neutral gas as far as $\gtrsim 1$ Mpc away \citep[e.g.][]{Faucher-Giguere:2008aa, Zahedy:2021aa, Dorigo:2022aa}.
These additional sources of radiation can modify the ionization structure in the cool CGM, showing detectable deviation from UVB-only photoionization models.

In the CUBSz1 sample, we found that photoionization models based on the fiducial HM05 UVB cannot explain the component with the highest $N(\mbox{\ion{H}{I}})/N(\mbox{\ion{He}{I}})$ ratio (i.e., absorption component s6c2).
This extremely high ratio of $\log N$(\ion{H}{I})/$N$(\ion{He}{I}) $\approx 2.5$ may be reproduced by adding a blackbody component emitting at $E \approx 10\rm~ Ryd$ and leading to a hard field at $1-2$ Ryd (Figure \ref{fig:HI_HeI_bb}). 
At the same time, photoionization models with the additional blackbody over-predict the $N$(\ion{H}{I})/$N$(\ion{He}{I}) ratio for at least three other components, which have upper limits of $\log N$(\ion{H}{I})/$N$(\ion{He}{I}) $\approx 1.5$.
Among these three components, two are in the same system (i.e., System s6) hosting the component with the highest $N$(\ion{H}{I})/$N$(\ion{He}{I}) ratio.
The distinction between different components in the same system may indicate that large fluctuations exist in the local radiation field at $d_{\rm proj} \lesssim 160$ kpc.

As stated in Section \ref{sec:obs_he1}, there are two possible ways of modifying the ionizing spectrum to produce the observed high $N$(\ion{H}{I})/$N$(\ion{He}{I}) ratio, a hard radiation field at $1-2$ Ryd and additional sources emitting at $> 4$ Ryd.
Various processes could produce harder ionizing fields at $1-2$ Ryd than the standard HM05 UVB, including attenuation of the UVB by a nearby optically-thick cloud, elevated leakage of ionizing photons from binary populations \citep{Ma:2016aa}, and harder intrinsic UV emission in low-metallicity galaxies \citep{Steidel:2014aa}.
The latter two possibilities require galaxies to be close to the absorber (i.e., $\lesssim 50$ kpc) to dominate the ionizing field over the UVB assuming an escape fraction of $\lesssim 5 \%$ (see discussion in \citealt{Chen:2017cc}).
Therefore, if the observed high $N$(\ion{H}{I})/$N$(\ion{He}{I}) ratio is due to the escaping flux from nearby galaxies, there should be a low-mass galaxy closer to the absorbers, which is still undetected in the CUBS survey.  We also test whether an AGN-like ionizing spectrum can produce the observation.
Adopting the AGN template in Cloudy and allowing the temperature as a free parameter, we find it impossible to simultaneously match \ion{H}{I}/\ion{He}{I} and \ion{O}{II}/\ion{O}{III} within the 2-$\sigma$ uncertainties.  This is understood as due to the relatively strong emission at $1-4$ Ryd, which has a significant impact on the low ions.

Alternatively, we consider the possibility of an additional blackbody at $\log T_{\rm bb}/{\rm K}\approx 5.6$ in Section \ref{sec:obs_he1}.
Incorporating other ions into the consideration and varying the blackbody temperature, a blackbody model with $\log T_{\rm bb}/{\rm K}= 5.6_{-0.1}^{+0.2}$ can produce a $N$(\ion{H}{I})/$N$(\ion{He}{I}) ratio to within $2\sigma$ of the observed value while keeping the predicted metal column densities to within 1\,$\sigma$ of the observed values.

Considering the necessary blackbody temperature of $\log\,T_{\rm bb}/{\rm K}\approx 5.6$, supersoft x-ray sources are possible candidates (e.g., \citealt{Upton-Sanderbeck:2018aa}).
They are a population of white dwarfs accreting from nearby main sequence or red giant companions, which have blackbody emission with temperatures of $\approx 10^5$ K \citep{Kahabka:1997aa}.
However, observations on supersoft X-ray sources are currently limited, especially for low-mass white dwarfs, due to temperatures being too low to be detected by the current facilities.
A single supersoft x-ray source typically has a luminosity of $\approx 10^{38}\rm~ erg~ s^{-1}$ \citep{Kahabka:2006aa}.
At the same time, a luminosity of at least $\approx 10^{44}\rm~ erg~ s^{-1}$ is required 
to produce the observed high $N$(\ion{H}{I})/$N$(\ion{He}{I}) ratio if the supersoft x-ray source resides in the closest galaxy found at $d_{\rm proj} = 72$ kpc in the CUBSz1 galaxy sample.
To reproduce the observation, an unrealistically large population of supersoft x-ray sources in this galaxy or a faint galaxy closer to the high $N$(\ion{H}{I})/$N$(\ion{He}{I}) component would be needed to match the required spectral index at $1-4$ Ryd for the ionizing radiation. 
Although we cannot rule out the presence of faint galaxies lost in the glare of the QSO light, the lack of [\ion{O}{II}] emission in the quasar spectrum at the absorber redshift places a 3-$\sigma$ upper limit on the SFR of $\lesssim 0.4\,M_\odot\,{\rm yr}^{-1}$ following the relation of \cite{Moustakas2006} for dwarf galaxies with $M_B > -16$.

To summarize, the observed high $N$(\ion{H}{I})/$N$(\ion{He}{I}) ratio can be explained by variations of incident ionizing fields through various scenarios.
These scenarios all suggest a local variation of radiation field induced by a nearby galaxy.
Currently, with only one such system, the origin of this variation is still inconclusive, but more data could significantly improve the understanding of the incident field.

\subsection{Implications for the pressure balance in the multiphase CGM}
\label{sec:dis_density}

\begin{figure}
\begin{center}
\includegraphics[width=0.48\textwidth]{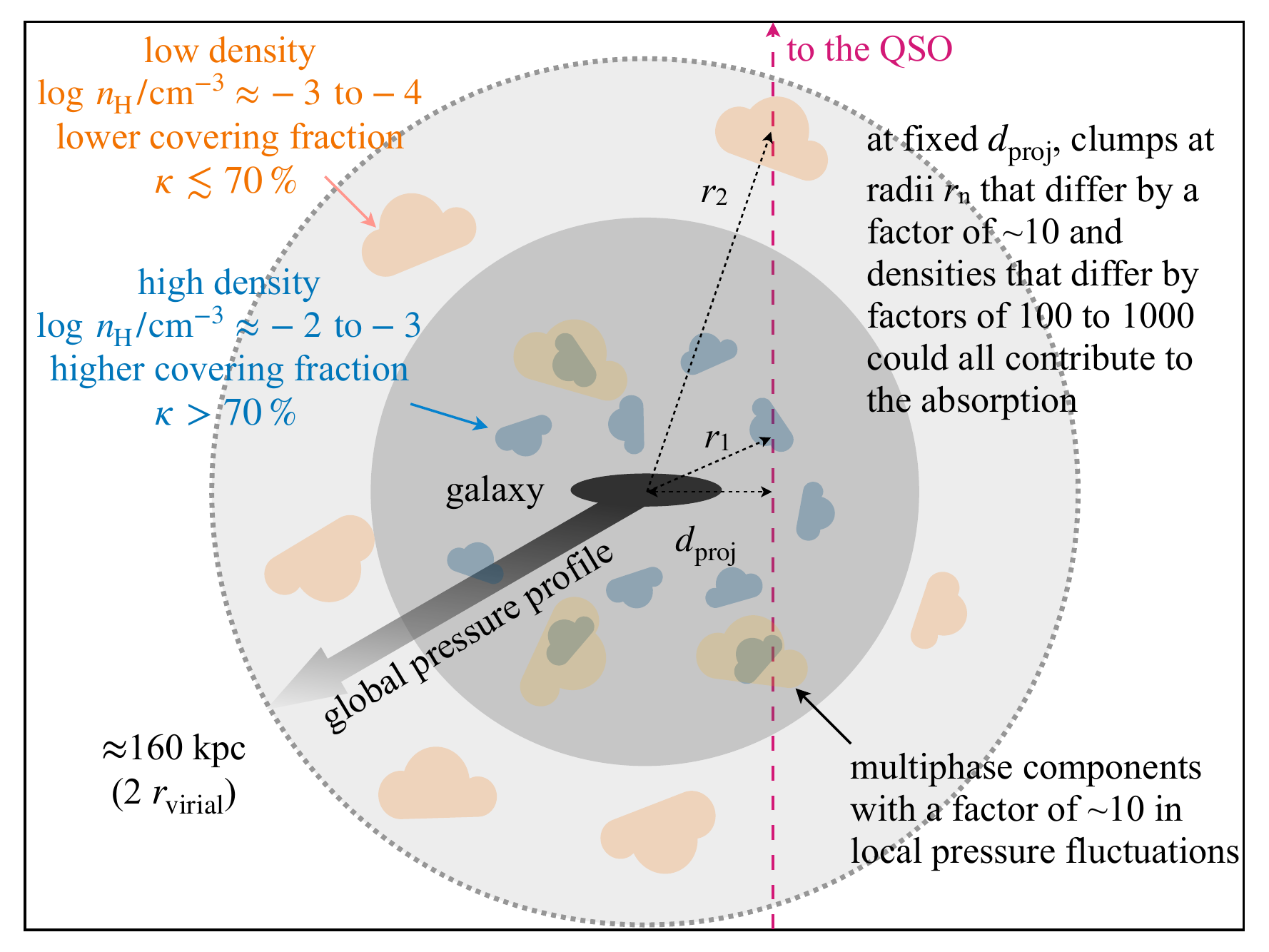}
\end{center}
\caption{
A cartoon illustration of the pressure variation in the cool CGM (not-to-scale).
The observed density variation of the cool CGM ($n_{\rm H}$) can be attributed to two different factors, the global gradient of the pressure profile and local fluctuations (see the text in \S\ \ref{sec:dis_density} for details).
The global pressure profile derived based on the cool gas is consistent with that of the hot CGM (Figure \ref{fig:n_dist} and Section \ref{sec:obs_density}).
Specifically, with a power-law density profile of slope $2-3$, the observed factor of 100 to 1000 scatter in gas density can be explained by a factor of 10 difference in 3D radii probed at $d_{\rm proj} \lesssim 0.3\,r_{\rm virial}$.
At the same time, the multiphase absorption components detected in the inner halo suggest that local pressure fluctuations can be as large as a factor of $\sim\,10$ \citepalias{CUBSV}.
}
\label{fig:illustration}
\end{figure}

For the cool CGM, the highest gas density along each sightline is found to depend on the projected distance to the galaxy or the closest galaxy in a galaxy group (Figure \ref{fig:n_dist}).
At the same time, components with two to three orders of magnitude lower densities are also present along the same sightlines.  
Because the cool CGM has a narrow range in temperature ($\log T/{\rm K}\approx 4.3\pm0.3$) under the PIE model \citepalias{CUBSV}, the observed density distribution along individual sightlines at different projected distances provides unique insights into pressure fluctuations in the cool CGM.
We present in Figure \ref{fig:illustration} a simple schematic diagram to illustrate the spatial distribution of different components relative to the host galaxy/galaxy groups.  Various thermodynamic processes, such as tidal interaction, ram pressure stripping, or additional ionizing field, may contribute to seeding the dense, cool clumps and the formation of multi-phase gas in the CGM.

In Section \ref{sec:obs_density}, we find that the relation between the maximum density along each sightline and the projected distance is consistent with the pressure profile of the hot CGM, suggesting that these high-density clumps are in pressure equilibrium with the hot gas.
A remaining question is whether or not the low-density gas is also in pressure balance with the hot CGM.
The answer to this question depends on where the low-density gas is located.
If these cool, low-density clumps reside in the inner halo, then their pressure would be significantly lower than that of the hot gas (and the associated high-density components) and therefore they are expected to be short-lived. 
On the other hand, if these low-density clumps are located in the halo outskirts and only appear at small impact parameters by projection, then they can be in pressure balance with the hot CGM.

To explore whether the scenario of pressure balance is consistent with observations, we first determine a power-law slope ($\Delta \log P / \Delta \log r$) of the pressure profile using the low-density components.
The scatter in gas density $\Delta\,n \propto \Delta\,P/T$ is determined using the derived gas densities along sightlines at small projected distances, and the maximum range in halo radius $\Delta r$ is set by the largest projected distance where absorbing gas is detected.
As shown in Figure \ref{fig:n_dist}, the densities of individual components detected at $d_{\rm proj}=0.1-0.3~ r_{\rm virial}$ span two to three decades (i.e., $\Delta \log P \approx 2-3$).
The halo radii where the detected absorption components are located span approximately one decade ($ \log r/{r_{\rm virial}} \approx -1$ to 0).  
Together, the constraints in $\Delta\log\,P$ and $\Delta\,r$ lead to a power law slope of $\approx 2-3$ for the global pressure profile of these low-density clumps.
This power law slope is slightly higher than the best fit ($1.9\pm 0.3$) derived for the $d_{\rm proj}$-dependence of the maximum density components but the slopes are still consistent within the measurement uncertainty.
This would suggest the low-density gas at small $d_{\rm proj}$ could be located at large physical distances.

We further compare the detection rates of the low-density gas at small and large projected distances.
If the detection rate of low-density gas is significantly higher at smaller impact parameters, then it would suggest that a considerable amount of cool, low-density gas resides in the inner halo with unbalanced pressure.
In the CUBSz1 sample, three of the four systems ($\approx 75$ per cent) at $d_{\rm proj}\lesssim r_{\rm virial}$ have low-density components with $\log n_{\rm H}/\cc < -3$.
In comparison, seven out of 11 systems ($\approx 64$ per cent) in the COS LRG sample exhibit a low-density component.
Adopting the best-fit spatial profile for the maximum-density component (Figure \ref{fig:n_dist}), we expect that these low-density components should occur at $r \approx 1-2\,r_{\rm virial}$.
Over $d_{\rm proj}=1-2\,r_{\rm virial}$, we find that the detection rate of the low-density gas is 5/7 $\approx 70$ per cent, consistent with the rate of incidence of low-density components at small impact parameters.
We, therefore, argue that most of the detected low-density clumps are located in the halo outskirts and only appear at small impact parameters by projection and that they are likely in pressure balance with the hot CGM.

However, we cannot rule out the possibility that a small fraction of the low-density clumps are located in the inner halo and are out of pressure balance with the hot halo.  In particular, we have shown in \citetalias{CUBSV} that multiphase components are present in the inner halo with an order-of-magnitude difference in pressure between different phases.
This difference in pressure is consistent with the intrinsic scatter of $\sigma_{\rm p} \approx 0.4$ in the relation between the highest densities and projected distances.  Therefore, it is possible that local pressure fluctuations do contribute to the observed scatter of the density along individual sightlines.

In conclusion, the scatter of gas densities over three decades observed along individual sightlines at small projected distances can be explained by a combination of both the global profile and local fluctuations of the pressure.
On average, the cool CGM is in pressure balance with the hot CGM, and a global radial profile with a power law slope of $1.5-2$ can account for about two orders of magnitude of the observed density scatter.
Any remaining fluctuations can be attributed to local pressure imbalance between different phases.

\section{Summary}
In this work, we report galaxy properties and absorption features of the CUBSz1 sample, which contains sub-$L^*$ and $L^*$ star-forming galaxies at $0.89<z<1.21$ (Figure \ref{fig:gal_pro}).  Comparisons of the stellar mass and SFR of these galaxies show that they are consistent with typical main-sequence star-forming galaxies at $z\approx 1$ (Figure \ref{fig:sm_sfr}).  We investigate the impact of the galaxy environment on the cool photoionized CGM.
The CUBSz1 sample together with COS LRG sample covers a wide range in projected distance from $\approx 0.1$ to 3 $r_{\rm virial}$ and in total projected gravitational potential from $-\psi/G\approx 3\times 10^9$ to $10^{12}~{\rm M_\odot/kpc}$.
Our major findings are summarized below.
\begin{itemize}
    \item In the CUBSz1 sample, galaxies in dense environments exhibit a higher detection rate of absorption features (Figure \ref{fig:gal_environment}).
    Specifically, all galaxy systems with the closest galaxy having $M_{\rm star}>10^{10} M_\odot$ have associated absorption features.
    \item When multiple galaxies are found near an absorber, the distance to the closest galaxy (in projection), rather than the distance to the most massive galaxy of the group or the centre of the mass for galaxy groups, appears to be most correlated with the presence and absorption properties of the gas (Section \ref{sec:obs_inc} and Figure \ref{fig:comp_env}).
    \item Low ionization state ions (e.g., \ion{O}{II}) are only detected in inner halos ($\lesssim 0.8\,r_{\rm virial}$).
    Intermediate- and high-ionization state ions (e.g., \ion{O}{IV} and \ion{Ne}{VIII}) are more spatially extended and can be detected up to $2\,r_{\rm virial}$ (Figure \ref{fig:ion_dist}).
    \item The presence, ionization state, and gas density of the cool (i.e., $log T/{\rm K}\approx 4$) CGM are found to correlate strongly with $d_{\rm proj}/r_{\rm virial}$ and with $\psi$ (Figure \ref{fig:comp_env}, \ref{fig:ion_dist} and \ref{fig:n_dist}).
    Specifically, we found that the gas density is linearly correlated with $\psi$ following a power law slope of $0.95_{-0.14}^{+0.15}$.
    Given the narrow temperature range of the cool CGM, our finding, therefore, shows that the high-pressure cool CGM is confined in the deep gravitational potential.
    \item The large fluctuation seen in the observed $\log N(\mbox{\ion{H}{I}})/N(\mbox{\ion{He}{I}})$ ratio may be explained by the presence of local variations in the ionizing flux, particularly in the soft x-ray band, which may be explained by escaped ionizing photons from nearby galaxies or QSOs (Figure \ref{fig:HI_HeI_bb}; Section \ref{sec:dis_h1_he1}).
\end{itemize}

\section*{Acknowledgements}
The authors thank the referee for a timely and thoughtful report.
ZQ acknowledges partial support from HST-GO-15163.001A, NSF AST-1715692 grants, and NASA ADAP grant 80NSSC22K0481.
HWC and MCC acknowledge partial support from HST-GO-15163.001A and NSF AST-1715692 grants.
GCR acknowledges partial support from HST-GO-15163.015A.
FSZ acknowledges the support of a Carnegie Fellowship from the Observatories of the Carnegie Institution for Science.
SDJ acknowledges partial support from HST-GO-15280.009A.
EB acknowledges partial support from NASA under award No. 80GSFC21M0002.
KLC acknowledges partial support from NSF AST-1615296.
SC gratefully acknowledges support from the European Research Council (ERC) under the European Union’s Horizon 2020 Research and Innovation programme grant agreement No 864361.
CAFG was supported by NSF through grants AST-1715216, AST-2108230, and CAREER award AST-1652522; by NASA through grants 17-ATP17-006 7 and 21-ATP21-0036; by STScI through grants HST-AR-16124.001-A and HST-GO-16730.016-A; by CXO through grant TM2-23005X; and by the Research Corporation for Science Advancement through a Cottrell Scholar Award.
SL was funded by FONDECYT grant number 1231187.
This work is based on observations made with ESO Telescopes at the Paranal Observatory under program ID 0104.A-0147(A), observations made with the 6.5m Magellan Telescopes located at Las Campanas Observatory, and spectroscopic data gathered under the HST-GO-15163.01A program using the NASA/ESA Hubble Space Telescope operated by the Space Telescope Science Institute and the Association of Universities for Research in Astronomy, Inc., under NASA contract NAS 5-26555. This research has made use of the services of the ESO Science Archive Facility and the Astrophysics Data Service (ADS)\footnote{\url{https://ui.adsabs.harvard.edu}}. The analysis in this work was greatly facilitated by the following \texttt{python} packages:  \texttt{Numpy} \citep{Numpy}, \texttt{Scipy} \citep{Scipy}, \texttt{Astropy} \citep{astropy:2013,astropy:2018}, \texttt{Matplotlib} \citep{Matplotlib}, and \texttt{emcee} \citep{Foreman-Mackey:2013aa}. 

\section*{Data Availability}

The data underlying this article will be shared on reasonable request to the corresponding author.



\bibliographystyle{mnras}
\bibliography{ms} 




\appendix

\section{Absorption Features and Galaxy Spectra}
In Table \ref{tab:ions}, we report measurements and uncertainties of the column density ($\log N$), line width ($b$), and line centroid ($v_{\rm c}$) for all detected ions in the CUBSz1 sample.
The full table of all measured ions with upper limits will be provided in a machine-readable format based on a request to the authors.

From Figure \ref{fig:a1} to Figure \ref{fig:a20}, the absorption features of each absorption system are shown with the model resulting in the obtained best-fit parameters.

From Figure \ref{fig:g1} to Figure \ref{fig:g20}, we show the spectra for member galaxies in each galaxy system.
All of these galaxies have securely determined redshifts, which have at least two confirmed spectral features.
Particularly, if the [OII] doublet is resolved in the MUSE spectra, we also consider it a secure measurement.

\renewcommand\thefigure{A.\arabic{figure}}
\renewcommand\thetable{A.\arabic{table}}

\begin{table}
    \caption{Summary of detected ions$^a$}
    \label{tab:ions}
    \centering
    \begin{tabular}{llcccc}
\hline
\hline
QSO & z & ion & $\log N$ & $b$ & $v^b$ \\
& &  & (/cm$^{-2}$) & (km s$^{-1}$) & (km s$^{-1}$)\\
\hline
\multicolumn{6}{c}{Absorption ID: s2c1$^c$}\\
\multicolumn{6}{c}{$\log (n_{\rm H}/\cc) = -3.74_{-0.16}^{+0.23}$; $\log (T/{\rm K}) = 4.41_{-0.20}^{+0.13}$}\\
\hline
J0114 &  0.8963 & \ion{H}{I}       & $16.00_{-0.10}^{+0.09}$        & $21.7_{-3.8}^{+3.7}$           & $-46$          \\
J0114 &  0.8963 & \ion{O}{III}     & $>14.3$                        & $<9.7$                         & $-52.5_{-1.1}^{+1.5}$          \\
J0114 &  0.8963 & \ion{O}{IV}      & $>14.1$                        & $<15.8$                        & $-45.5_{-2.8}^{+2.4}$          \\
J0114 &  0.8963 & \ion{O}{V}       & $>13.9$                        & $<18.7$                        & $-46$          \\
\hline
\multicolumn{6}{c}{Absorption ID: s2c2$^c$}\\
\multicolumn{6}{c}{$\log (n_{\rm H}/\cc) = -3.27_{-0.19}^{+0.18}$; $\log (T/{\rm K}) <4.6$}\\
\hline
J0114 &  0.8963 & \ion{H}{I}       & $15.66_{-0.21}^{+0.14}$        & $29.7_{-7.1}^{+6.2}$           & $-21$          \\
J0114 &  0.8963 & \ion{O}{III}     & $14.17_{-0.19}^{+0.15}$        & $27.6_{-12.3}^{+8.5}$          & $-21$          \\
J0114 &  0.8963 & \ion{O}{IV}      & $14.41_{-0.10}^{+0.10}$        & $24.4_{-6.0}^{+6.2}$           & $-21.1_{-3.2}^{+3.3}$          \\
J0114 &  0.8963 & \ion{O}{V}       & $>13.8$                        & $<33.2$                        & $-21$          \\
\hline
\multicolumn{6}{c}{Absorption ID: s2c3}\\
\multicolumn{6}{c}{$\log (n_{\rm H}/\cc) = -2.53_{-0.11}^{+0.10}$; $\log (T/{\rm K}) <5.2$}\\
\hline
J0114 &  0.8963 & \ion{O}{III}     & $>14.3$                        & $<21.5$                        & $329$          \\
J0114 &  0.8963 & \ion{O}{IV}      & $13.83_{-0.24}^{+0.15}$        & $16.6_{-3.8}^{+2.5}$           & $329$          \\
J0114 &  0.8963 & \ion{Mg}{II}     & $11.86_{-0.06}^{+0.06}$        & $9.1_{-3.3}^{+3.5}$            & $328.6_{-1.1}^{+1.4}$          \\
J0114 &  0.8963 & \ion{S}{III}     & $12.90_{-0.12}^{+0.11}$        & $21.5_{-8.8}^{+6.0}$           & $324.8_{-2.0}^{+3.4}$          \\
\hline
\multicolumn{6}{c}{Absorption ID: s2c4}\\
\multicolumn{6}{c}{phase 1: $\log (n_{\rm H}/\cc) = -0.90_{-0.56}^{+0.58}$; $\log (T/{\rm K}) = 4.34_{-0.26}^{+0.24}$}\\
\multicolumn{6}{c}{phase 2: $\log (n_{\rm H}/\cc) = -3.19_{-0.17}^{+0.14}$; $\log (T/{\rm K}) <5.1$}\\
\hline
J0114 &  0.8963 & \ion{H}{I}       & $15.98_{-0.22}^{+0.15}$        & $29.1_{-9.8}^{+6.9}$           & $368$          \\
J0114 &  0.8963 & \ion{O}{III}     & $14.23_{-0.17}^{+0.31}$        & $9.5_{-3.7}^{+10.9}$           & $368$          \\
J0114 &  0.8963 & \ion{O}{IV}      & $14.01_{-0.21}^{+0.18}$        & $22.0_{-8.4}^{+5.8}$           & $368$          \\
J0114 &  0.8963 & \ion{O}{V}       & $13.71_{-0.16}^{+0.36}$        & $10.6_{-4.3}^{+7.7}$           & $368$          \\
J0114 &  0.8963 & \ion{Mg}{II}     & $12.12_{-0.03}^{+0.03}$        & $3.8_{-0.6}^{+0.8}$            & $368.0_{-0.5}^{+0.5}$          \\
J0114 &  0.8963 & \ion{Fe}{II}     & $11.80_{-0.08}^{+0.09}$        & $12.6_{-2.0}^{+3.8}$           & $368$          \\
\hline
\multicolumn{6}{c}{Absorption ID: s2c5}\\
\hline
J0114 &  0.8963 & \ion{H}{I}       & $15.94_{-0.21}^{+0.15}$        & $31.5_{-5.7}^{+4.6}$           & $390.2_{-3.0}^{+3.8}$          \\
\hline
\multicolumn{6}{c}{Absorption ID: s2c6$^c$}\\
\hline
J0114 &  0.8963 & \ion{H}{I}       & $16.68_{-0.05}^{+0.05}$        & $24.0_{-1.7}^{+1.7}$           & $600.9_{-0.6}^{+1.0}$          \\
J0114 &  0.8963 & \ion{He}{I}      & $>14.7$                        & $<23.7$                        & $601.2_{-3.6}^{+3.0}$          \\
\hline
\multicolumn{6}{c}{Absorption ID: s5c1$^c$}\\
\hline
J2245 &  0.9352 & \ion{He}{I}      & $>13.6$                        & $<41.2$                        & $120.1_{-3.3}^{+3.5}$          \\
\hline
\multicolumn{6}{c}{Absorption ID: s5c2$^c$}\\
\multicolumn{6}{c}{$\log (n_{\rm H}/\cc) = -3.54_{-0.18}^{+0.15}$; $\log (T/{\rm K}) = 4.72_{-0.29}^{+0.24}$}\\
\hline
J2245 &  0.9352 & \ion{H}{I}       & $16.39_{-0.05}^{+0.05}$        & $35.2_{-5.4}^{+6.9}$           & $180.6_{-3.0}^{+2.8}$          \\
J2245 &  0.9352 & \ion{He}{I}      & $>14.5$                        & $<42.6$                        & $179.8_{-2.9}^{+3.3}$          \\
J2245 &  0.9352 & \ion{O}{III}     & $14.01_{-0.08}^{+0.08}$        & $19.3_{-5.9}^{+7.7}$           & $176.4_{-0.7}^{+1.5}$          \\
J2245 &  0.9352 & \ion{O}{IV}      & $14.09_{-0.08}^{+0.07}$        & $23.5_{-7.7}^{+8.7}$           & $178.3_{-2.1}^{+2.8}$          \\
J2245 &  0.9352 & \ion{O}{V}       & $13.71_{-0.13}^{+0.25}$        & $22.3_{-12.2}^{+11.9}$         & $179.1_{-2.6}^{+3.5}$          \\
J2245 &  0.9352 & \ion{S}{IV}      & $12.86_{-0.17}^{+0.12}$        & $29.1_{-10.1}^{+7.3}$          & $178.7_{-2.4}^{+3.9}$          \\
\hline
\multicolumn{6}{c}{Absorption ID: s6c1$^c$}\\
\multicolumn{6}{c}{phase 1: $\log (n_{\rm H}/\cc) = -2.21_{-0.09}^{+0.09}$; $\log (T/{\rm K}) = 4.00_{-0.15}^{+0.10}$}\\
\multicolumn{6}{c}{phase 2: $\log (n_{\rm H}/\cc) = -3.95_{-0.11}^{+0.12}$; $\log (T/{\rm K}) <4.5$}\\
\hline
J0333 &  0.9373 & \ion{H}{I}       & $16.18_{-0.16}^{+0.14}$        & $13.2_{-4.6}^{+4.5}$           & $-62$          \\
J0333 &  0.9373 & \ion{He}{I}      & $>14.7$                        & $<22.2$                        & $-62$          \\
J0333 &  0.9373 & \ion{C}{II}      & $13.84_{-0.05}^{+0.05}$        & $19.0_{-4.0}^{+3.9}$           & $-63.1_{-2.3}^{+2.9}$          \\
J0333 &  0.9373 & \ion{N}{II}      & $13.02_{-0.21}^{+0.14}$        & $10.9_{-4.3}^{+6.8}$           & $-61.6_{-3.1}^{+3.2}$          \\
J0333 &  0.9373 & \ion{N}{III}     & $14.08_{-0.05}^{+0.05}$        & $22.3_{-3.8}^{+3.6}$           & $-60.6_{-2.5}^{+2.4}$          \\
J0333 &  0.9373 & \ion{N}{IV}      & $14.15_{-0.04}^{+0.04}$        & $39.5_{-2.6}^{+2.6}$           & $-61.2_{-2.8}^{+2.4}$          \\
J0333 &  0.9373 & \ion{O}{II}      & $14.03_{-0.12}^{+0.11}$        & $12.9_{-4.8}^{+6.5}$           & $-60.8_{-2.7}^{+2.5}$          \\
J0333 &  0.9373 & \ion{O}{III}     & $>15.1$                        & $<32.2$                        & $-62$          \\
J0333 &  0.9373 & \ion{O}{IV}      & $15.52_{-0.10}^{+0.21}$        & $32.9_{-2.3}^{+2.1}$           & $-62$          \\
J0333 &  0.9373 & \ion{O}{V}       & $16.06_{-0.29}^{+0.27}$        & $27.2_{-1.5}^{+2.0}$           & $-62$          \\
J0333 &  0.9373 & \ion{Ne}{V}      & $14.41_{-0.22}^{+0.18}$        & $33.1_{-16.4}^{+11.8}$         & $-62$          \\
J0333 &  0.9373 & \ion{Mg}{II}     & $12.70_{-0.02}^{+0.02}$        & $8.5_{-0.4}^{+0.4}$            & $-61.4_{-0.3}^{+0.3}$          \\
J0333 &  0.9373 & \ion{S}{III}     & $13.49_{-0.07}^{+0.05}$        & $18.5_{-4.3}^{+4.1}$           & $-59.7_{-2.5}^{+1.9}$          \\
J0333 &  0.9373 & \ion{S}{IV}      & $13.43_{-0.07}^{+0.05}$        & $22.1_{-6.8}^{+5.2}$           & $-62.0_{-3.1}^{+3.1}$          \\
J0333 &  0.9373 & \ion{S}{V}       & $13.11_{-0.07}^{+0.06}$        & $27.3_{-4.1}^{+4.3}$           & $-60.1_{-3.2}^{+2.2}$          \\
J0333 &  0.9373 & \ion{Fe}{II}     & $12.36_{-0.03}^{+0.03}$        & $5.1_{-1.1}^{+1.2}$            & $-61.8_{-0.5}^{+0.5}$          \\
J0333 &  0.9373 & \ion{Fe}{III}$^d$    & $13.88_{-0.09}^{+0.09}$        & $21.6_{-9.8}^{+5.7}$           & $-65.6_{-1.1}^{+2.4}$          \\
\hline
    \end{tabular}
\end{table}

\begin{table}
    \contcaption{Summary of detected ions}
    \centering
    \begin{tabular}{llcccc}
\hline
\hline

\multicolumn{6}{c}{Absorption ID: s6h}\\
\hline
J0333 &  0.9373 & \ion{Ne}{VIII}   &
$14.38_{-0.06}^{+0.07}$        & $93.4_{-14.1}^{+16.0}$          & $-38.3_{-7.6}^{+9.1}$          \\
\hline
\multicolumn{6}{c}{Absorption ID: s6c2$^c$}\\
\multicolumn{6}{c}{phase 1: $\log (n_{\rm H}/\cc) = -1.93_{-0.08}^{+0.07}$; $\log (T/{\rm K}) = 4.57_{-0.11}^{+0.11}$}\\
\multicolumn{6}{c}{phase 2: $\log (n_{\rm H}/\cc) = -3.29_{-0.18}^{+0.14}$; $\log (T/{\rm K}) <5.1$}\\
\hline
J0333 &  0.9373 & \ion{H}{I}       & $16.76_{-0.05}^{+0.05}$        & $23.7_{-2.8}^{+3.3}$           & $-20$          \\
J0333 &  0.9373 & \ion{He}{I}      & $14.27_{-0.15}^{+0.15}$        & $24.2_{-10.6}^{+8.0}$          & $-20$          \\
J0333 &  0.9373 & \ion{C}{II}      & $14.25_{-0.17}^{+0.13}$        & $10.5_{-1.7}^{+3.3}$           & $-22.4_{-1.2}^{+1.3}$          \\
J0333 &  0.9373 & \ion{N}{II}      & $13.75_{-0.05}^{+0.05}$        & $9.1_{-1.8}^{+2.4}$            & $-19.9_{-1.2}^{+1.3}$          \\
J0333 &  0.9373 & \ion{N}{III}     & $14.48_{-0.05}^{+0.07}$        & $11.7_{-1.4}^{+1.3}$           & $-22.7_{-1.0}^{+1.0}$          \\
J0333 &  0.9373 & \ion{N}{IV}      & $>13.6$                        & $<15.5$                        & $-23.4_{-1.2}^{+1.9}$          \\
J0333 &  0.9373 & \ion{O}{II}      & $>14.4$                        & $<15.6$                        & $-24.7_{-1.4}^{+1.6}$          \\
J0333 &  0.9373 & \ion{O}{III}     & $>13.7$                        & $<19.5$                        & $-20$          \\
J0333 &  0.9373 & \ion{Mg}{II}     & $13.53_{-0.07}^{+0.12}$        & $8.1_{-0.5}^{+0.4}$            & $-20.5_{-0.1}^{+0.2}$          \\
J0333 &  0.9373 & \ion{S}{II}      & $12.98_{-0.08}^{+0.06}$        & $11.5_{-4.0}^{+4.3}$           & $-23.8_{-0.9}^{+1.6}$          \\
J0333 &  0.9373 & \ion{S}{III}     & $13.76_{-0.05}^{+0.06}$        & $12.7_{-1.6}^{+1.7}$           & $-21.5_{-1.6}^{+1.5}$          \\
J0333 &  0.9373 & \ion{S}{IV}      & $13.68_{-0.04}^{+0.05}$        & $14.8_{-2.2}^{+2.6}$           & $-19.2_{-1.9}^{+1.8}$          \\
J0333 &  0.9373 & \ion{S}{V}       & $13.25_{-0.14}^{+0.43}$        & $8.4_{-2.7}^{+5.6}$            & $-21.8_{-1.7}^{+1.8}$          \\
J0333 &  0.9373 & \ion{Fe}{II}     & $13.10_{-0.01}^{+0.01}$        & $8.6_{-0.3}^{+0.3}$            & $-20.5_{-0.2}^{+0.2}$          \\
\hline
\multicolumn{6}{c}{Absorption ID: s6c3$^c$}\\
\multicolumn{6}{c}{$\log (n_{\rm H}/\cc) = -3.12_{-0.08}^{+0.05}$; $\log (T/{\rm K}) < 5.0$}\\
\hline
J0333 &  0.9373 & \ion{He}{I}      & $13.48_{-0.39}^{+0.19}$        & $26.1_{-3.9}^{+5.4}$           & $34.4_{-3.0}^{+3.8}$           \\
J0333 &  0.9373 & \ion{C}{II}      & $12.95_{-0.15}^{+0.14}$        & $23.3_{-5.3}^{+8.0}$           & $34.4_{-3.1}^{+3.6}$           \\
J0333 &  0.9373 & \ion{N}{III}     & $13.16_{-0.15}^{+0.11}$        & $29.5_{-8.6}^{+7.3}$           & $34.4_{-3.2}^{+3.5}$           \\
J0333 &  0.9373 & \ion{N}{IV}      & $13.21_{-0.10}^{+0.08}$        & $40.0_{-10.5}^{+7.1}$          & $36.7_{-3.7}^{+2.4}$           \\
J0333 &  0.9373 & \ion{O}{III}     & $14.51_{-0.03}^{+0.03}$        & $27.9_{-3.0}^{+3.7}$           & $34.5_{-1.7}^{+1.9}$           \\
J0333 &  0.9373 & \ion{O}{IV}      & $14.72_{-0.02}^{+0.02}$        & $28.4_{-2.7}^{+3.1}$           & $35.5_{-0.7}^{+0.3}$           \\
J0333 &  0.9373 & \ion{O}{V}       & $14.17_{-0.09}^{+0.10}$        & $23.0_{-2.1}^{+3.3}$           & $35.0_{-0.1}^{+0.1}$           \\
J0333 &  0.9373 & \ion{S}{III}     & $12.67_{-0.14}^{+0.12}$        & $28.7_{-8.6}^{+7.7}$           & $35.7_{-3.4}^{+2.9}$           \\
J0333 &  0.9373 & \ion{S}{IV}      & $12.85_{-0.10}^{+0.10}$        & $27.1_{-8.7}^{+12.9}$          & $32.5_{-1.9}^{+3.4}$           \\
\hline
\multicolumn{6}{c}{Absorption ID: s12c1}\\
\multicolumn{6}{c}{$\log (n_{\rm H}/\cc) = -3.62_{-0.17}^{+0.11}$; $\log (T/{\rm K}) < 5.4$}\\
\hline
J0333 &  0.9988 & \ion{N}{IV}      & $13.25_{-0.07}^{+0.07}$        & $28.0_{-6.5}^{+8.5}$           & $-316.8_{-2.3}^{+3.6}$         \\
J0333 &  0.9988 & \ion{O}{IV}      & $13.88_{-0.11}^{+0.11}$        & $28.9_{-8.7}^{+10.6}$          & $-315.3_{-3.1}^{+3.5}$         \\
J0333 &  0.9988 & \ion{O}{V}       & $13.61_{-0.05}^{+0.05}$        & $24.5_{-4.6}^{+5.6}$           & $-317.6_{-1.8}^{+2.7}$         \\
\hline
\multicolumn{6}{c}{Absorption ID: s12c2}\\
\multicolumn{6}{c}{$\log (n_{\rm H}/\cc) = -3.69_{-0.11}^{+0.08}$; $\log (T/{\rm K}) < 5.2$}\\
\hline
J0333 &  0.9988 & \ion{O}{III}     & $13.37_{-0.14}^{+0.10}$        & $24.4_{-5.7}^{+6.7}$           & $-269.0_{-3.0}^{+3.8}$         \\
J0333 &  0.9988 & \ion{O}{IV}      & $13.91_{-0.10}^{+0.08}$        & $22.5_{-1.8}^{+3.9}$           & $-270.3_{-2.0}^{+3.0}$         \\
J0333 &  0.9988 & \ion{O}{V}       & $13.83_{-0.03}^{+0.03}$        & $23.8_{-2.5}^{+2.6}$           & $-265.1_{-1.7}^{+1.3}$         \\
\hline
\multicolumn{6}{c}{Absorption ID: s13c1}\\
\multicolumn{6}{c}{$\log (n_{\rm H}/\cc) = -1.61_{-0.10}^{+0.08}$; $\log (T/{\rm K}) = 4.42_{-0.23}^{+0.18}$}\\
\hline
J2339 &  1.0032 & \ion{He}{I}      & $>14.5$                        & $<18.0$                        & $-60.8_{-1.1}^{+1.0}$          \\
J2339 &  1.0032 & \ion{C}{II}      & $14.58_{-0.10}^{+0.21}$        & $12.1_{-1.9}^{+1.6}$           & $-63.4_{-1.0}^{+0.9}$          \\
J2339 &  1.0032 & \ion{N}{II}      & $13.90_{-0.06}^{+0.07}$        & $15.1_{-4.4}^{+3.3}$           & $-60.6_{-2.9}^{+2.7}$          \\
J2339 &  1.0032 & \ion{N}{III}     & $14.08_{-0.05}^{+0.05}$        & $15.5_{-2.4}^{+2.5}$           & $-56.8_{-0.9}^{+0.5}$          \\
J2339 &  1.0032 & \ion{N}{IV}      & $13.27_{-0.06}^{+0.05}$        & $17.8_{-3.6}^{+4.3}$           & $-59.5_{-2.4}^{+2.2}$          \\
J2339 &  1.0032 & \ion{O}{II}      & $14.68_{-0.09}^{+0.13}$        & $14.4_{-3.2}^{+3.0}$           & $-57.3_{-1.9}^{+0.9}$          \\
J2339 &  1.0032 & \ion{O}{III}     & $>14.5$                        & $<20.0$                        & $-57.4_{-0.8}^{+0.8}$          \\
J2339 &  1.0032 & \ion{O}{V}       & $13.53_{-0.05}^{+0.05}$        & $30.4_{-7.2}^{+6.2}$           & $-61.9_{-2.6}^{+3.0}$          \\
J2339 &  1.0032 & \ion{Mg}{I}      & $10.77_{-0.15}^{+0.12}$        & $14.1_{-2.9}^{+3.8}$           & $-61.1_{-1.3}^{+1.3}$          \\
J2339 &  1.0032 & \ion{Mg}{II}     & $13.01_{-0.01}^{+0.01}$        & $7.8_{-0.1}^{+0.2}$            & $-60.8_{-0.1}^{+0.1}$          \\
J2339 &  1.0032 & \ion{S}{II}      & $13.08_{-0.05}^{+0.07}$        & $10.7_{-2.8}^{+3.1}$           & $-57.8_{-1.2}^{+1.1}$          \\
J2339 &  1.0032 & \ion{S}{III}     & $13.21_{-0.04}^{+0.04}$        & $12.4_{-1.7}^{+1.7}$           & $-50.8_{-0.9}^{+0.6}$          \\
J2339 &  1.0032 & \ion{Fe}{II}     & $12.61_{-0.01}^{+0.01}$        & $7.2_{-0.4}^{+0.4}$            & $-61.8_{-0.2}^{+0.2}$          \\
\hline
\multicolumn{6}{c}{Absorption ID: s13c2}\\
\multicolumn{6}{c}{$\log (n_{\rm H}/\cc) = -1.12_{-0.16}^{+0.19}$; $\log (T/{\rm K}) = 4.03_{-0.32}^{+0.24}$}\\
\hline
J2339 &  1.0032 & \ion{He}{I}      & $>14.3$                        & $<11.1$                        & $2.5_{-0.7}^{+0.4}$            \\
J2339 &  1.0032 & \ion{C}{II}      & $13.82_{-0.12}^{+0.13}$        & $7.1_{-1.5}^{+2.9}$            & $0.9_{-1.7}^{+1.7}$            \\
J2339 &  1.0032 & \ion{O}{II}      & $13.88_{-0.14}^{+0.13}$        & $15.1_{-7.2}^{+3.6}$           & $3.5_{-3.0}^{+1.8}$            \\
J2339 &  1.0032 & \ion{Mg}{II}     & $12.29_{-0.04}^{+0.13}$        & $3.1_{-0.6}^{+0.6}$            & $0.7_{-0.2}^{+0.3}$            \\
J2339 &  1.0032 & \ion{Fe}{II}     & $12.13_{-0.04}^{+0.09}$        & $3.3_{-0.7}^{+0.7}$            & $0.9_{-0.4}^{+0.5}$            \\
\hline
    \end{tabular}
\end{table}

\begin{table}
    \contcaption{Summary of detected ions}
    \centering
    \begin{tabular}{llcccc}
\hline
\hline

\multicolumn{6}{c}{Absorption ID: s15c1}\\
\multicolumn{6}{c}{$\log (n_{\rm H}/\cc) = -3.07_{-0.04}^{+0.04}$; $\log (T/{\rm K}) = 4.60_{-0.43}^{+0.26}$}\\
\hline
J0154 &  1.0116 & \ion{He}{I}      & $13.81_{-0.04}^{+0.05}$        & $21.8_{-4.2}^{+5.0}$           & $-76.3_{-2.5}^{+2.1}$          \\
J0154 &  1.0116 & \ion{C}{II}      & $13.41_{-0.07}^{+0.07}$        & $16.3_{-4.1}^{+2.5}$           & $-79.9_{-2.1}^{+3.2}$          \\
J0154 &  1.0116 & \ion{N}{III}     & $14.66_{-0.05}^{+0.08}$        & $14.7_{-1.1}^{+1.1}$           & $-81.4_{-0.6}^{+0.7}$          \\
J0154 &  1.0116 & \ion{N}{IV}      & $>14.5$                        & $<27.5$                        & $-82.2_{-0.6}^{+1.1}$          \\
J0154 &  1.0116 & \ion{O}{III}     & $14.87_{-0.14}^{+0.51}$        & $14.4_{-3.2}^{+1.8}$           & $-80.3_{-0.7}^{+0.6}$          \\
J0154 &  1.0116 & \ion{O}{IV}      & $15.01_{-0.09}^{+0.18}$        & $19.3_{-2.7}^{+2.5}$           & $-79.5_{-1.2}^{+1.2}$          \\
J0154 &  1.0116 & \ion{O}{V}       & $>14.4$                        & $<27.1$                        & $-74.9_{-1.3}^{+1.3}$          \\
J0154 &  1.0116 & \ion{Ne}{IV}     & $>14.5$                        & $<31.3$                        & $-77.7_{-3.3}^{+3.4}$          \\
J0154 &  1.0116 & \ion{Ne}{V}      & $14.70_{-0.04}^{+0.04}$        & $25.8_{-3.2}^{+3.4}$           & $-81.2_{-1.2}^{+1.7}$          \\
J0154 &  1.0116 & \ion{Ne}{VI}     & $14.28_{-0.14}^{+0.13}$        & $22.7_{-11.6}^{+12.4}$         & $-77.2_{-3.7}^{+2.9}$          \\
J0154 &  1.0116 & \ion{Mg}{II}     & $11.80_{-0.06}^{+0.06}$        & $15.6_{-2.5}^{+2.6}$           & $-77.9_{-1.8}^{+1.9}$          \\
J0154 &  1.0116 & \ion{S}{III}     & $13.30_{-0.03}^{+0.03}$        & $16.3_{-2.1}^{+2.0}$           & $-81.6_{-0.9}^{+1.1}$          \\
J0154 &  1.0116 & \ion{S}{IV}      & $13.64_{-0.03}^{+0.04}$        & $16.6_{-1.5}^{+1.5}$           & $-74.3_{-1.0}^{+0.8}$          \\
J0154 &  1.0116 & \ion{S}{V}$^d$       & $13.31_{-0.05}^{+0.08}$        & $15.9_{-3.1}^{+3.1}$           & $-80.3_{-1.3}^{+1.4}$          \\
\hline

\multicolumn{6}{c}{Absorption ID: s15c2}\\
\multicolumn{6}{c}{$\log (n_{\rm H}/\cc) = -3.81_{-0.16}^{+0.14}$; $\log (T/{\rm K}) < 5.5$}\\
\hline
J0154 &  1.0116 & \ion{N}{IV}      & $13.72_{-0.08}^{+0.06}$        & $23.4_{-7.0}^{+4.8}$           & $-21.7_{-2.4}^{+1.2}$          \\
J0154 &  1.0116 & \ion{O}{IV}      & $13.77_{-0.14}^{+0.10}$        & $16.6_{-4.4}^{+2.4}$           & $-24.9_{-3.5}^{+3.1}$          \\
J0154 &  1.0116 & \ion{O}{V}       & $13.75_{-0.10}^{+0.08}$        & $14.6_{-3.4}^{+3.8}$           & $-26.6_{-2.2}^{+3.0}$          \\
\hline
\multicolumn{6}{c}{Absorption ID: s15c3}\\
\multicolumn{6}{c}{phase 1: $\log (n_{\rm H}/\cc) = -2.04_{-0.14}^{+0.11}$; $\log (T/{\rm K}) = 4.65_{-0.19}^{+0.09}$}\\
\multicolumn{6}{c}{phase 2: $\log (n_{\rm H}/\cc) = -3.60_{-0.19}^{+0.11}$; $\log (T/{\rm K}) = <5.4$}\\
\hline
J0154 &  1.0116 & \ion{He}{I}      & $>14.1$                        & $<16.5$                        & $28.6_{-1.2}^{+1.1}$           \\
J0154 &  1.0116 & \ion{C}{II}      & $13.60_{-0.09}^{+0.09}$        & $7.2_{-1.6}^{+2.7}$            & $24.8_{-1.5}^{+1.8}$           \\
J0154 &  1.0116 & \ion{N}{III}     & $13.96_{-0.06}^{+0.08}$        & $7.8_{-1.3}^{+1.5}$            & $25.3_{-1.0}^{+1.0}$           \\
J0154 &  1.0116 & \ion{N}{IV}      & $13.67_{-0.07}^{+0.19}$        & $14.8_{-6.4}^{+5.8}$           & $26.0_{-2.2}^{+2.2}$           \\
J0154 &  1.0116 & \ion{O}{II}      & $13.83_{-0.08}^{+0.07}$        & $14.2_{-5.6}^{+7.4}$           & $25.3_{-2.0}^{+2.9}$           \\
J0154 &  1.0116 & \ion{O}{III}     & $14.47_{-0.10}^{+0.15}$        & $10.1_{-1.4}^{+1.9}$           & $27.9_{-0.8}^{+0.6}$           \\
J0154 &  1.0116 & \ion{O}{IV}      & $14.21_{-0.09}^{+0.14}$        & $15.0_{-4.6}^{+4.4}$           & $22.8_{-0.6}^{+1.2}$           \\
J0154 &  1.0116 & \ion{O}{V}       & $13.89_{-0.03}^{+0.03}$        & $29.6_{-4.0}^{+3.7}$           & $28.1_{-2.6}^{+2.3}$           \\
J0154 &  1.0116 & \ion{Mg}{II}     & $12.22_{-0.02}^{+0.02}$        & $5.3_{-0.6}^{+0.7}$            & $27.4_{-0.3}^{+0.3}$           \\
J0154 &  1.0116 & \ion{S}{II}      & $12.92_{-0.09}^{+0.09}$        & $29.7_{-7.4}^{+11.2}$          & $29.1_{-3.1}^{+2.1}$           \\
J0154 &  1.0116 & \ion{S}{III}     & $13.03_{-0.05}^{+0.06}$        & $8.6_{-2.1}^{+2.5}$            & $29.3_{-1.1}^{+1.1}$           \\
\hline
\multicolumn{6}{c}{Absorption ID: s17c1}\\
\hline
J0333 &  1.0550 & \ion{O}{V}       & $13.53_{-0.08}^{+0.07}$        & $28.6_{-5.4}^{+5.6}$           & $19.6_{-3.6}^{+3.2}$           \\
\hline
\multicolumn{6}{c}{Absorption ID: s17c2}\\
\multicolumn{6}{c}{$\log (n_{\rm H}/\cc) = -3.48_{-0.13}^{+0.10}$; $\log (T/{\rm K}) < 5.1$}\\
\hline
J0333 &  1.0550 & \ion{N}{III}     & $13.00_{-0.17}^{+0.12}$        & $15.4_{-3.5}^{+3.2}$           & $51.8_{-3.4}^{+3.0}$           \\
J0333 &  1.0550 & \ion{O}{III}     & $13.70_{-0.09}^{+0.08}$        & $15.4_{-4.3}^{+3.2}$           & $49.3_{-2.2}^{+3.1}$           \\
J0333 &  1.0550 & \ion{O}{IV}      & $14.16_{-0.12}^{+0.22}$        & $19.8_{-9.0}^{+20.2}$          & $50.9_{-2.8}^{+3.1}$           \\
J0333 &  1.0550 & \ion{O}{V}       & $>14.0$                        & $<19.1$                        & $51.8_{-2.3}^{+2.2}$           \\
\hline
\multicolumn{6}{c}{Absorption ID: s17c3}\\
\hline
J0333 &  1.0550 & \ion{O}{V}       & $13.86_{-0.09}^{+0.09}$        & $15.1_{-5.8}^{+6.4}$           & $85.9_{-2.9}^{+3.0}$           \\
\hline
\multicolumn{6}{c}{Absorption ID: s17h}\\
\hline
J0333 &  1.0550 & \ion{Ne}{VI}     & $14.02_{-0.08}^{+0.07}$        & $46.1_{-5.2}^{+2.8}$           & $87.6_{-3.9}^{+2.0}$           \\
J0333 &  1.0550 & \ion{Ne}{VIII}   & $13.98_{-0.08}^{+0.07}$        & $39.6_{-9.8}^{+11.3}$          & $79.6_{-5.8}^{+6.2}$           \\
\hline
\multicolumn{6}{c}{Absorption ID: s17c4}\\
\multicolumn{6}{c}{$\log (n_{\rm H}/\cc) = -3.03_{-0.14}^{+0.10}$; $\log (T/{\rm K}) < 5.0$}\\
\hline
J0333 &  1.0550 & \ion{O}{III}     & $13.76_{-0.06}^{+0.07}$        & $9.2_{-2.7}^{+3.6}$            & $142.7_{-1.6}^{+1.9}$          \\
J0333 &  1.0550 & \ion{O}{IV}      & $13.95_{-0.05}^{+0.05}$        & $16.0_{-3.8}^{+3.9}$           & $143.6_{-2.0}^{+2.4}$          \\
J0333 &  1.0550 & \ion{O}{V}       & $13.75_{-0.04}^{+0.04}$        & $25.1_{-4.9}^{+6.0}$           & $145.3_{-2.3}^{+2.1}$          \\
J0333 &  1.0550 & \ion{S}{IV}      & $12.34_{-0.16}^{+0.12}$        & $15.2_{-3.5}^{+3.3}$           & $147.6_{-3.3}^{+1.9}$          \\
\hline
\multicolumn{6}{c}{Absorption ID: s18c1}\\
\multicolumn{6}{c}{$\log (n_{\rm H}/\cc) = -3.37_{-0.12}^{+0.10}$; $\log (T/{\rm K}) < 5.4$}\\
\hline
J0333 &  1.0631 & \ion{O}{III}     & $13.71_{-0.09}^{+0.09}$        & $26.8_{-3.8}^{+2.4}$           & $-19.1_{-3.8}^{+2.9}$          \\
J0333 &  1.0631 & \ion{O}{IV}      & $14.08_{-0.03}^{+0.03}$        & $26.6_{-3.6}^{+3.7}$           & $-23.6_{-1.0}^{+1.6}$          \\
J0333 &  1.0631 & \ion{O}{V}       & $14.11_{-0.04}^{+0.04}$        & $25.1_{-5.0}^{+5.3}$           & $-21.3_{-2.3}^{+2.2}$          \\
\hline
    \end{tabular}
\end{table}

\begin{table}
    \contcaption{Summary of detected ions}
    \centering
    \begin{tabular}{llcccc}
\hline
\hline
\multicolumn{6}{c}{Absorption ID: s18c2}\\
\multicolumn{6}{c}{$\log (n_{\rm H}/\cc) = -3.55_{-0.16}^{+0.14}$; $\log (T/{\rm K}) < 5.3$}\\
\hline
J0333 &  1.0631 & \ion{O}{IV}      & $13.51_{-0.08}^{+0.07}$        & $13.6_{-5.0}^{+6.2}$           & $68.0_{-2.0}^{+2.6}$           \\
J0333 &  1.0631 & \ion{O}{V}       & $13.14_{-0.19}^{+0.11}$        & $16.3_{-6.8}^{+8.7}$           & $70.0_{-3.2}^{+3.3}$           \\
\hline
\multicolumn{6}{c}{Absorption ID: s20c1}\\
\multicolumn{6}{c}{$\log (n_{\rm H}/\cc) = -4.24_{-0.05}^{+0.06}$; $\log (T/{\rm K}) < 4.9$}\\
\hline
J0154 &  1.2133 & CIV      & $13.52_{-0.10}^{+0.13}$        & $8.3_{-2.9}^{+3.2}$            & $55.4_{-1.7}^{+1.5}$           \\
J0154 &  1.2133 & \ion{O}{IV}      & $14.14_{-0.08}^{+0.09}$        & $10.7_{-4.2}^{+6.7}$           & $52.5_{-1.1}^{+1.9}$           \\
J0154 &  1.2133 & \ion{O}{V}       & $>14.2$                        & $<14.4$                        & $55.3_{-2.8}^{+3.6}$           \\
J0154 &  1.2133 & \ion{Ne}{V}      & $14.07_{-0.11}^{+0.15}$        & $11.6_{-5.3}^{+21.4}$          & $56.8_{-2.8}^{+2.7}$           \\
J0154 &  1.2133 & \ion{Ne}{VI}     & $14.07_{-0.10}^{+0.08}$        & $25.2_{-6.8}^{+7.3}$           & $56.6_{-3.3}^{+3.0}$           \\
\hline
\multicolumn{6}{c}{Absorption ID: s20h}\\
\hline
J0154 &  1.2133 & \ion{Ne}{VIII}   &
$14.03_{-0.10}^{+0.09}$        & $31.9_{-8.9}^{+10.2}$          & $68.0_{-5.2}^{+5.0}$           \\
\hline
\multicolumn{6}{c}{Absorption ID: s20c2}\\
\hline
J0154 &  1.2133 & \ion{O}{V}       & $13.80_{-0.19}^{+0.18}$        & $12.0_{-3.6}^{+5.4}$           & $84.8_{-2.5}^{+3.3}$       \\
\hline

    \end{tabular}

\begin{flushleft}
$^a$ For each absorption component with multiple metal transitions detected, we report the gas density determined from photoionization models and the temperature obtained from line width analysis (see Section \ref{sec:med_abs} and \citetalias{CUBSV}).
A single-phase photoionization model is a good representation for all but four absorption components.
For these four components, a two-phase model is sufficient to produce the observed ion ratios.  The derived density and temperature are noted for each of the two phases as phase 1 and 2 in this table.\\
$^b$ When the decomposition of components is not feasible for the targeted ion, we fix the line centroid to the best-constrained ions in the same component (e.g., \ion{Mg}{II}) in the Voigt profile analysis.
These components are shown as entries without uncertainties for the line centroid.\\
$^c$ Systems have been used in the analysis of \ion{H}{I}/\ion{He}{I} ratio.\\
$^d$ Although these ions have a matched line centroid with their assigned component, 
the best-fit photoionisation model suggests they are likely to be contaminated.
\end{flushleft}
\end{table}

\clearpage

\setcounter{figure}{0}  
\setcounter{table}{0}

\begin{figure*}
\begin{center}
\includegraphics[width=0.98\textwidth]{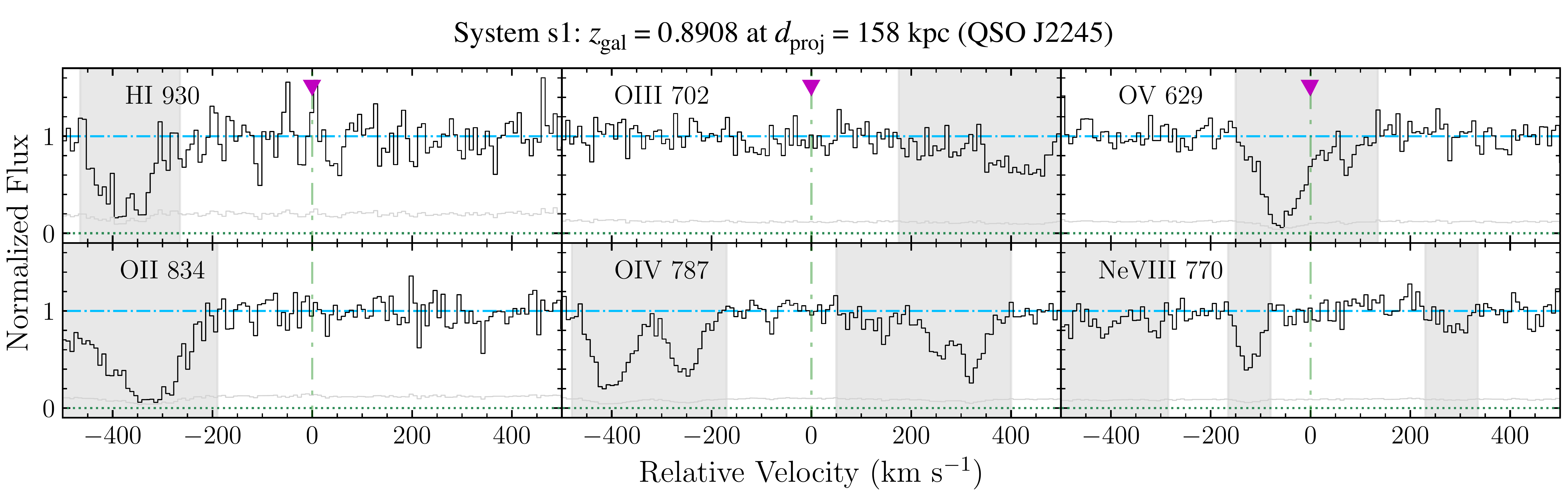}
\end{center}
\caption{System ID s1. Continuum normalized absorption profiles of different transitions of the galaxy system at $z=0.8908$ in the QSO field J2245.
The title of this plot indicates the redshift of the closest galaxy projected around the QSO, and its impact parameter.
The transition is identified in the top-left of each panel.
Particularly for this QSO-galaxy pair, no transition is detected in absorption.
Zero velocity marks the redshift of the closest galaxy in the galaxy system.
The magenta downward triangles and the pale green vertical lines mark the redshift of all member galaxies, with the closest galaxy marked as filled symbols.
The black step lines and the grey lines represent the normalized flux and the associated 1-$\sigma$ uncertainty.
Contamination features are shadowed by grey regions for clarity.
}
\label{fig:a1}
\end{figure*}

\begin{figure*}
\begin{center}
\includegraphics[width=0.98\textwidth]{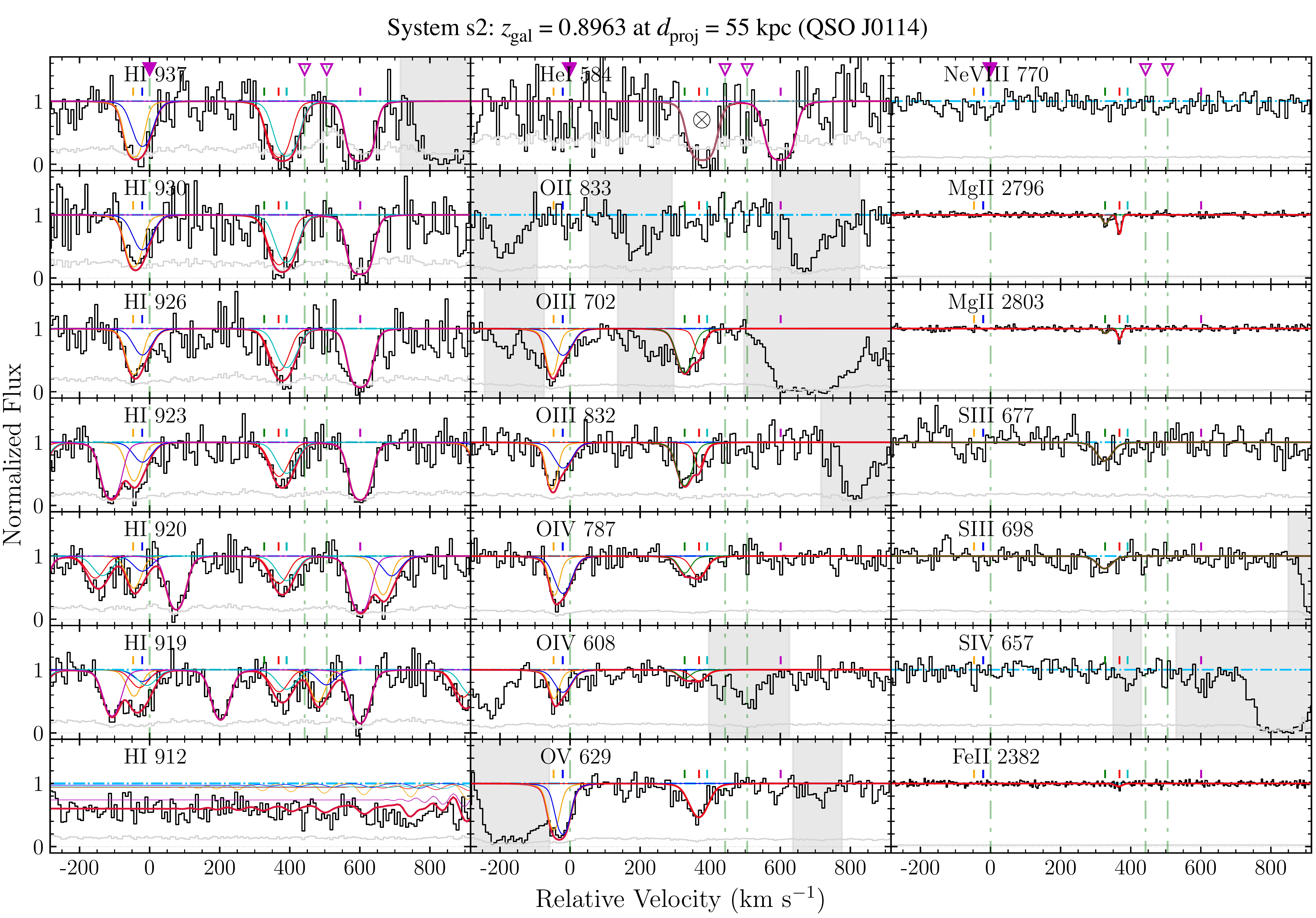}
\end{center}
\caption{System ID s2. Similar to Figure \ref{fig:a1}, but for the galaxy system at $z=0.8963$ towards QSO field J0114.
In this system, the six absorption components are detected, and marked in vertical colour bars above the normalized flux.
The best-fit Voigt profiles for each individual transition detected are plotted, both for the sum of all components (red curve) and for individual components (different-coloured curves).
Limited by the relatively low $S/N$ of the \ion{He}{I} $\lambda$ 584 transition, we cannot decompose the absorption feature at $v\approx 400 \kms$ into three components aligned with other ionic transitions (marked with the $\otimes$ symbol).
}
\label{fig:a2}
\end{figure*}

\begin{figure*}
\begin{center}
\includegraphics[width=0.98\textwidth]{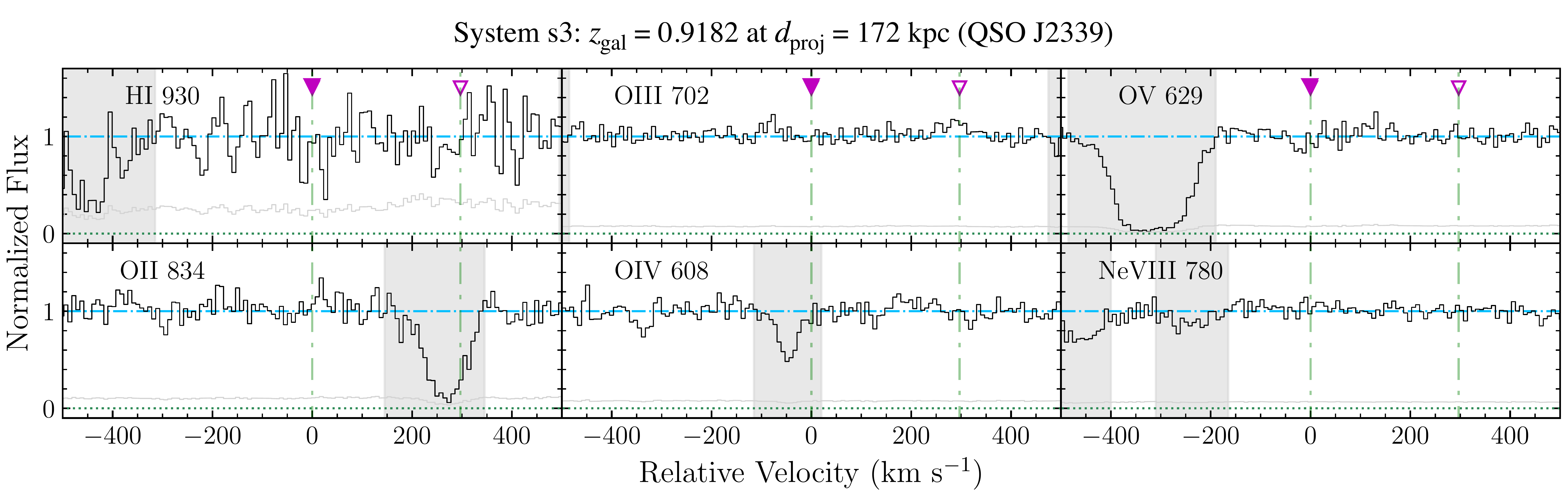}
\end{center}
\caption{System ID s3. Similar to Figure \ref{fig:a1}, but for the galaxy system at $z=0.9182$ towards QSO field J2339.
}
\label{fig:a3}
\end{figure*}

\begin{figure*}
\begin{center}
\includegraphics[width=0.98\textwidth]{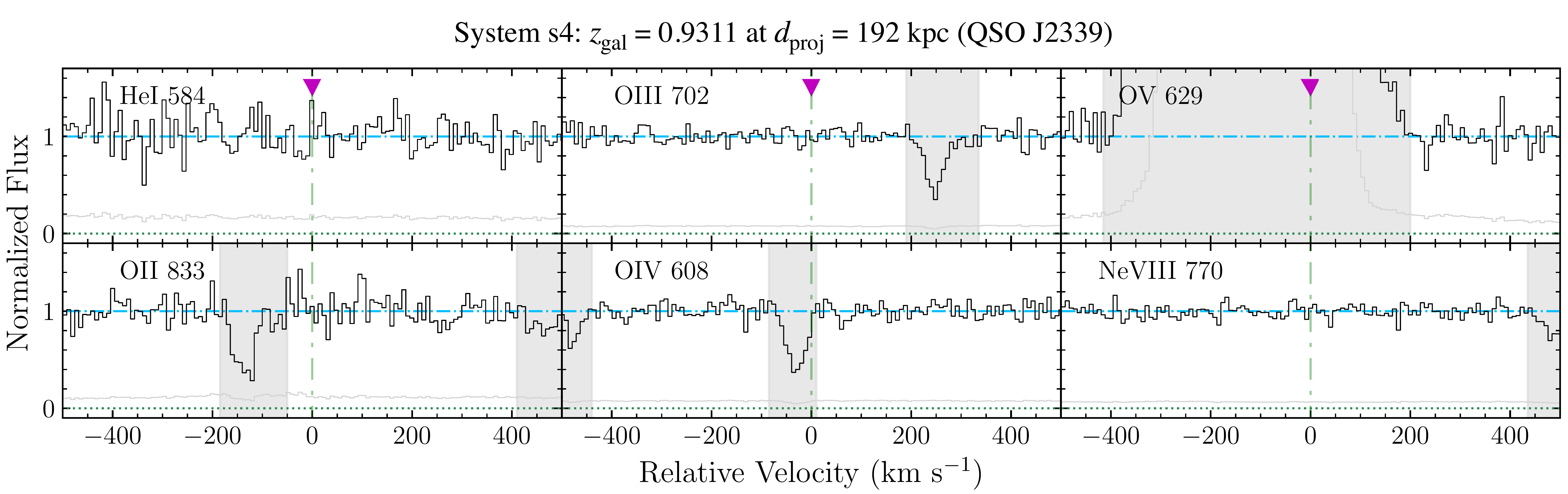}
\end{center}
\caption{System ID s4. Similar to Figure \ref{fig:a1}, but for the galaxy system at $z=0.9311$ towards QSO field J2339.
}
\label{fig:a4}
\end{figure*}

\begin{figure*}
\begin{center}
\includegraphics[width=0.98\textwidth]{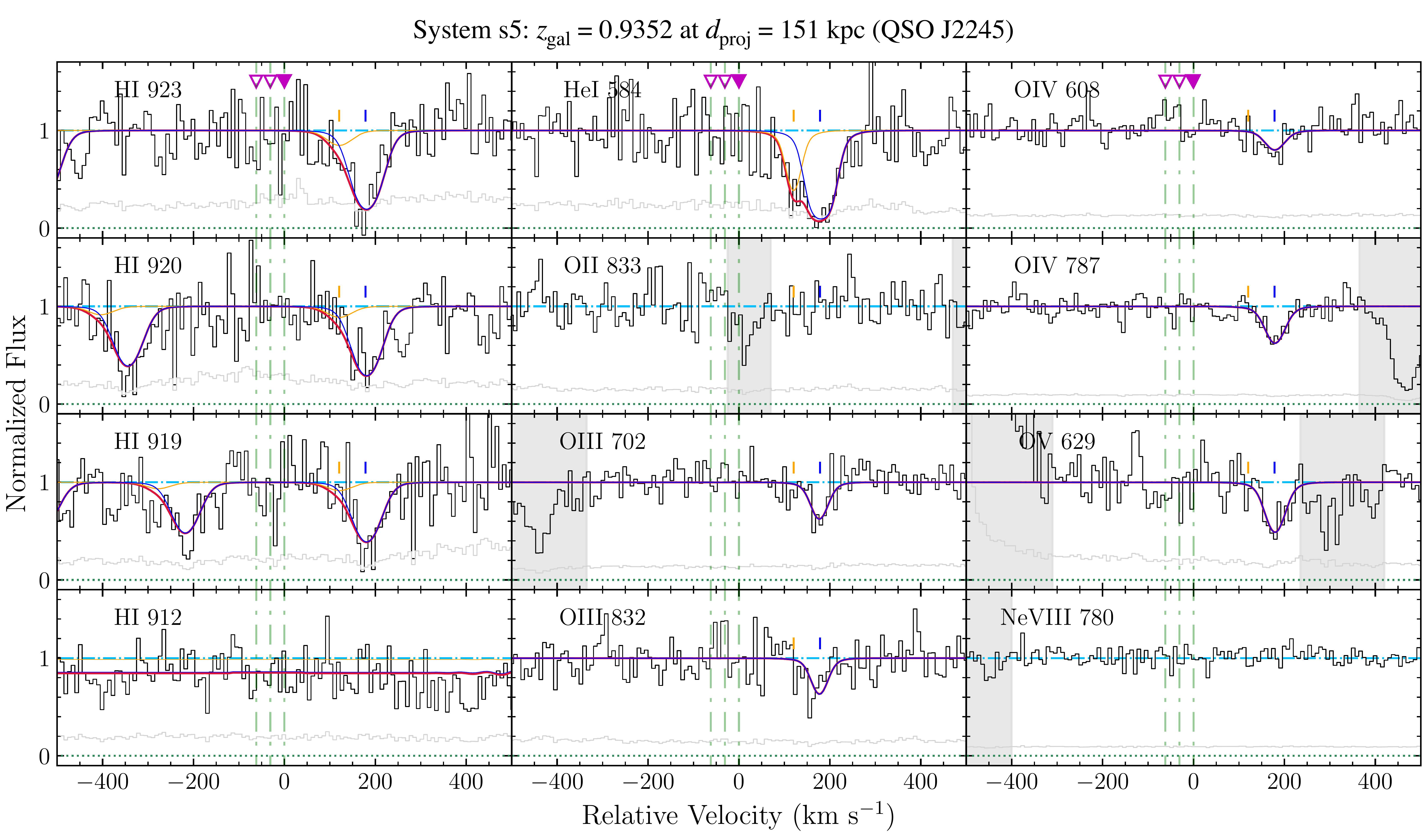}
\end{center}
\caption{System ID s5. Similar to Figure \ref{fig:a2}, but for the galaxy system at $z=0.9352$ towards QSO field J2245.
}
\label{fig:a5}
\end{figure*}

\begin{figure*}
\begin{center}
\includegraphics[width=0.98\textwidth]{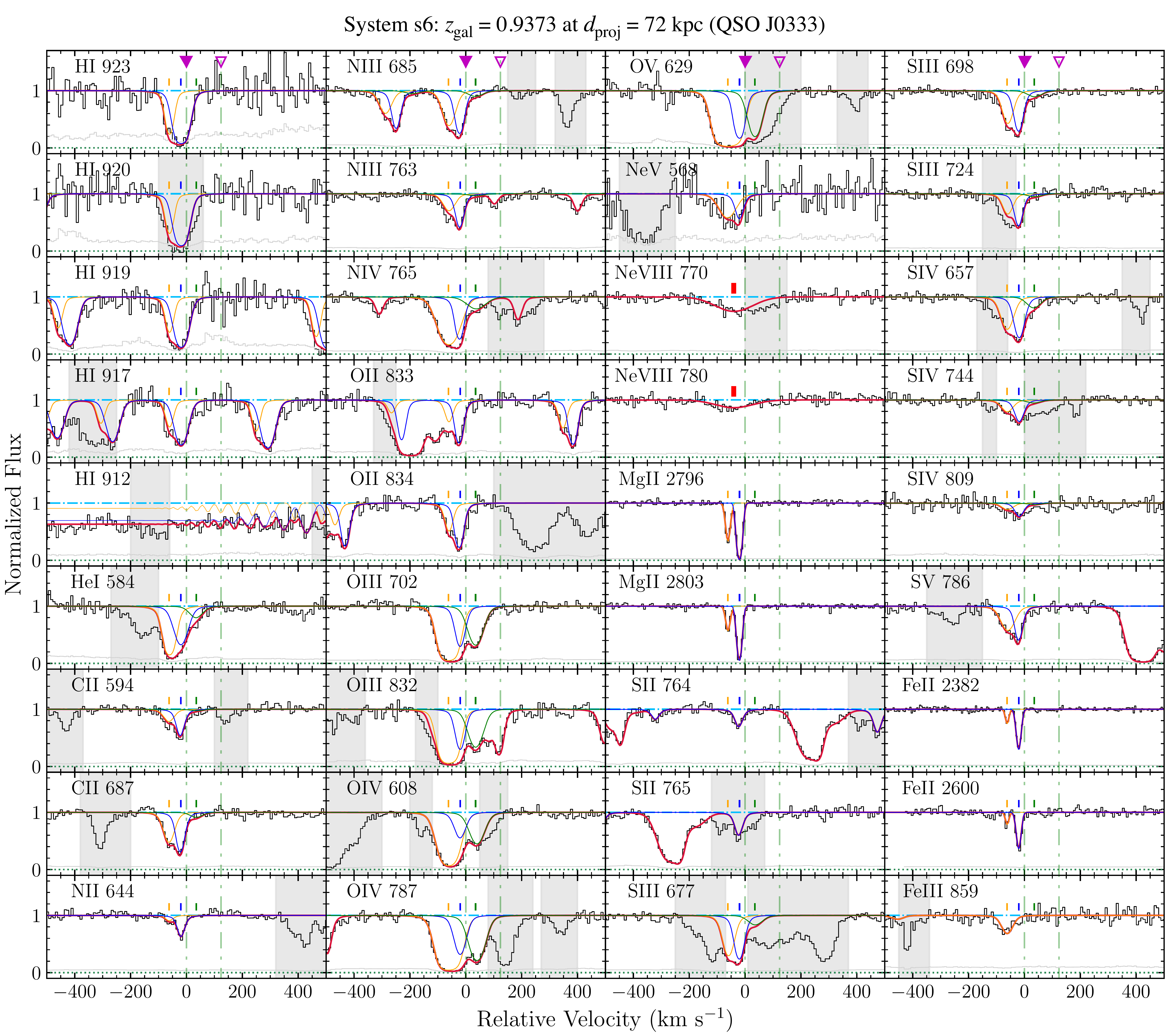}
\end{center}
\caption{System ID s6. Similar to Figure \ref{fig:a2}, but for the galaxy system at $z=0.9373$ towards QSO field J0333. The hot CGM is detected by the \ion{Ne}{VIII} doublet (marked with a thick red bar).
This system contains the component c2 with the extremely high $N$(\ion{H}{I})/$N$(\ion{He}{I}) ratio (Figure \ref{fig:HI_HeI_hm05}).
}
\label{fig:a6}
\end{figure*}

\begin{figure*}
\begin{center}
\includegraphics[width=0.98\textwidth]{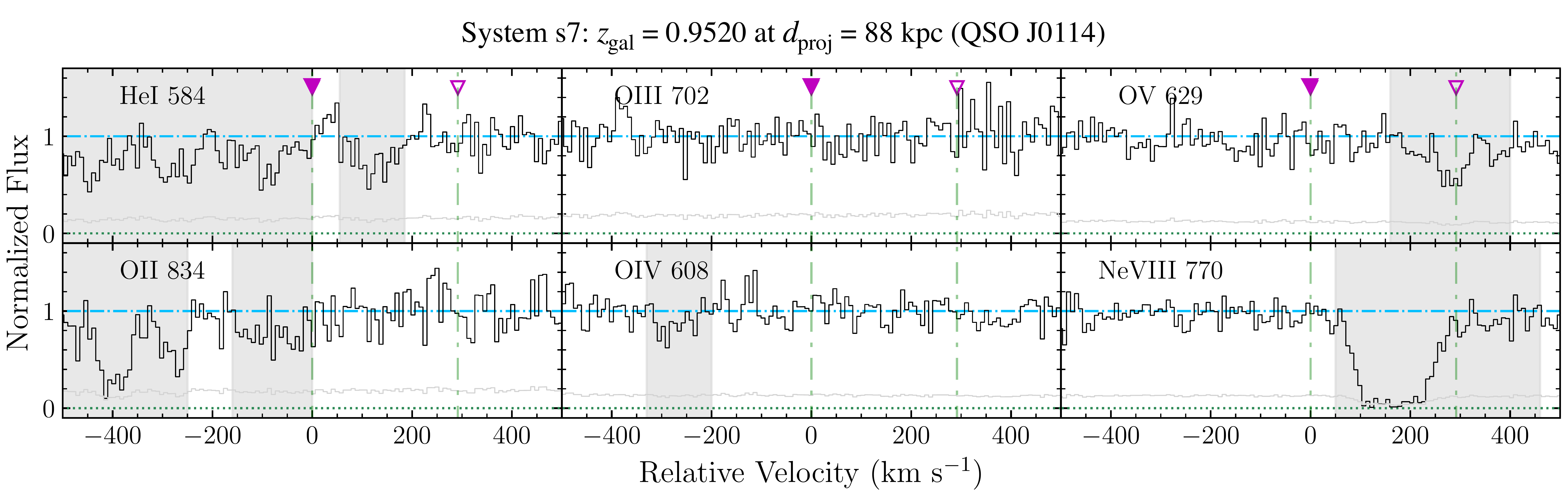}
\end{center}
\caption{System ID s7. Similar to Figure \ref{fig:a1}, but for the galaxy system at $z=0.9520$ towards QSO field J0114.
}
\label{fig:a7}
\end{figure*}

\begin{figure*}
\begin{center}
\includegraphics[width=0.98\textwidth]{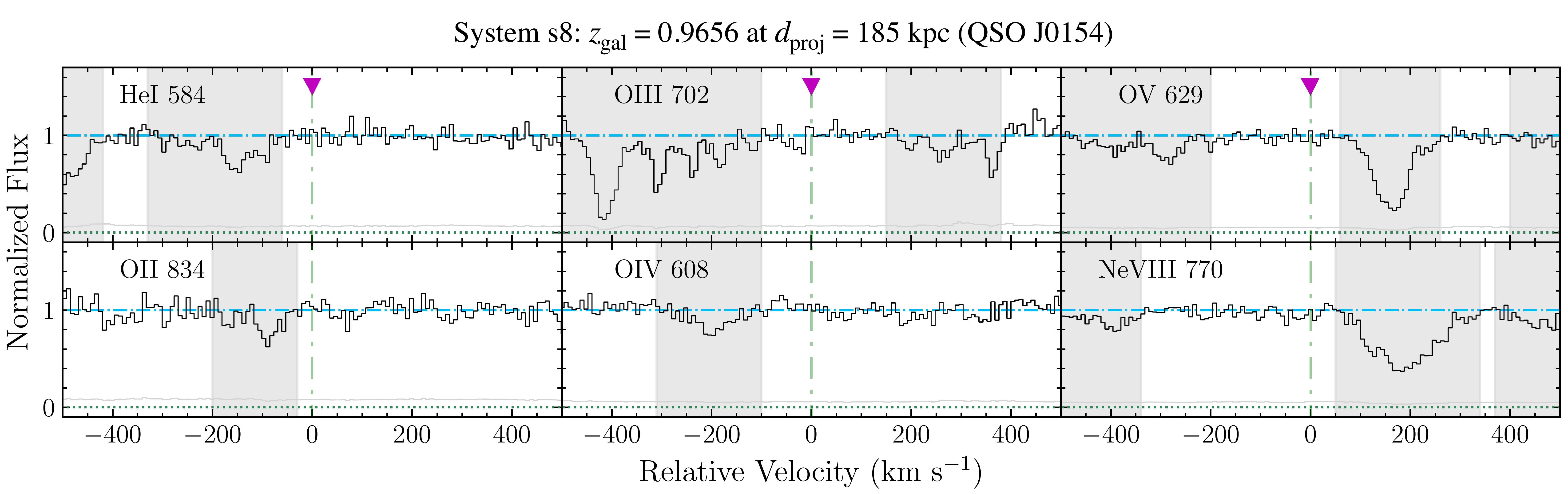}
\end{center}
\caption{System ID s8. Similar to Figure \ref{fig:a1}, but for the galaxy system at $z=0.9656$ towards QSO field J0154.
}
\label{fig:a8}
\end{figure*}

\begin{figure*}
\begin{center}
\includegraphics[width=0.98\textwidth]{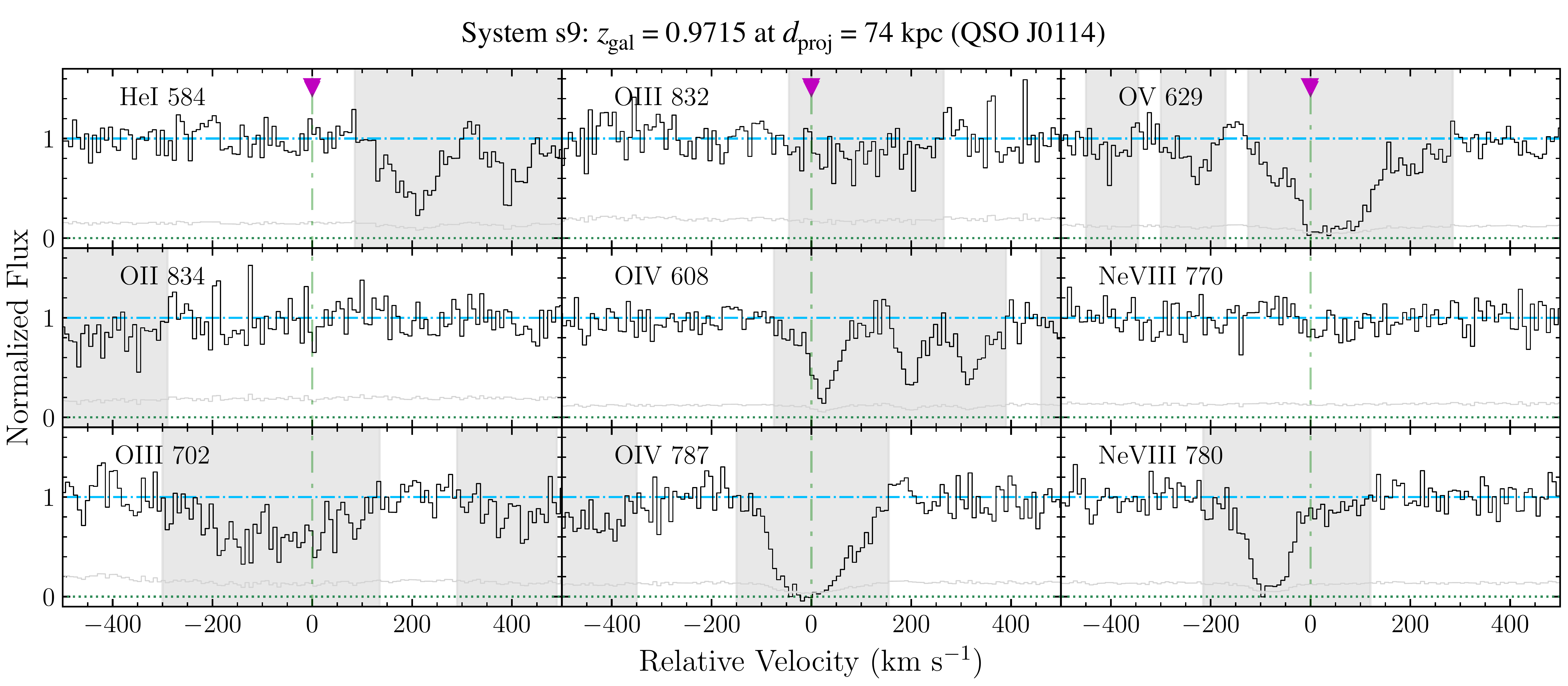}
\end{center}
\caption{System ID s9. Similar to Figure \ref{fig:a1}, but for the galaxy system at $z=0.9715$ towards QSO field J0114.
This system is excluded from absorption analysis, because no robust constraints can be obtained.
}
\label{fig:a9}
\end{figure*}

\begin{figure*}
\begin{center}
\includegraphics[width=0.98\textwidth]{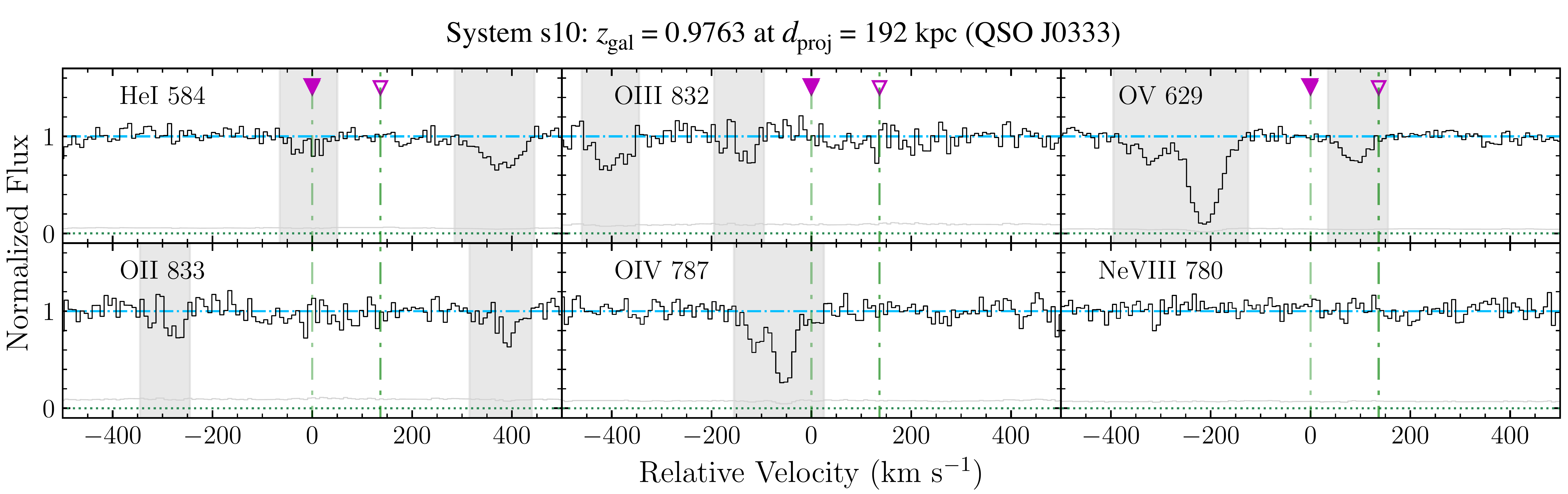}
\end{center}
\caption{System ID s10. Similar to Figure \ref{fig:a1}, but for the galaxy system at $z=0.9763$ towards QSO field J0333.
}
\label{fig:a10}
\end{figure*}

\begin{figure*}
\begin{center}
\includegraphics[width=0.98\textwidth]{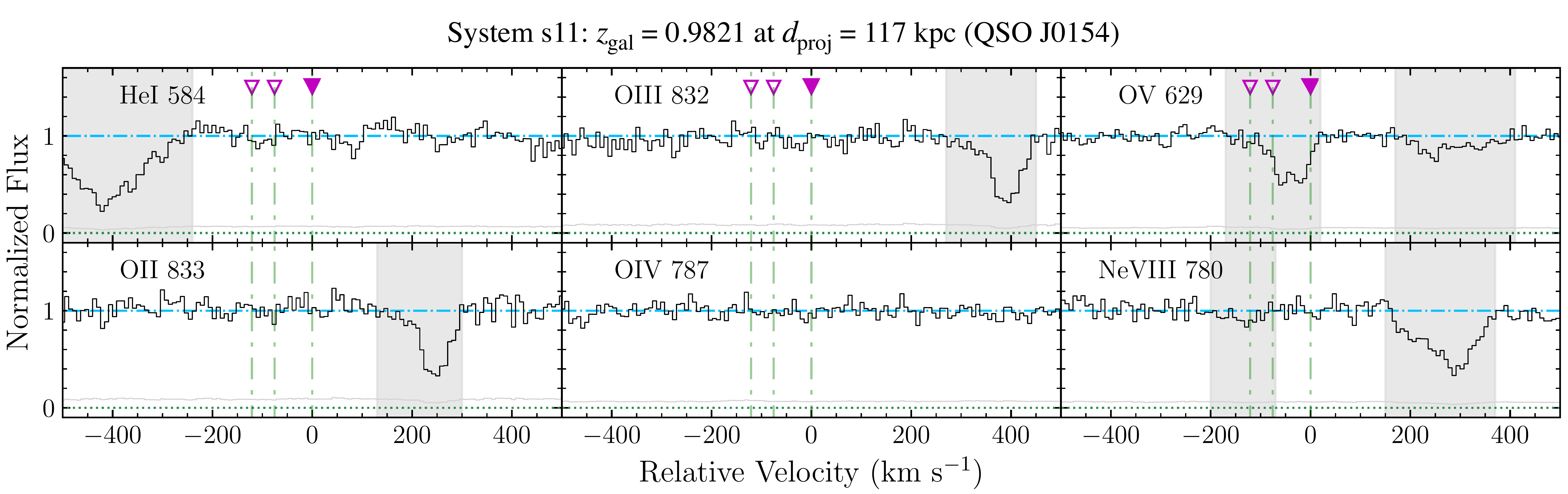}
\end{center}
\caption{System ID s11. Similar to Figure \ref{fig:a1}, but for the galaxy system at $z=0.9821$ towards QSO field J0154.
}
\label{fig:a11}
\end{figure*}

\begin{figure*}
\begin{center}
\includegraphics[width=0.98\textwidth]{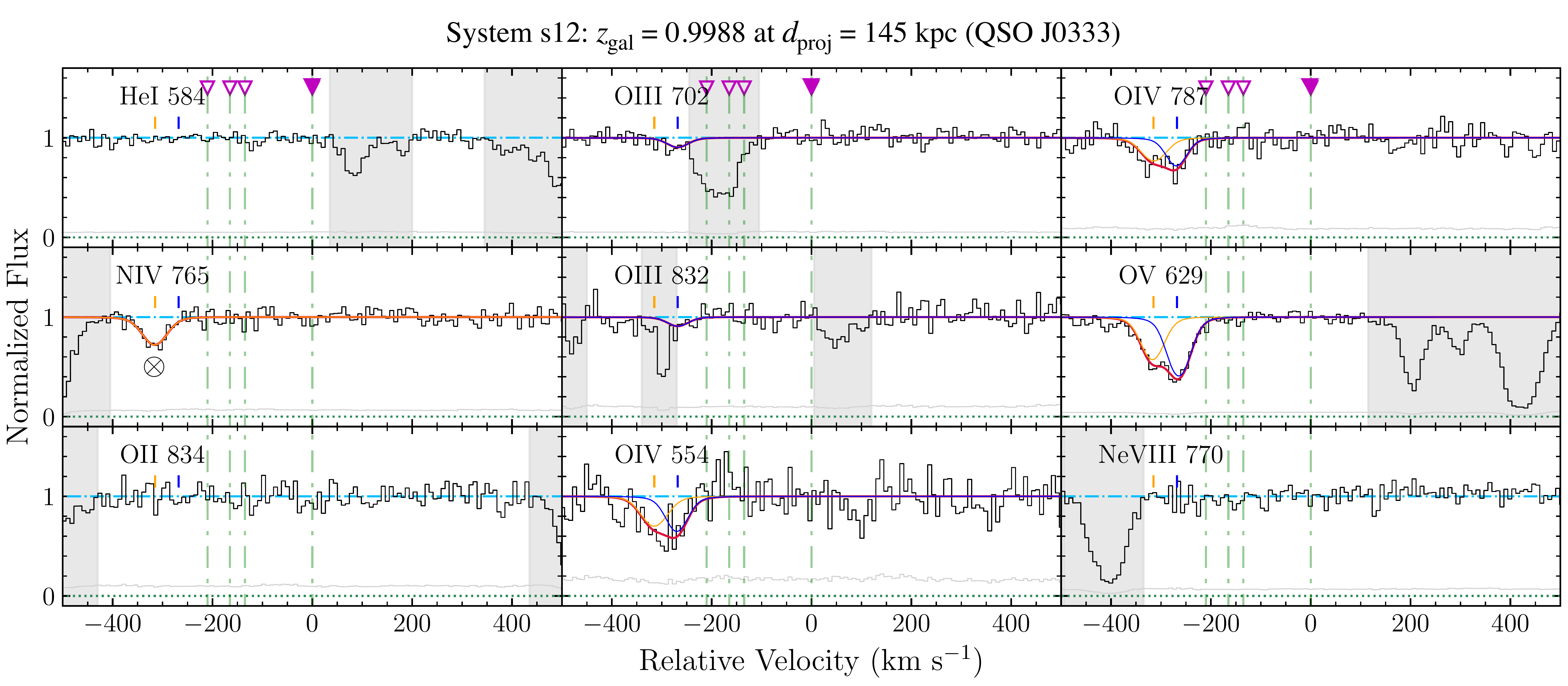}
\end{center}
\caption{System ID s12. Similar to Figure \ref{fig:a2}, but for the galaxy system at $z=0.9988$ towards QSO field J0333. The single transition of \ion{N}{IV} may be contaminated by unknown features consider the unusual nitrogen enhancement.
}
\label{fig:a12}
\end{figure*}

\begin{figure*}
\begin{center}
\includegraphics[width=0.98\textwidth]{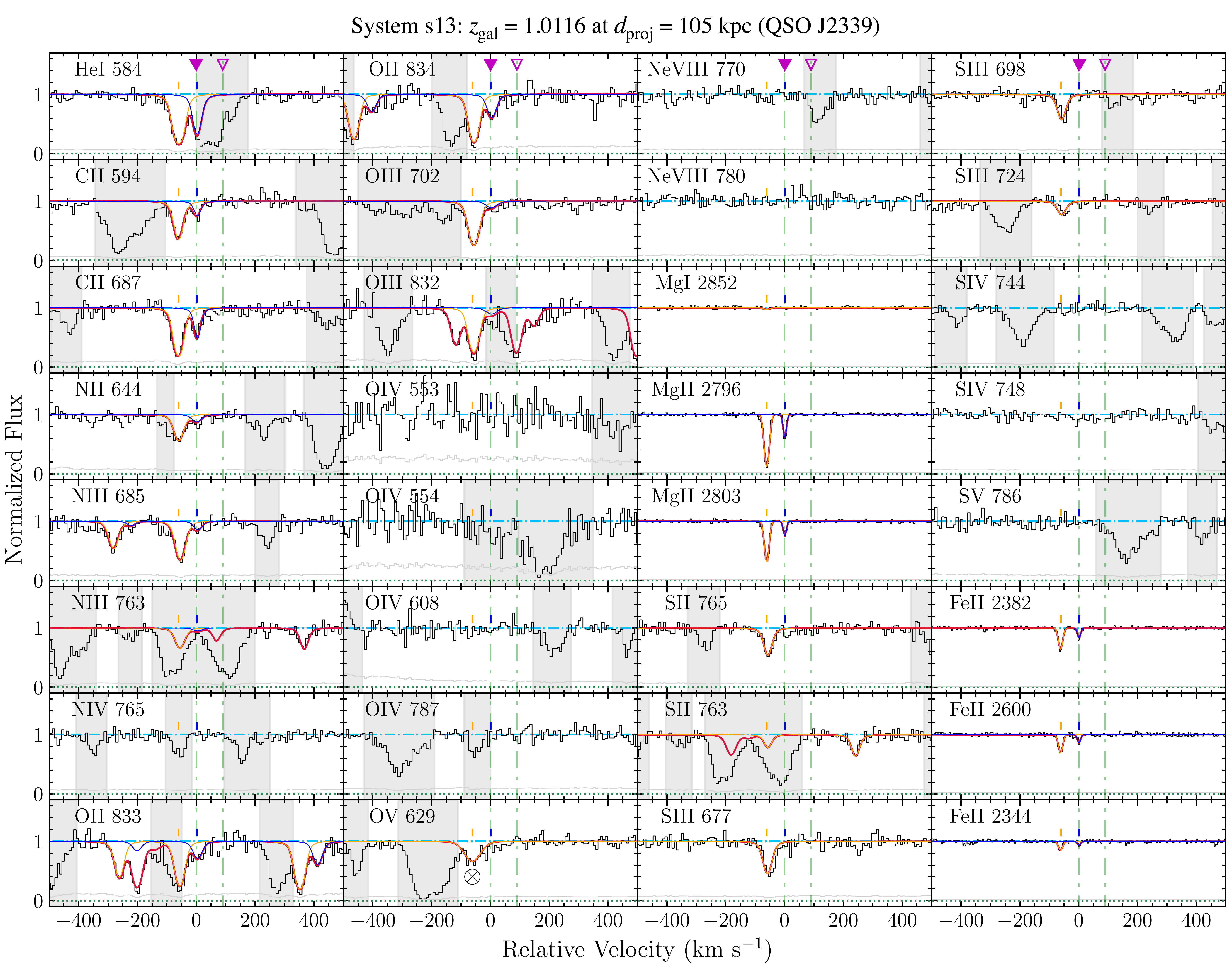}
\end{center}
\caption{System ID s13. Similar to Figure \ref{fig:a2}, but for the galaxy system at $z=1.0032$ towards QSO field J2339.
}
\label{fig:a13}
\end{figure*}

\begin{figure*}
\begin{center}
\includegraphics[width=0.98\textwidth]{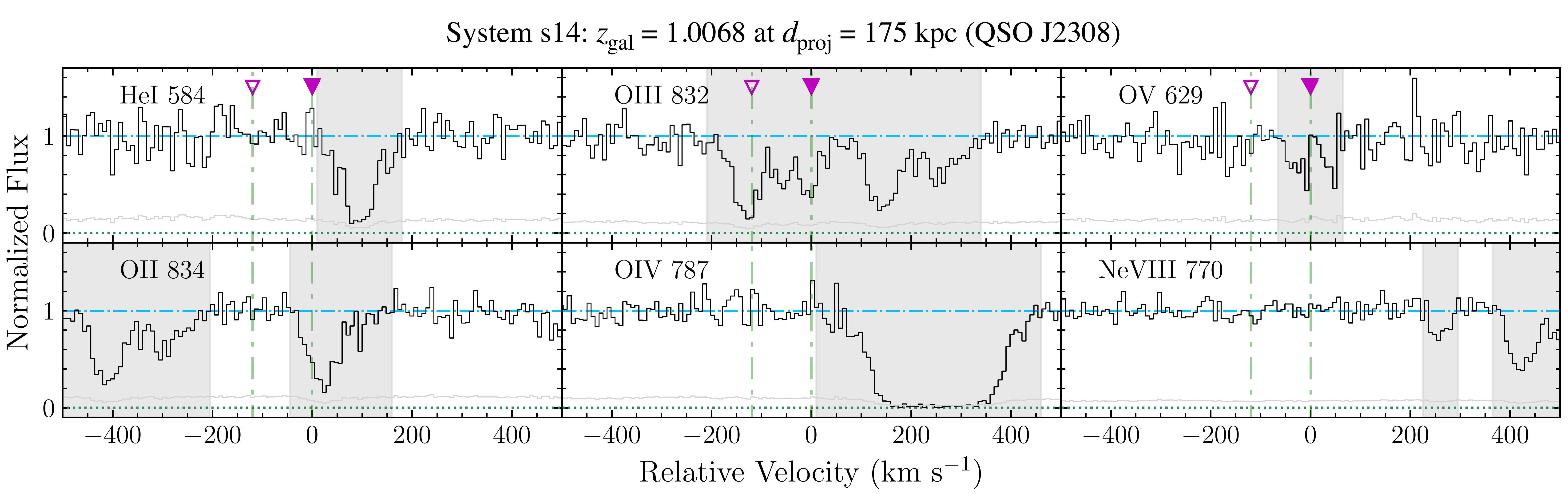}
\end{center}
\caption{System ID s14. Similar to Figure \ref{fig:a1}, but for the galaxy system at $z=1.0068$ towards QSO field J2308.
}
\label{fig:a14}
\end{figure*}

\begin{figure*}
\begin{center}
\includegraphics[width=0.98\textwidth]{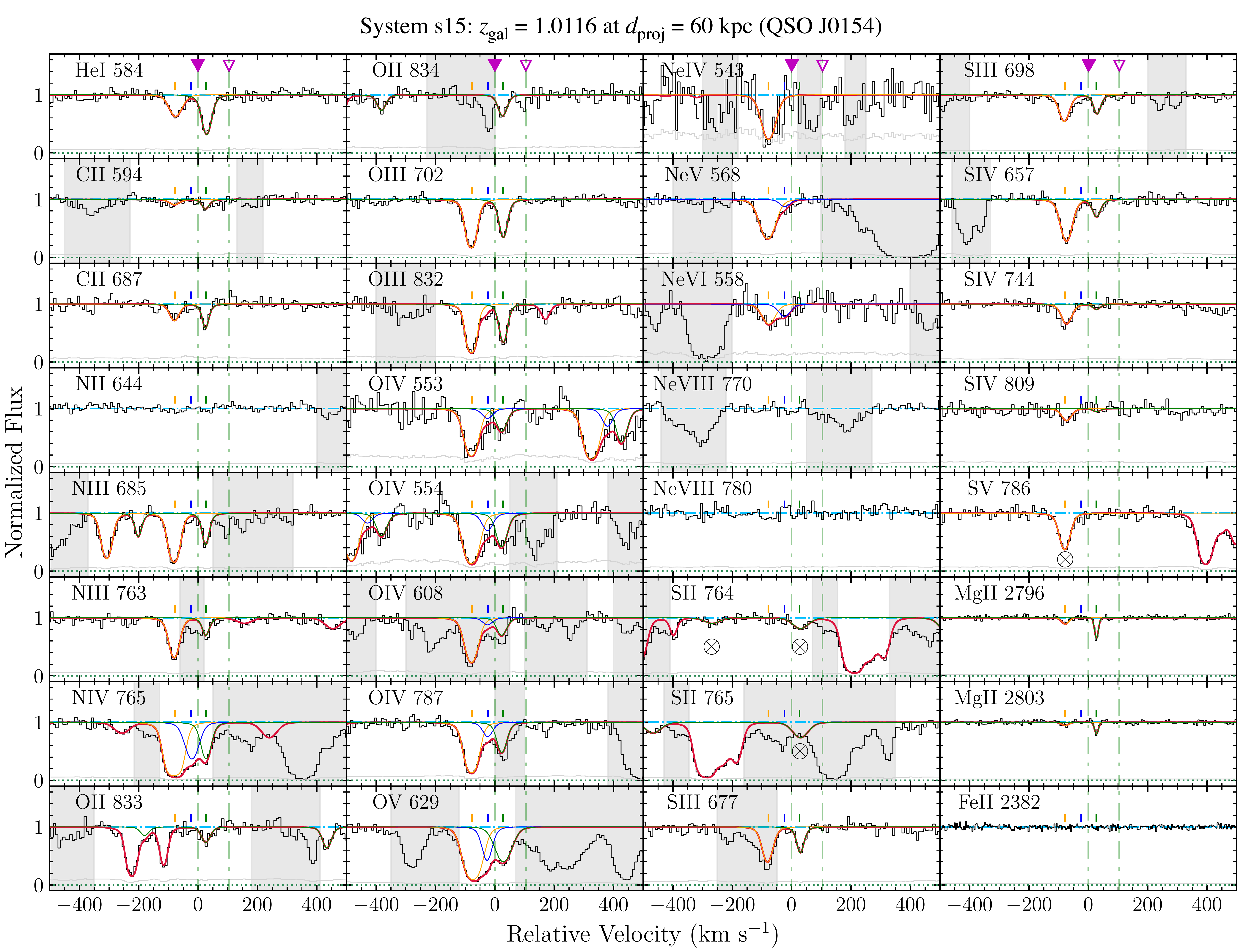}
\end{center}
\caption{System ID s15. Similar to Figure \ref{fig:a2}, but for the galaxy system at $z=1.0116$ towards QSO field J0154.
The \ion{S}{II} transitions marked with $\otimes$ symbols are suggested to be contaminated in the best-fit photoionisation model.
}
\label{fig:a15}
\end{figure*}

\begin{figure*}
\begin{center}
\includegraphics[width=0.98\textwidth]{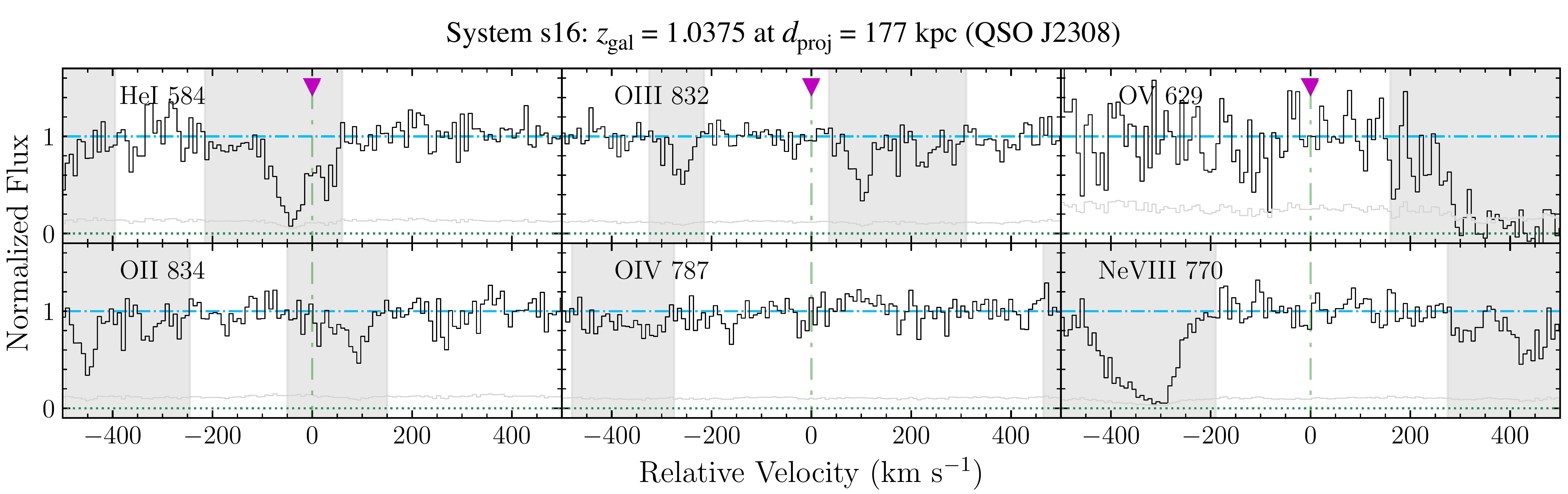}
\end{center}
\caption{System ID s16. Similar to Figure \ref{fig:a1}, but for the galaxy system at $z=1.0375$ towards QSO field J2308.
}
\label{fig:a16}
\end{figure*}

\begin{figure*}
\begin{center}
\includegraphics[width=0.98\textwidth]{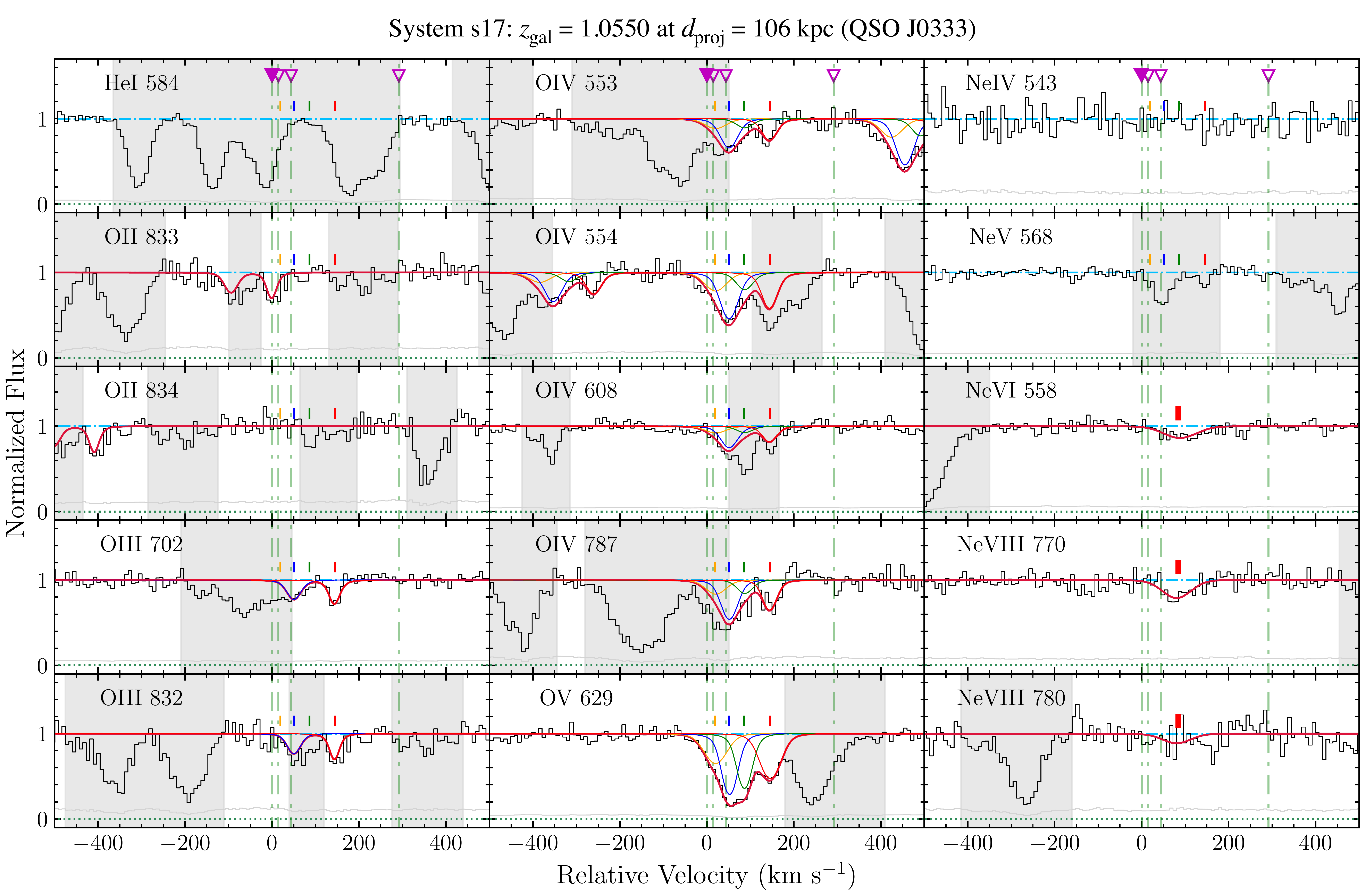}
\end{center}
\caption{System ID s17. Similar to Figure \ref{fig:a2}, but for the galaxy system at $z=1.0550$ towards QSO field J2339. The hot CGM is detected by \ion{Ne}{VI} $\lambda$ 558 and \ion{Ne}{VIII} $\lambda$ 770 (marked with a thick red bar), while the \ion{Ne}{VIII} $\lambda$ 780 transition is non-detected.
The two components at $v=20\kms$ (orange) and $86\kms$ (green) only have solid detections of \ion{O}{V}. However, we still consider them as absorption components, because \ion{O}{IV} of these two components are $\approx 2 \sigma$ detections.
}
\label{fig:a17}
\end{figure*}

\begin{figure*}
\begin{center}
\includegraphics[width=0.98\textwidth]{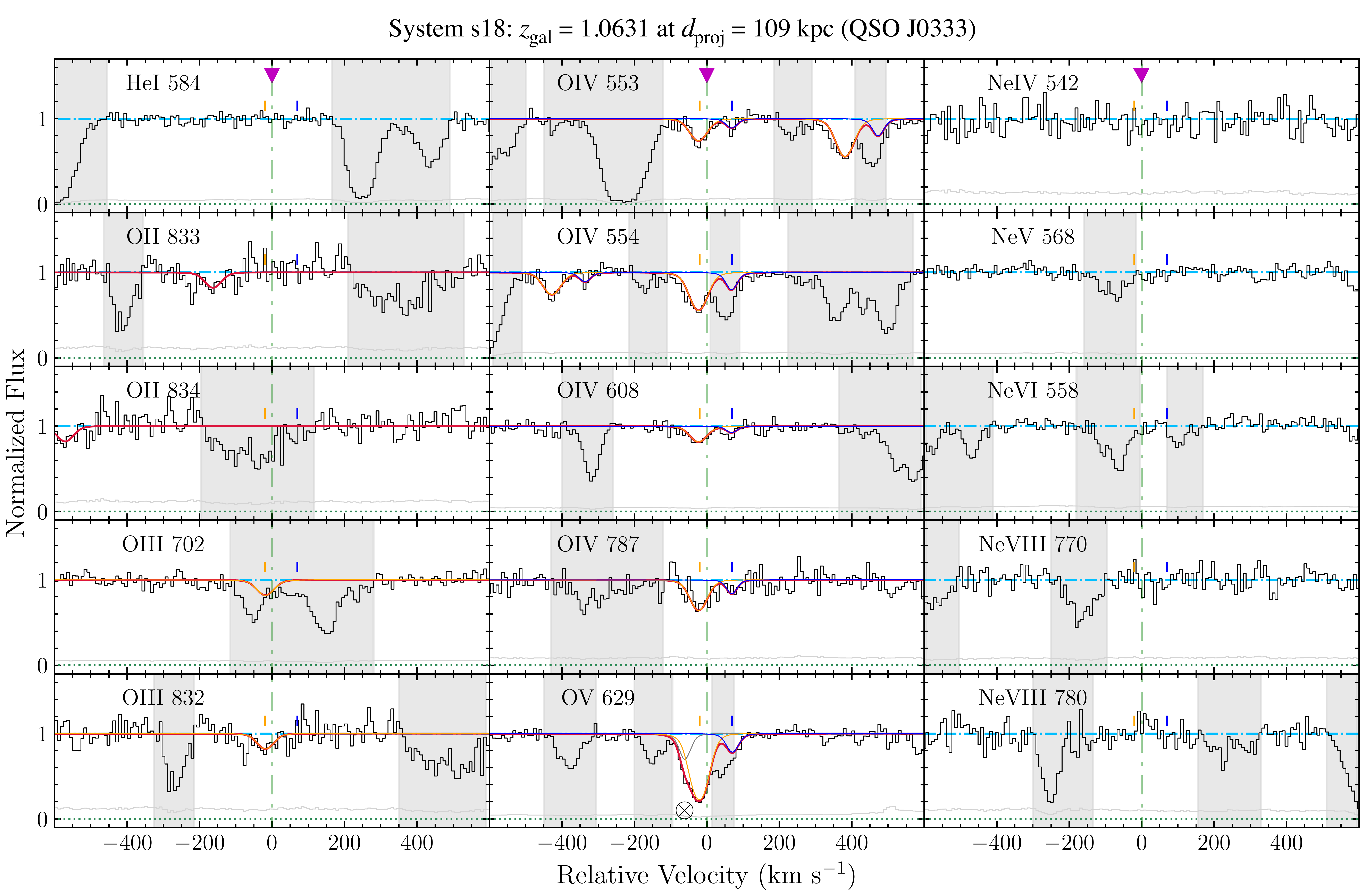}
\end{center}
\caption{System ID s18. Similar to Figure \ref{fig:a2}, but for the galaxy system at $z=1.0631$ towards QSO field J0333.
The leftmost component in the \ion{O}{V} transition is a contamination line included to obtain the best-fit voigt model, which is marked with the $\otimes$ symbol.
}
\label{fig:a18}
\end{figure*}

\begin{figure*}
\begin{center}
\includegraphics[width=0.98\textwidth]{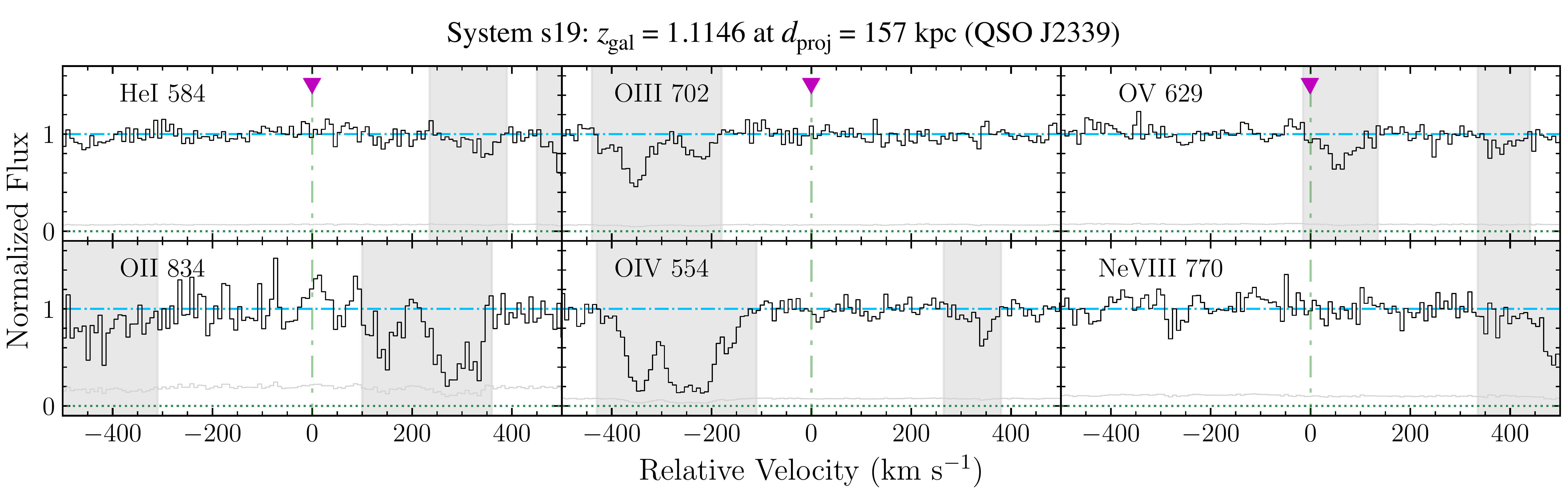}
\end{center}
\caption{System ID s19. Similar to Figure \ref{fig:a1}, but for the galaxy system at $z=1.1146$ towards QSO field J2339.
}
\label{fig:a19}
\end{figure*}

\begin{figure*}
\begin{center}
\includegraphics[width=0.98\textwidth]{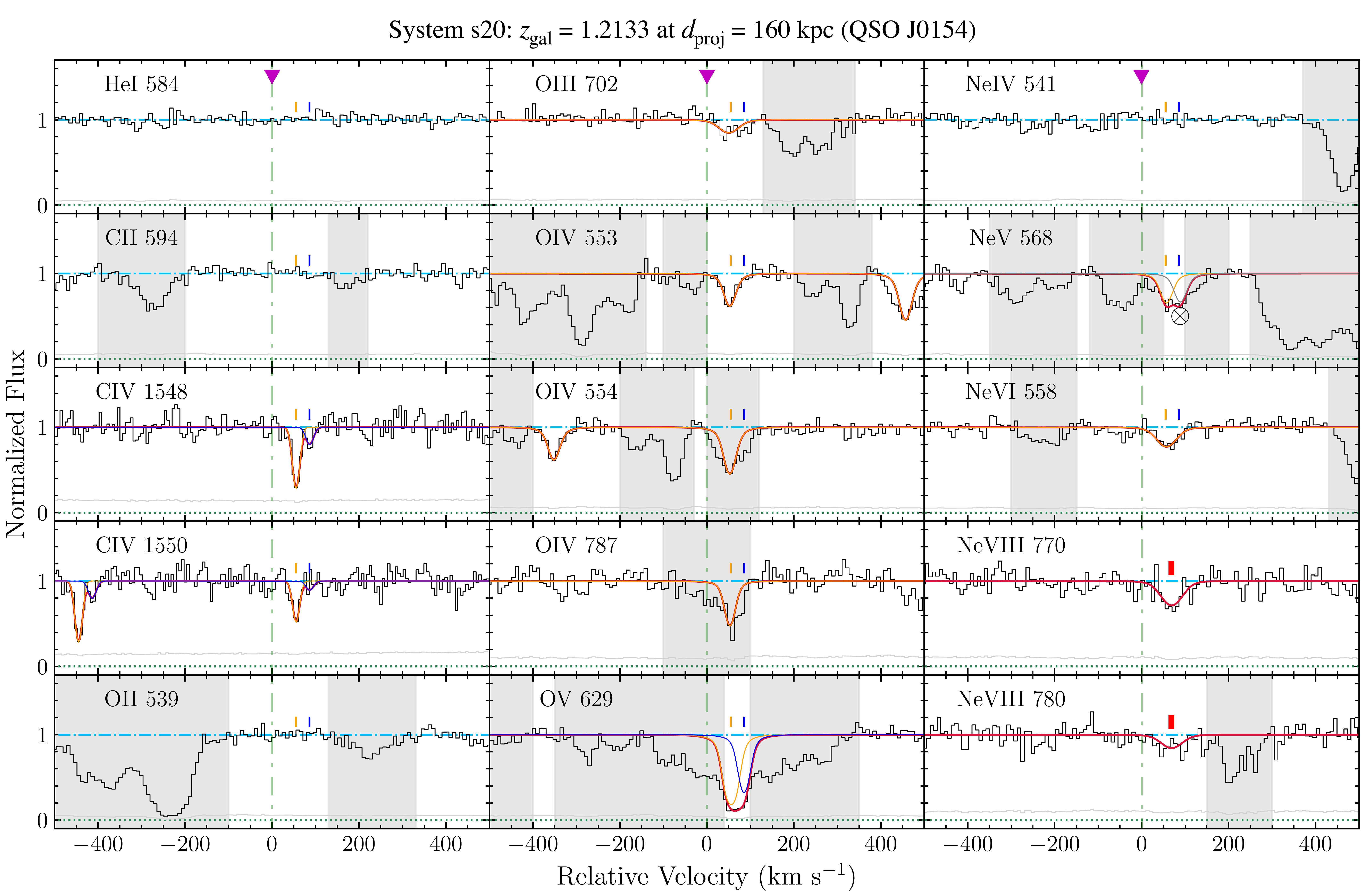}
\end{center}
\caption{System ID s20. Similar to Figure \ref{fig:a2}, but for the galaxy system at $z=1.2133$ towards QSO field J0154. The hot CGM is detected by the \ion{Ne}{VIII} doublet (marked with a thick red bar). The \ion{Ne}{VI} $\lambda$ 558 is consistent with the single photoionization phase producing \ion{O}{IV}, \ion{O}{V}, and \ion{Ne}{V}. The component at $v=86\kms$ (blue) only has one solid detection of \ion{O}{V}. However, we still consider it as a component, because \ion{C}{IV} is a $2.0\sigma$ detection.
The right component in the \ion{Ne}{V} transition is a contamination line included to obtain the best-fit voigt model, which is marked with the $\otimes$ symbol.
}
\label{fig:a20}
\end{figure*}

\renewcommand\thefigure{G.\arabic{figure}}

\setcounter{figure}{0}  
\setcounter{table}{0}

\begin{figure}
\begin{center}
\includegraphics[width=0.48\textwidth]{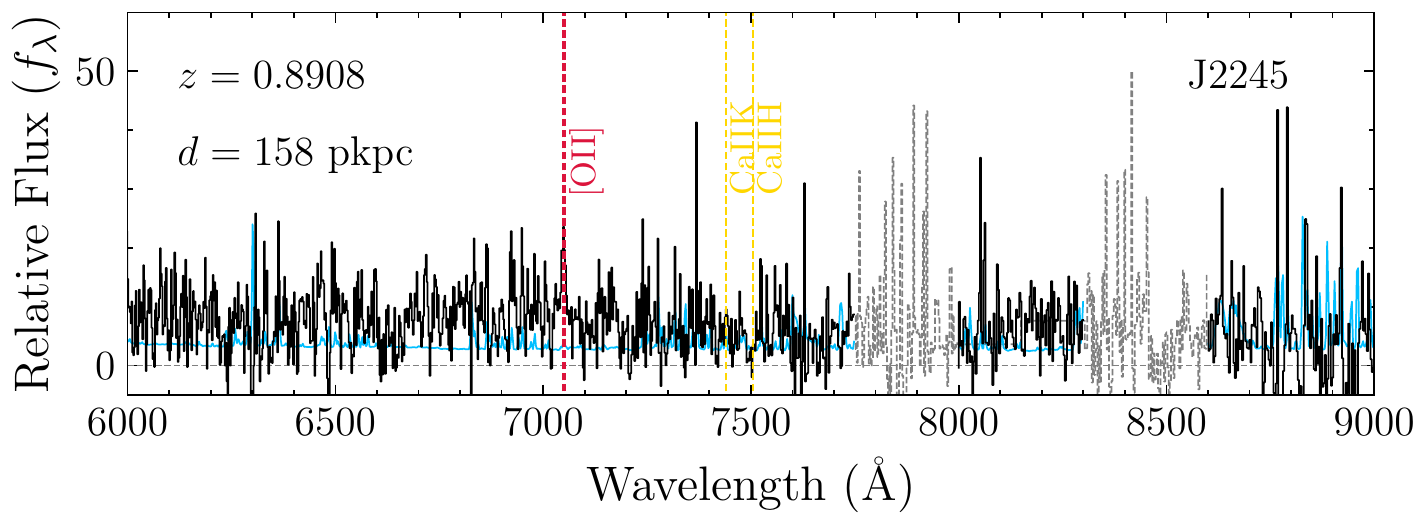}
\end{center}
\caption{The galaxy spectra of all member galaxies in System s1. Red and golden vertical lines represent emission and absorption features, respectively.
The black line is the relative galaxy flux, while the cyan line is the uncertainty associated with the flux. 
The gray dotted lines mark the regions that are contaminated by terrestrial features.
The [\ion{O}{II}] doublet $\lambda\lambda 3727, 3729$ can be resolved in the MUSE spectra.
Therefore, it is considered a secure determination of the redshift sometimes when only [\ion{O}{II}] is detected in the spectrum.
}
\label{fig:g1}
\end{figure}

\begin{figure}
\begin{center}
\includegraphics[width=0.48\textwidth]{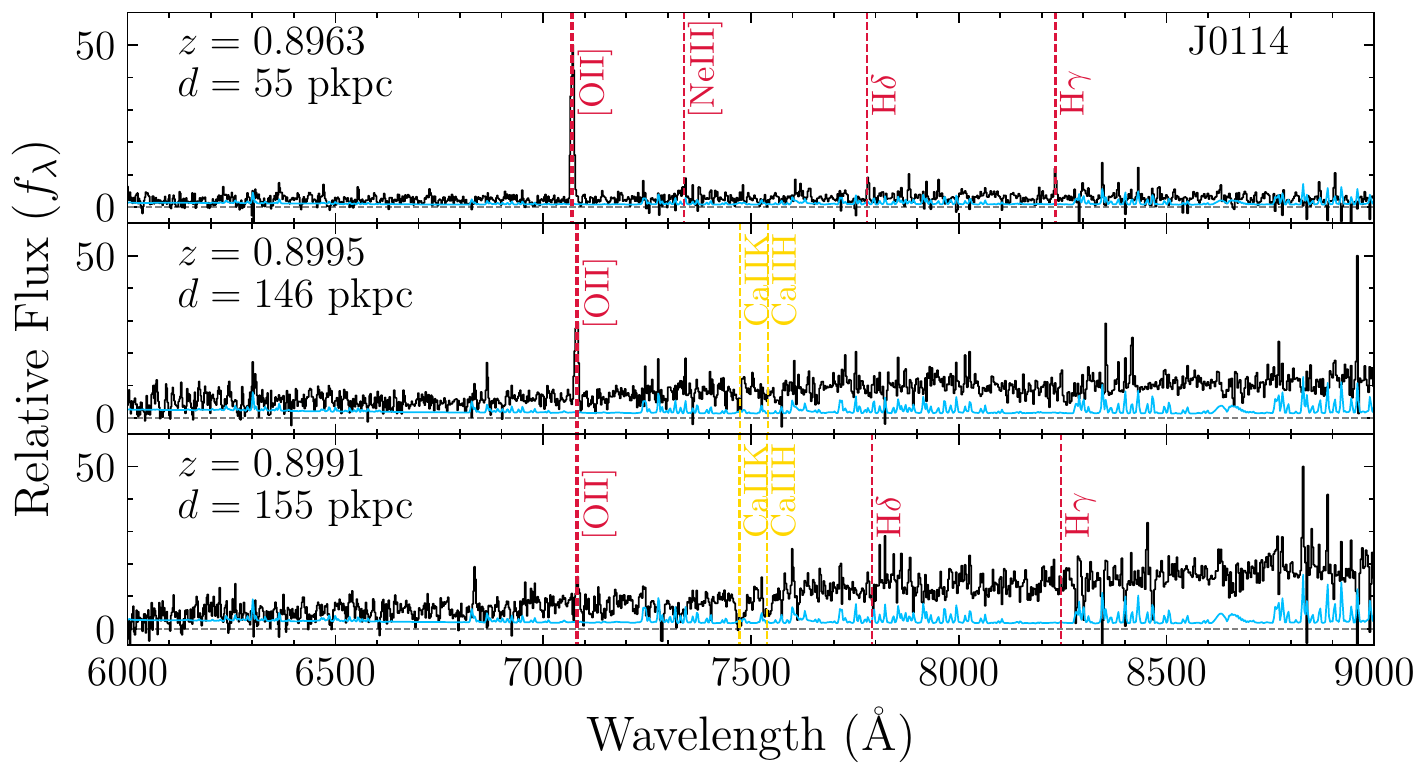}
\end{center}
\caption{Similar as Figure \ref{fig:g1} but for system ID: s2.
}
\label{fig:g2}
\end{figure}

\begin{figure}
\begin{center}
\includegraphics[width=0.48\textwidth]{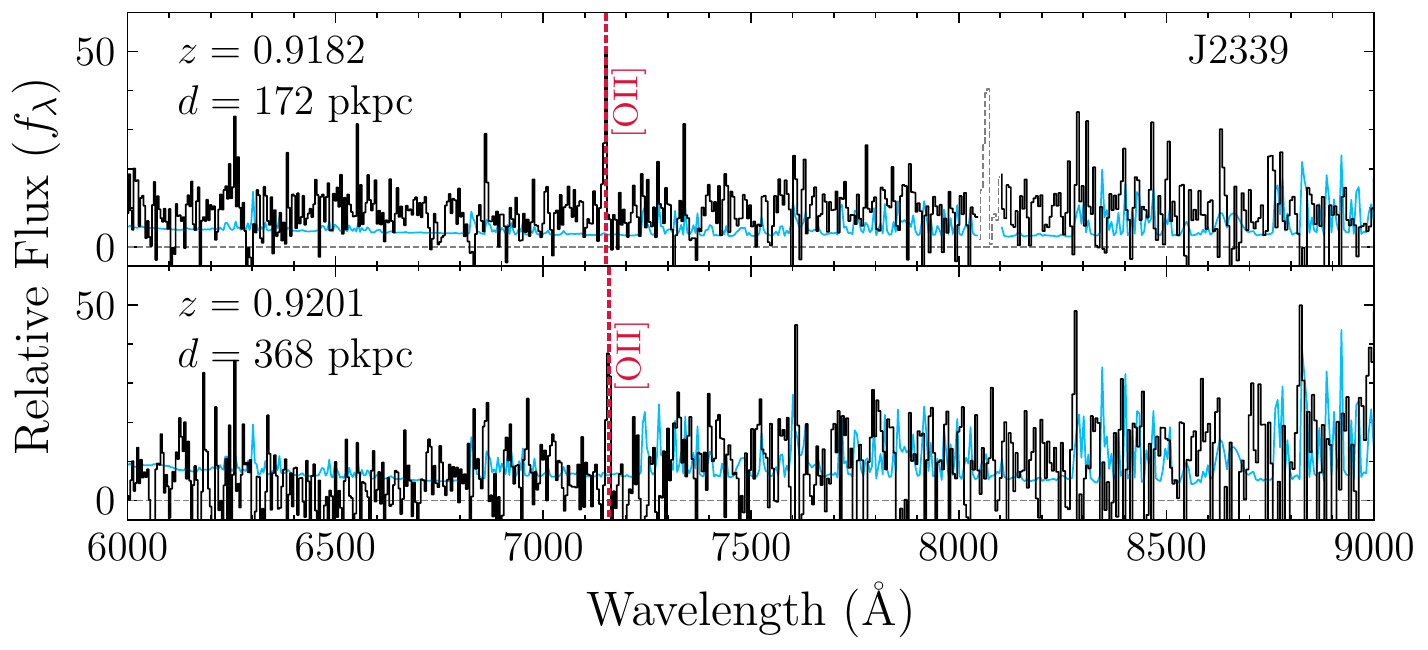}
\end{center}
\caption{Similar as Figure \ref{fig:g1} but for system ID: s3.
}
\label{fig:g3}
\end{figure}

\begin{figure}
\begin{center}
\includegraphics[width=0.48\textwidth]{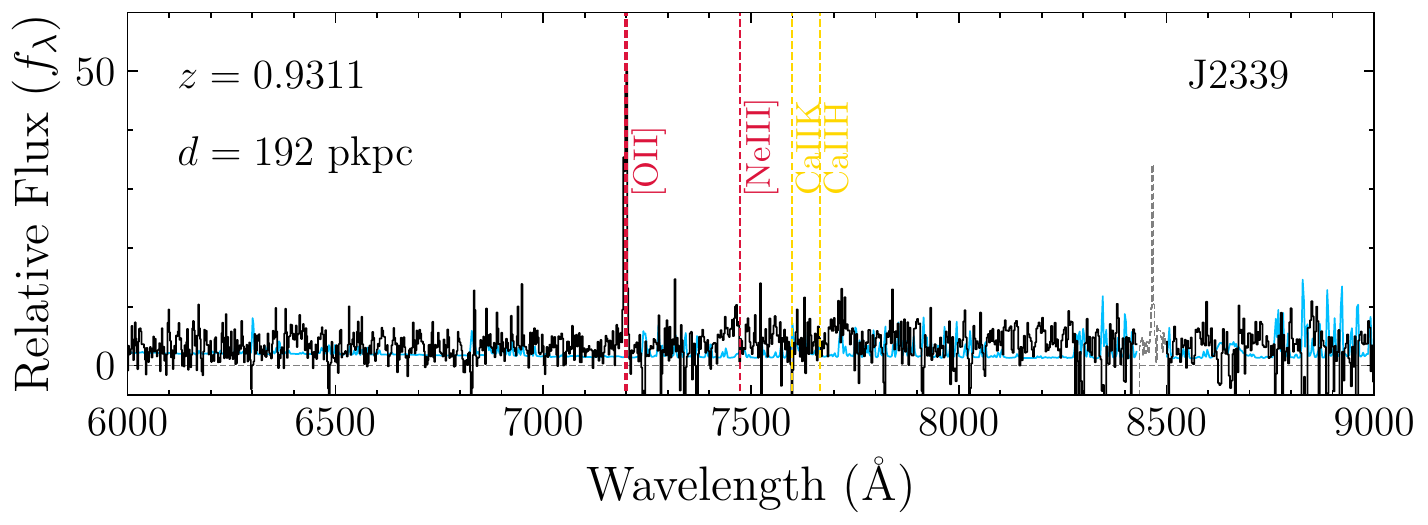}
\end{center}
\caption{Similar as Figure \ref{fig:g1} but for system ID: s4.
}
\label{fig:g4}
\end{figure}

\begin{figure}
\begin{center}
\includegraphics[width=0.48\textwidth]{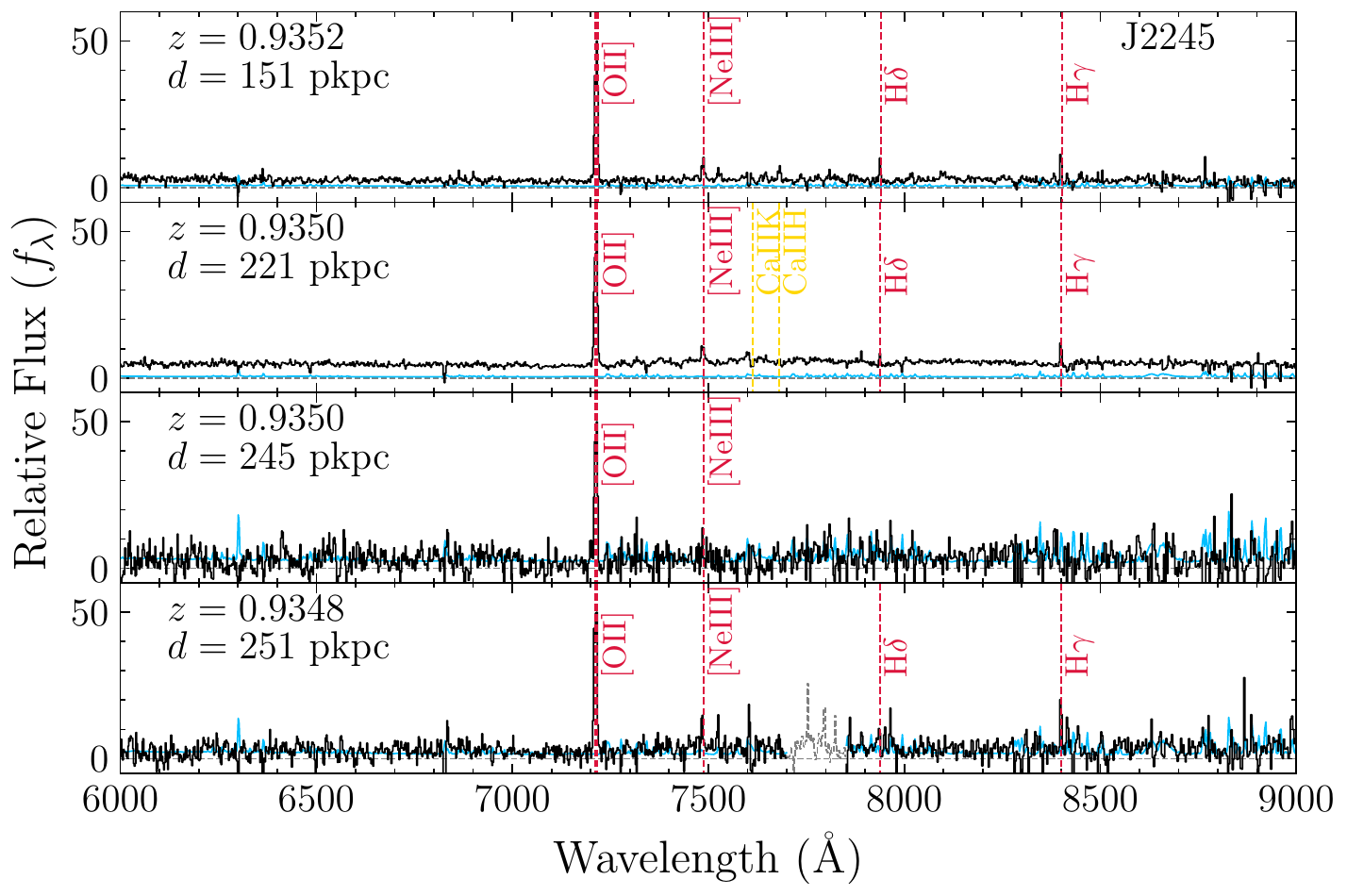}
\end{center}
\caption{Similar as Figure \ref{fig:g1} but for system ID: s5.
}
\label{fig:g5}
\end{figure}

\begin{figure}
\begin{center}
\includegraphics[width=0.48\textwidth]{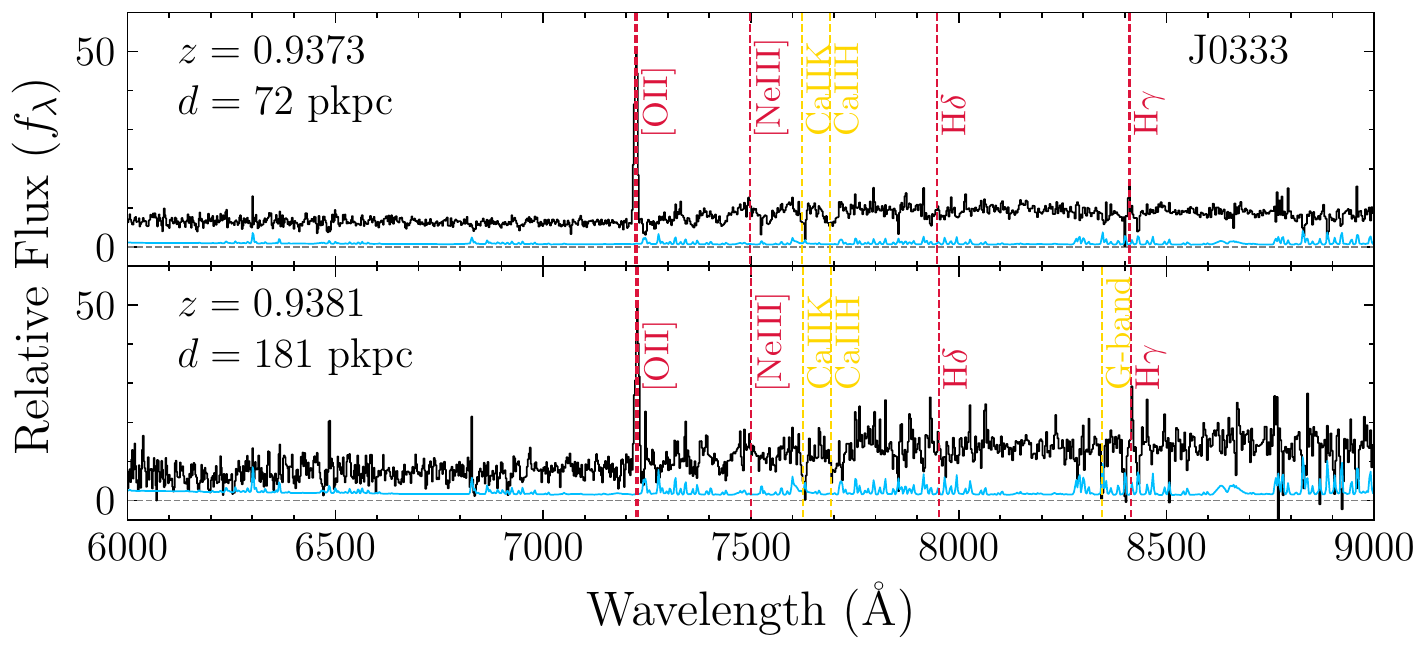}
\end{center}
\caption{Similar as Figure \ref{fig:g1} but for system ID: s6.
}
\label{fig:g6}
\end{figure}

\begin{figure}
\begin{center}
\includegraphics[width=0.48\textwidth]{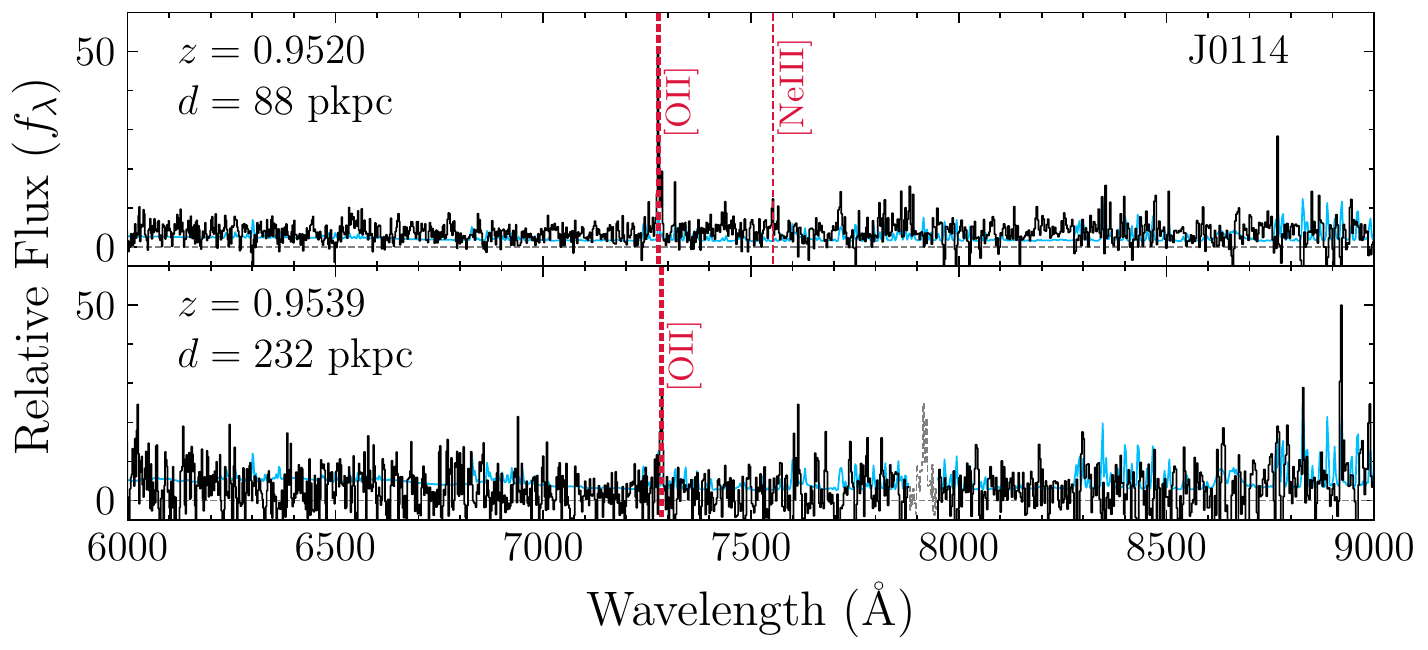}
\end{center}
\caption{Similar as Figure \ref{fig:g1} but for system ID: s7.
}
\label{fig:g7}
\end{figure}

\begin{figure}
\begin{center}
\includegraphics[width=0.48\textwidth]{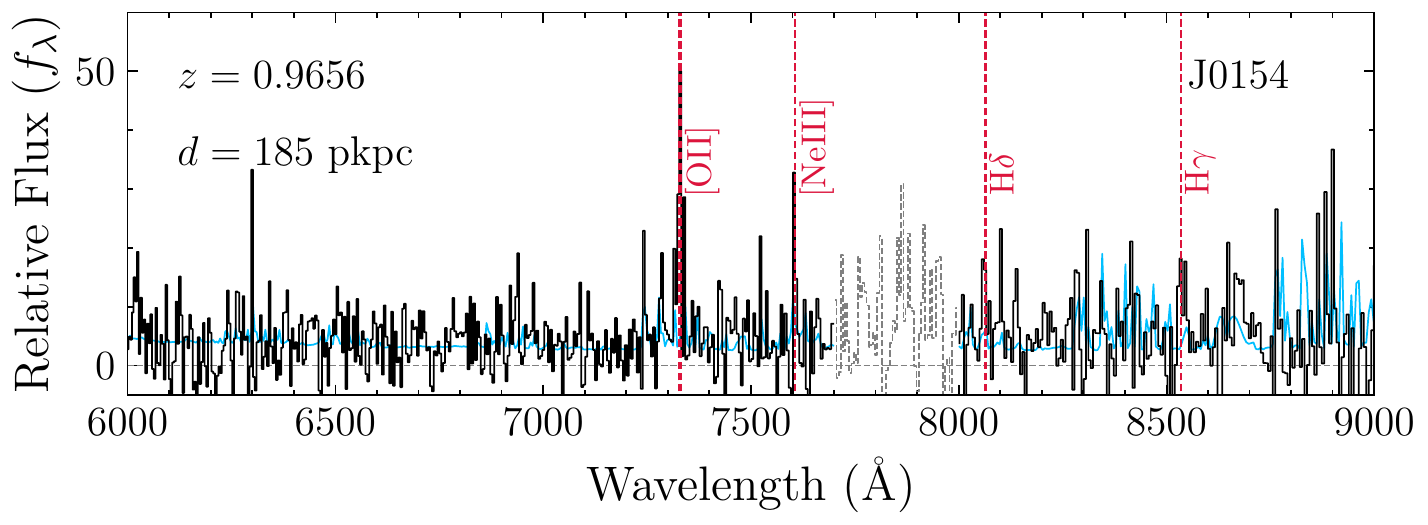}
\end{center}
\caption{Similar as Figure \ref{fig:g1} but for system ID: s8.
}
\label{fig:g8}
\end{figure}

\begin{figure}
\begin{center}
\includegraphics[width=0.48\textwidth]{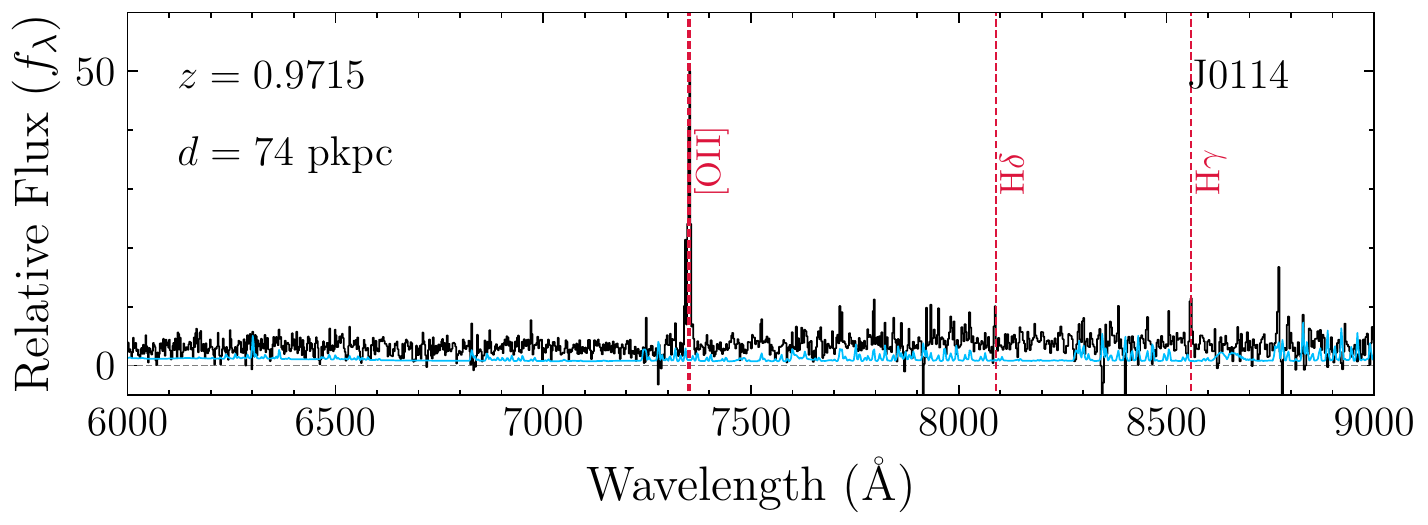}
\end{center}
\caption{Similar as Figure \ref{fig:g1} but for system ID: s9.
}
\label{fig:g9}
\end{figure}

\begin{figure}
\begin{center}
\includegraphics[width=0.48\textwidth]{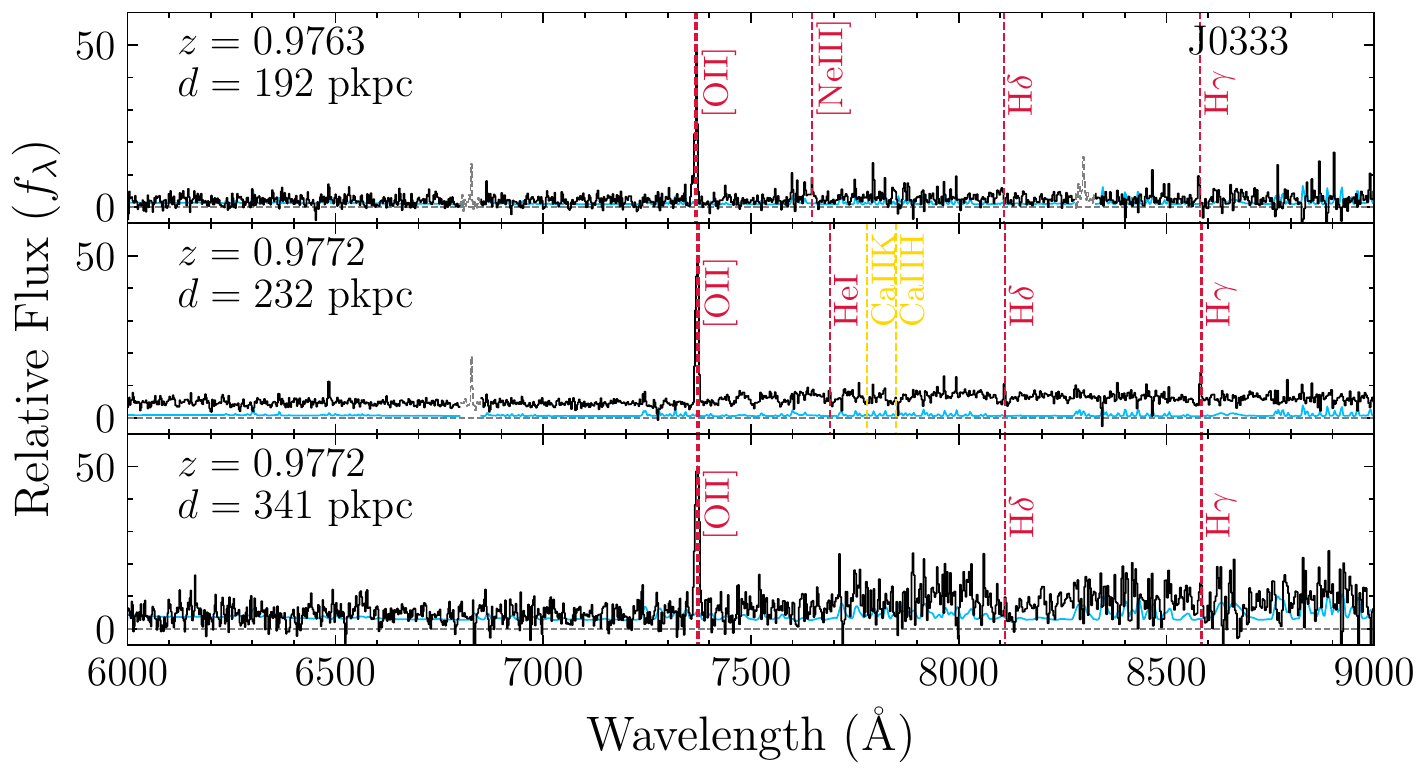}
\end{center}
\caption{Similar as Figure \ref{fig:g1} but for system ID: s10.
}
\label{fig:g10}
\end{figure}

\begin{figure}
\begin{center}
\includegraphics[width=0.48\textwidth]{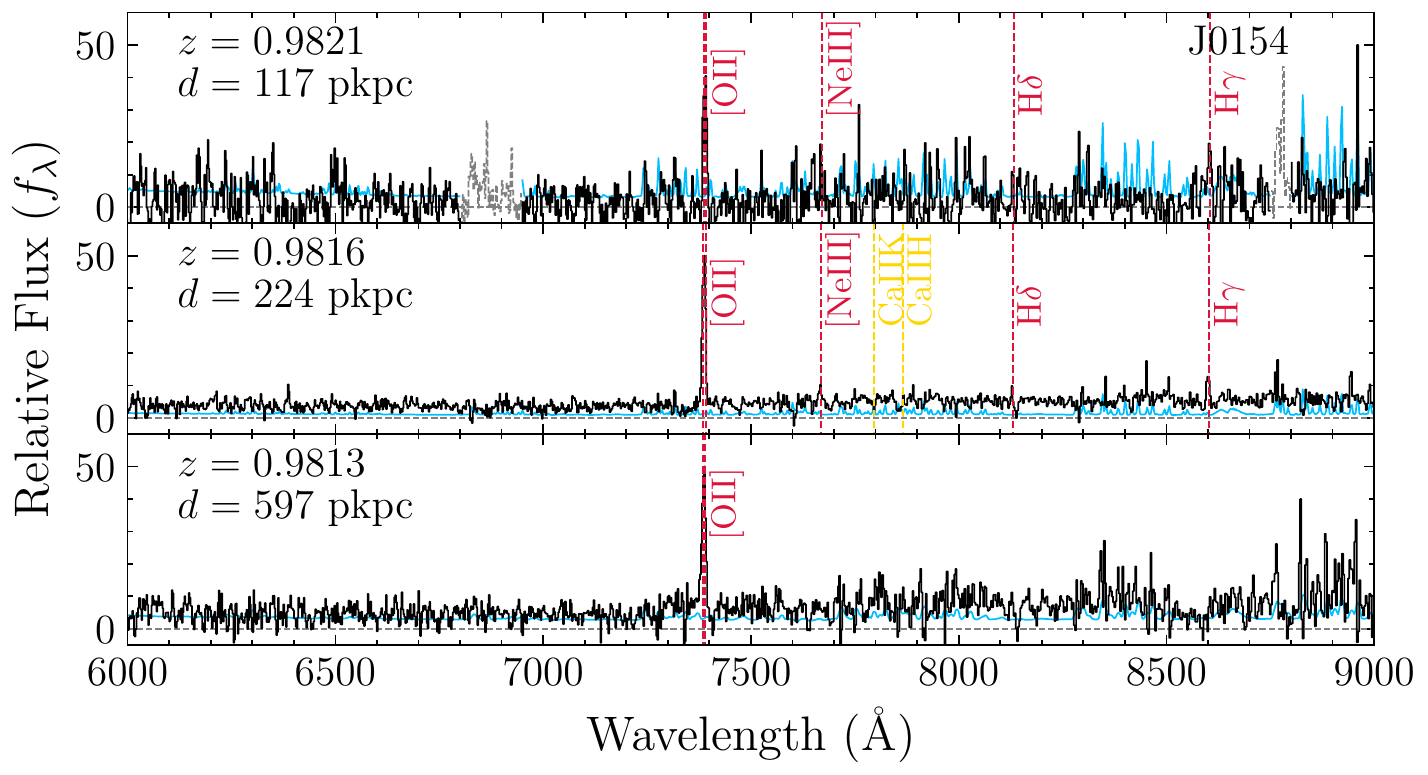}
\end{center}
\caption{Similar as Figure \ref{fig:g1} but for system ID: s11.
}
\label{fig:g11}
\end{figure}

\begin{figure}
\begin{center}
\includegraphics[width=0.48\textwidth]{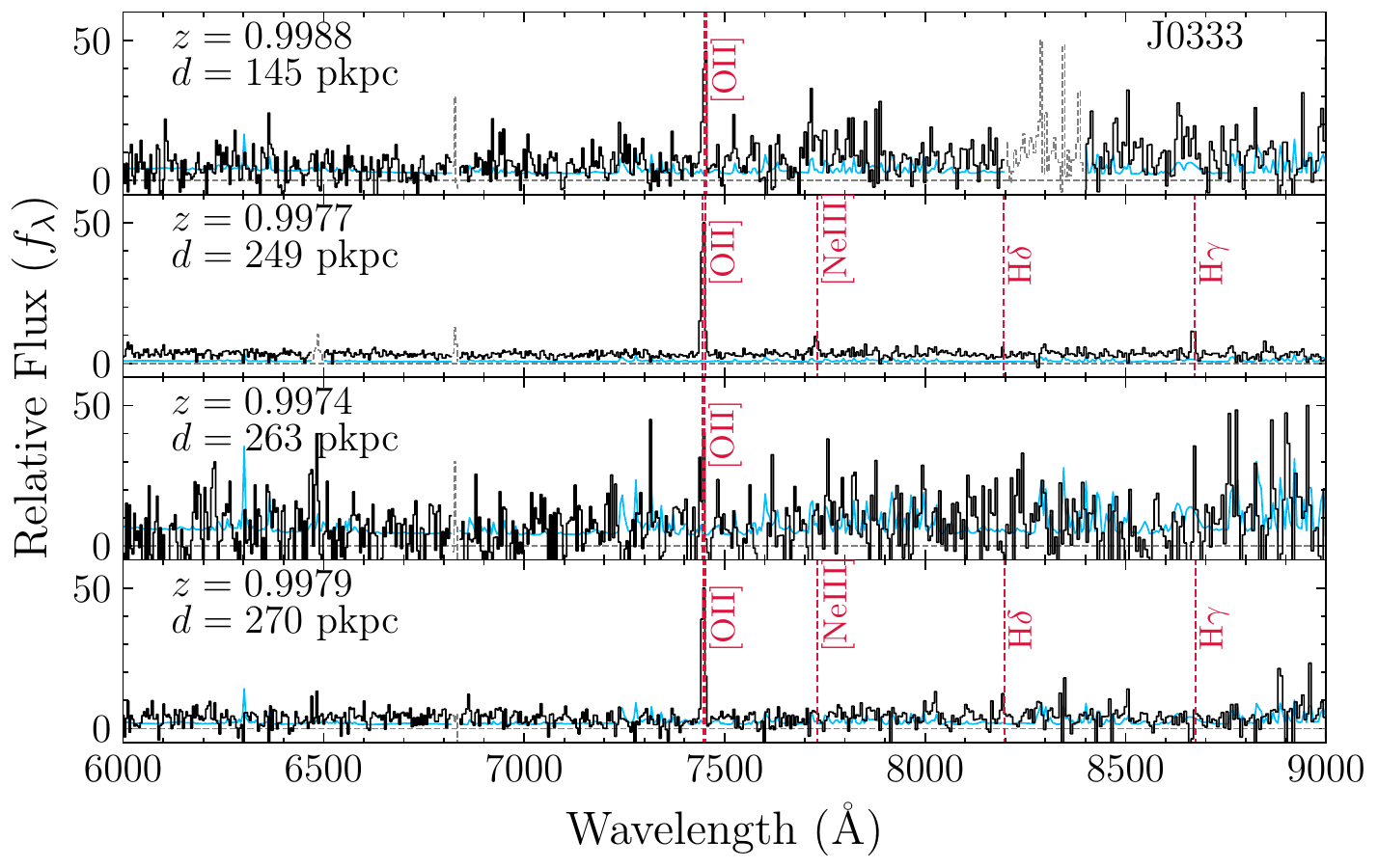}
\end{center}
\caption{Similar as Figure \ref{fig:g1} but for system ID: s12.
}
\label{fig:g12}
\end{figure}

\begin{figure}
\begin{center}
\includegraphics[width=0.48\textwidth]{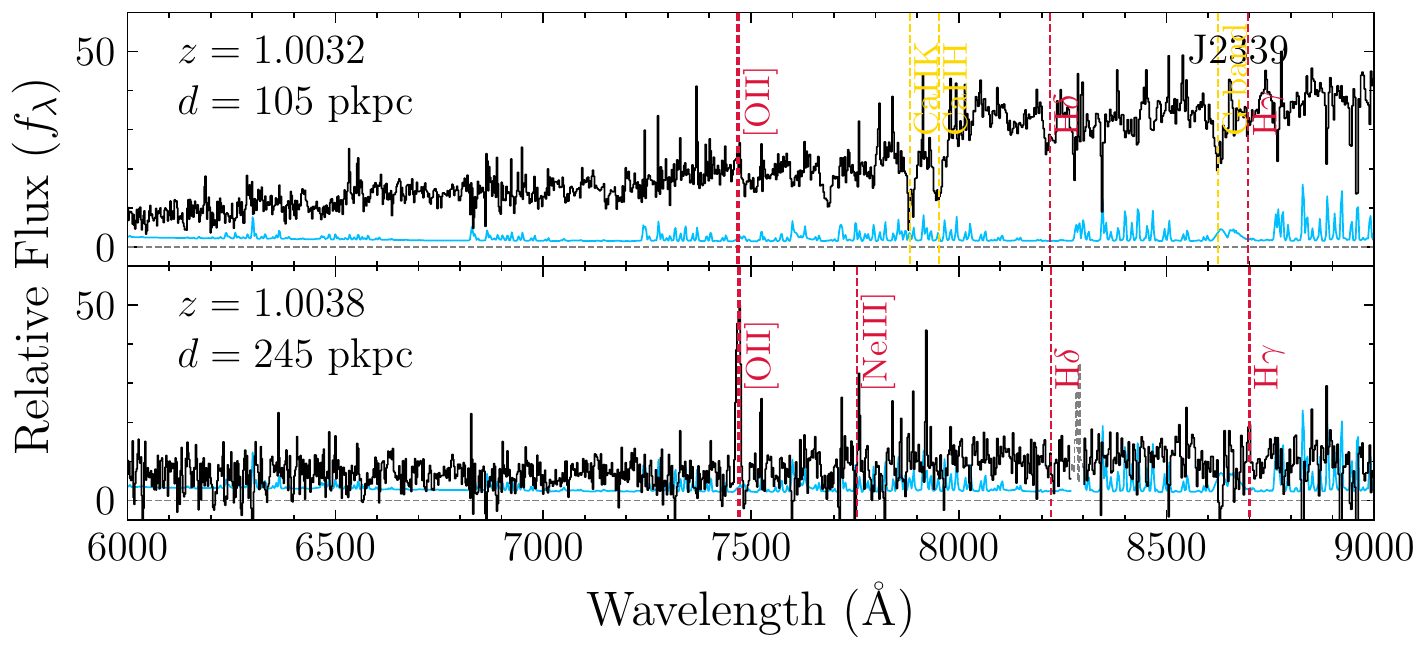}
\end{center}
\caption{Similar as Figure \ref{fig:g1} but for system ID: s13.
}
\label{fig:g13}
\end{figure}

\begin{figure}
\begin{center}
\includegraphics[width=0.48\textwidth]{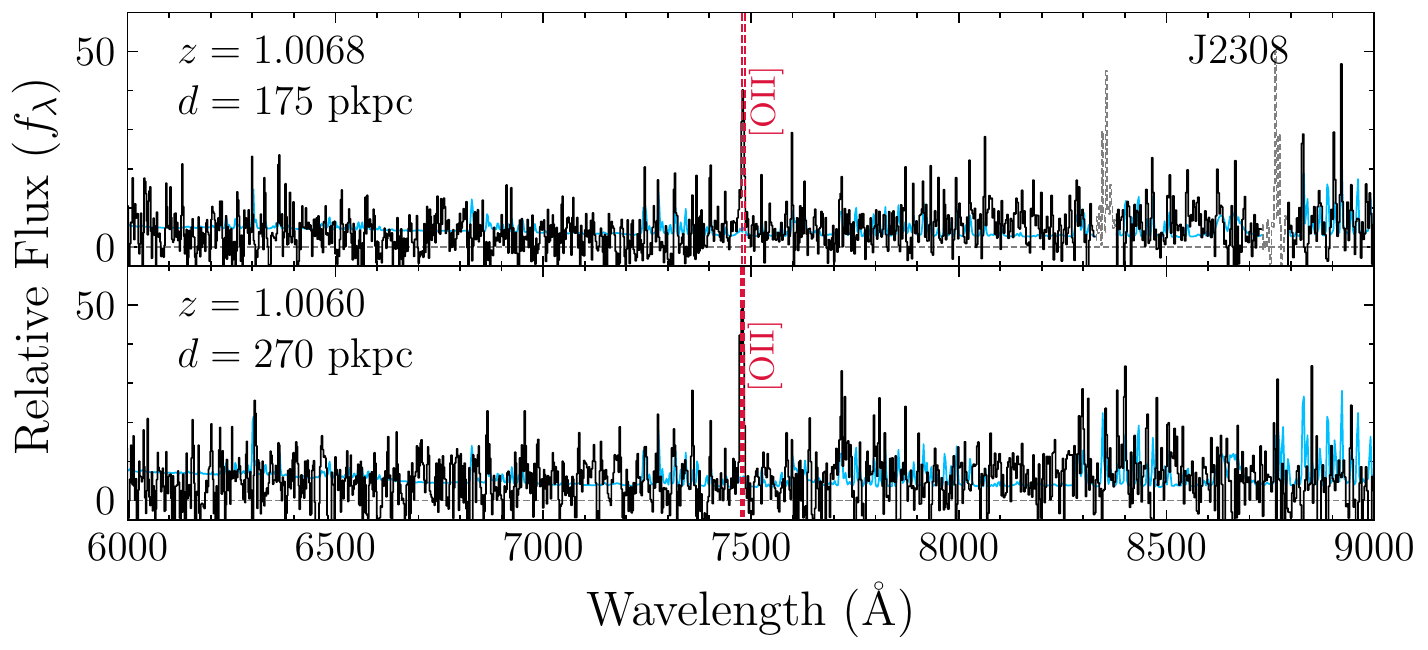}
\end{center}
\caption{Similar as Figure \ref{fig:g1} but for system ID: s14.
}
\label{fig:g14}
\end{figure}

\begin{figure}
\begin{center}
\includegraphics[width=0.48\textwidth]{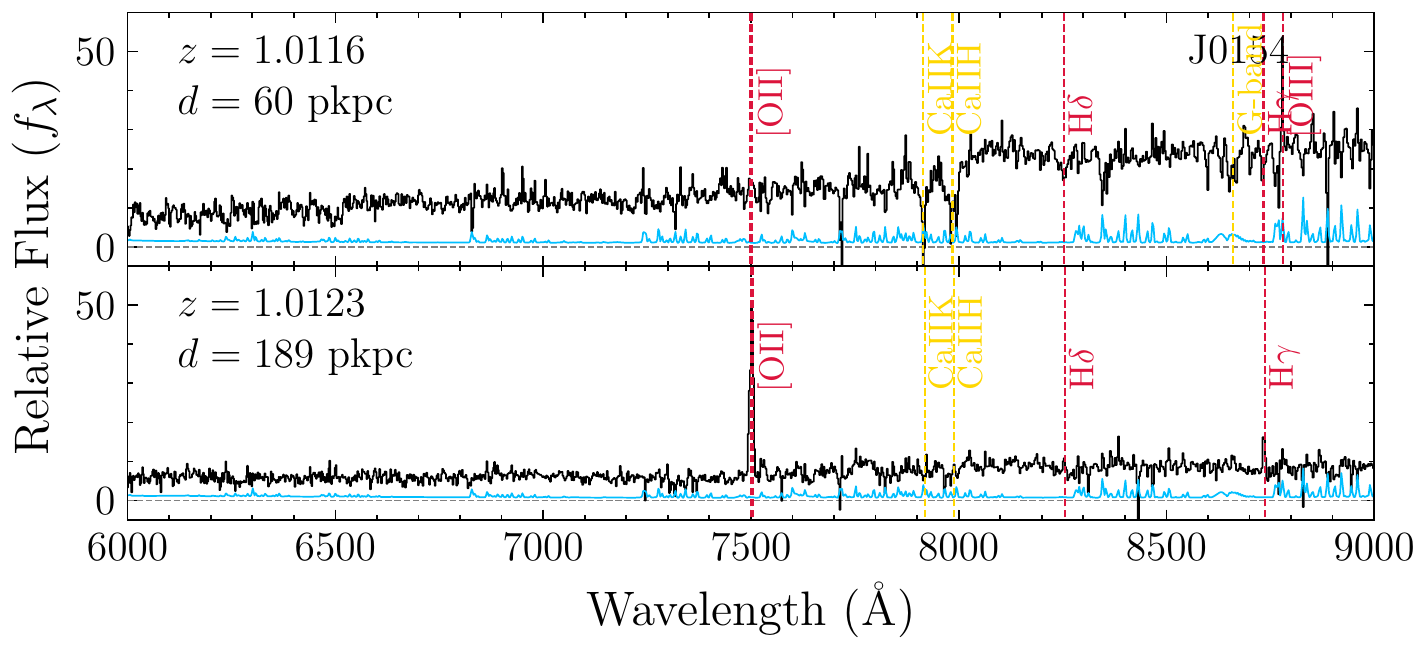}
\end{center}
\caption{Similar as Figure \ref{fig:g1} but for system ID: s15.
}
\label{fig:g15}
\end{figure}

\begin{figure}
\begin{center}
\includegraphics[width=0.48\textwidth]{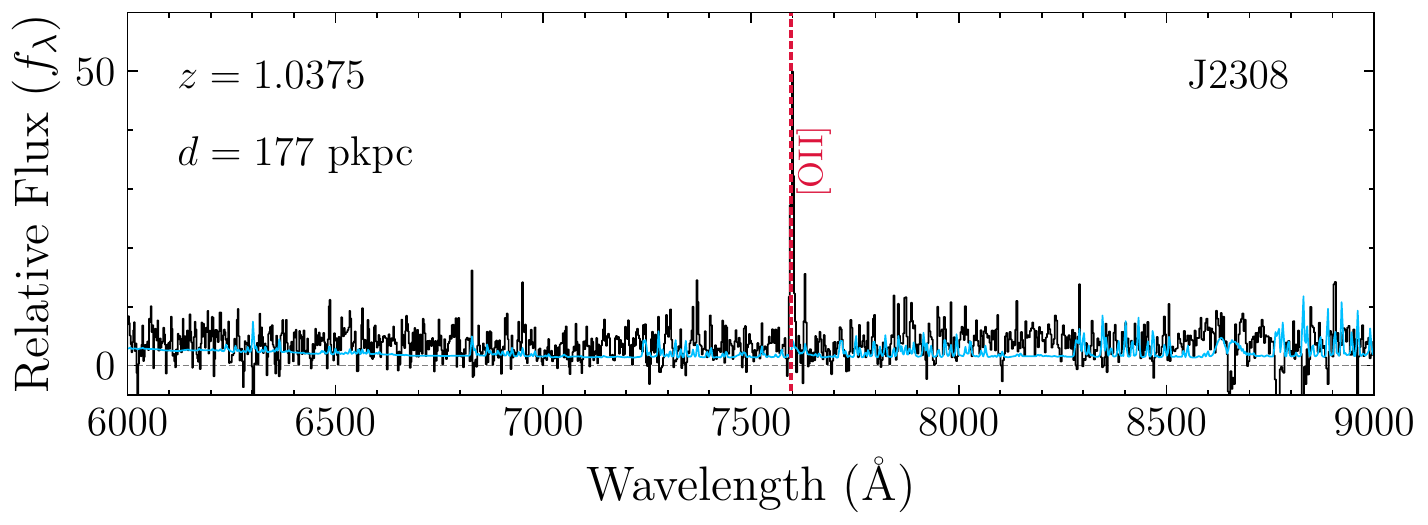}
\end{center}
\caption{Similar as Figure \ref{fig:g1} but for system ID: s16.
}
\label{fig:g16}
\end{figure}

\begin{figure}
\begin{center}
\includegraphics[width=0.48\textwidth]{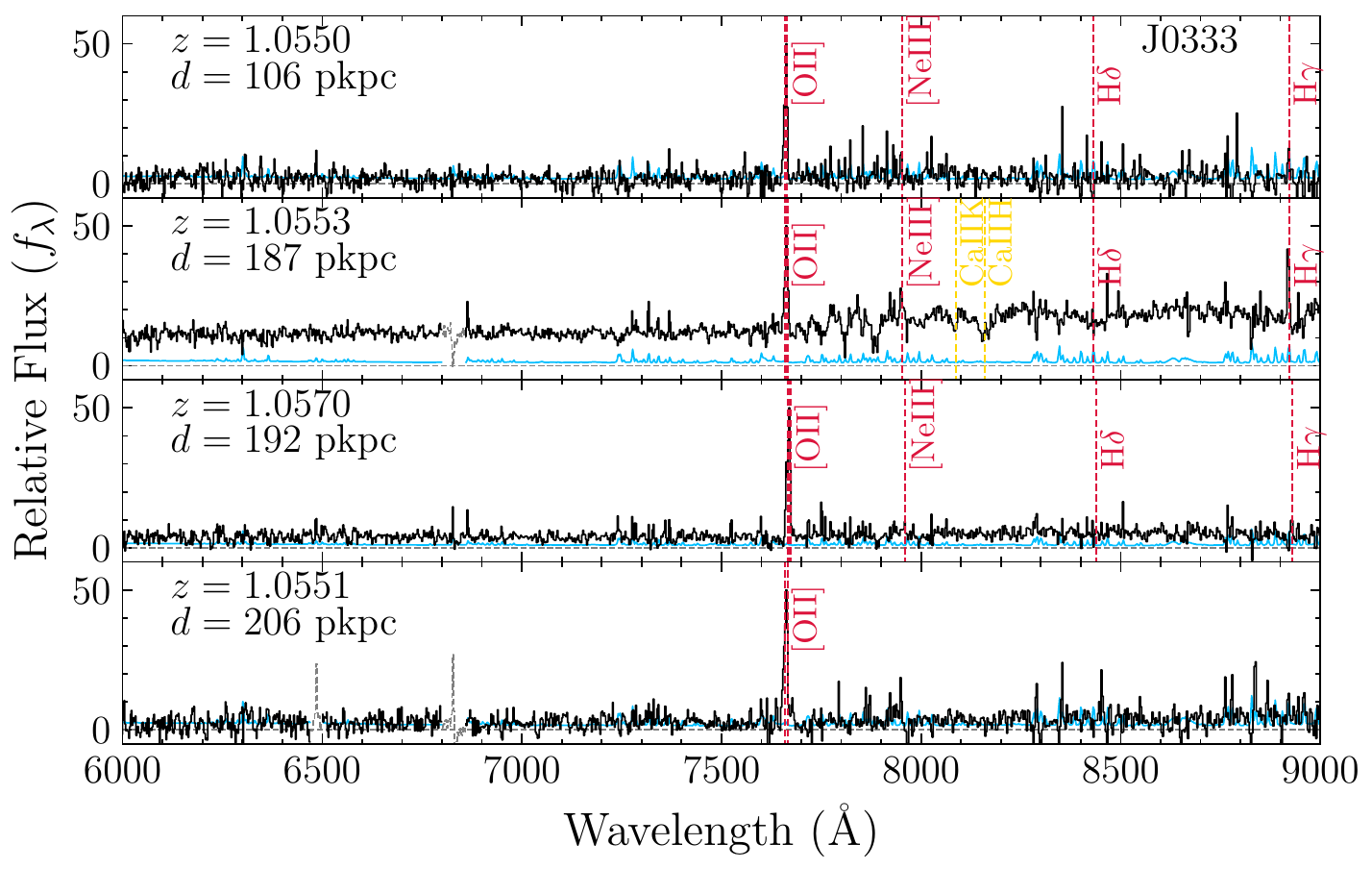}
\end{center}
\caption{Similar as Figure \ref{fig:g1} but for system ID: s17.
}
\label{fig:g17}
\end{figure}

\begin{figure}
\begin{center}
\includegraphics[width=0.48\textwidth]{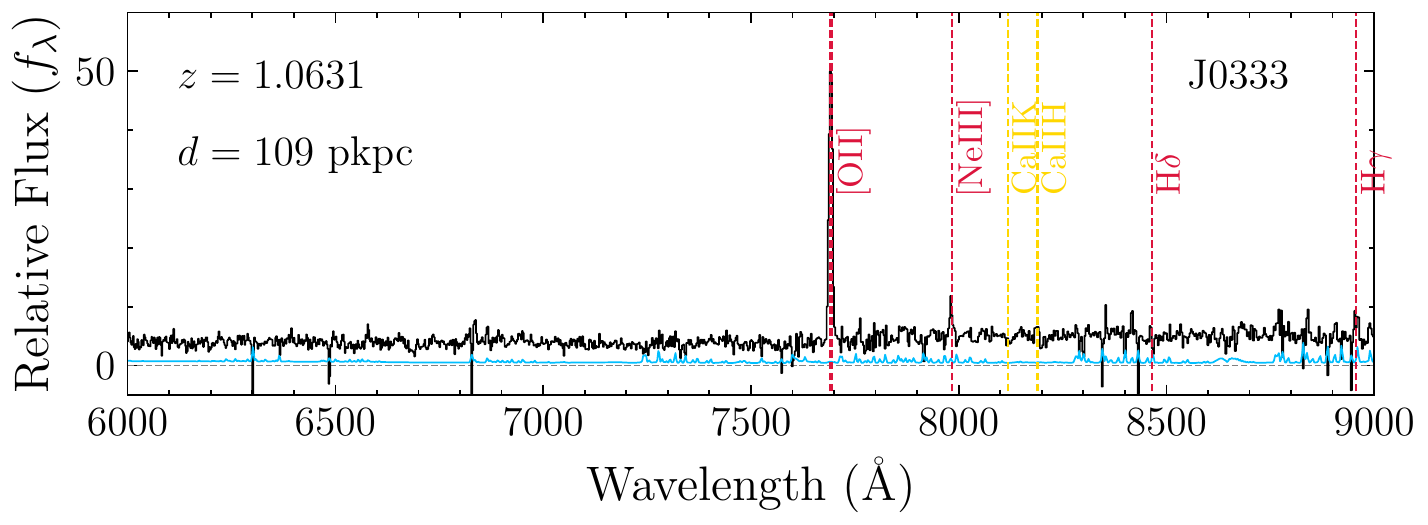}
\end{center}
\caption{Similar as Figure \ref{fig:g1} but for system ID: s18.
}
\label{fig:g18}
\end{figure}

\begin{figure}
\begin{center}
\includegraphics[width=0.48\textwidth]{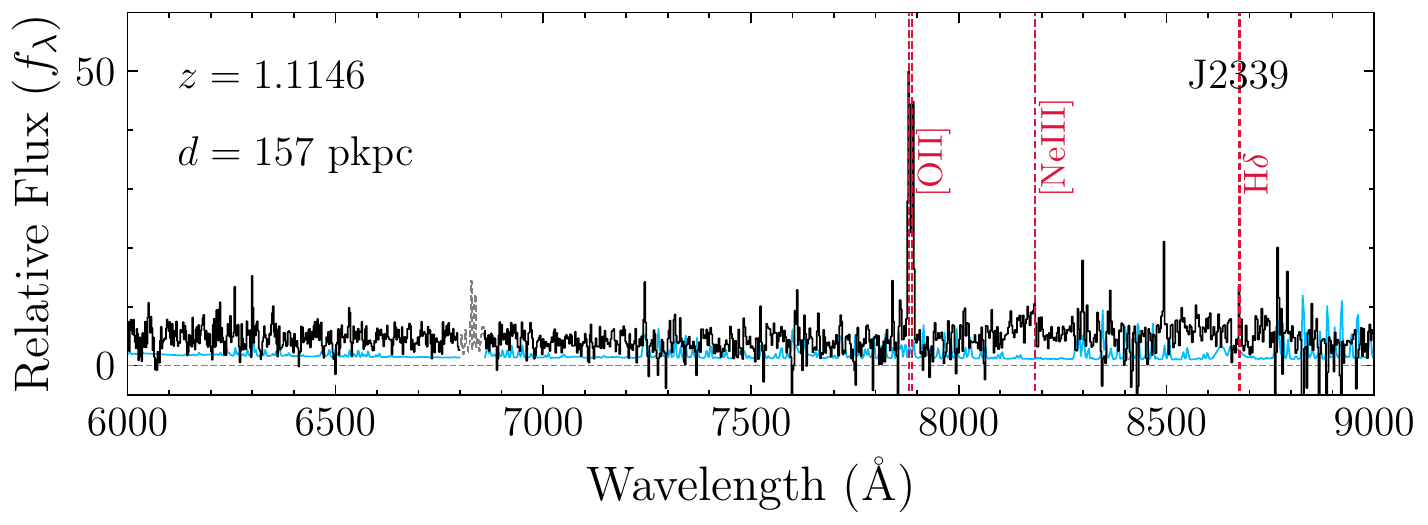}
\end{center}
\caption{Similar as Figure \ref{fig:g1} but for system ID: s19.
}
\label{fig:g19}
\end{figure}

\begin{figure}
\begin{center}
\includegraphics[width=0.48\textwidth]{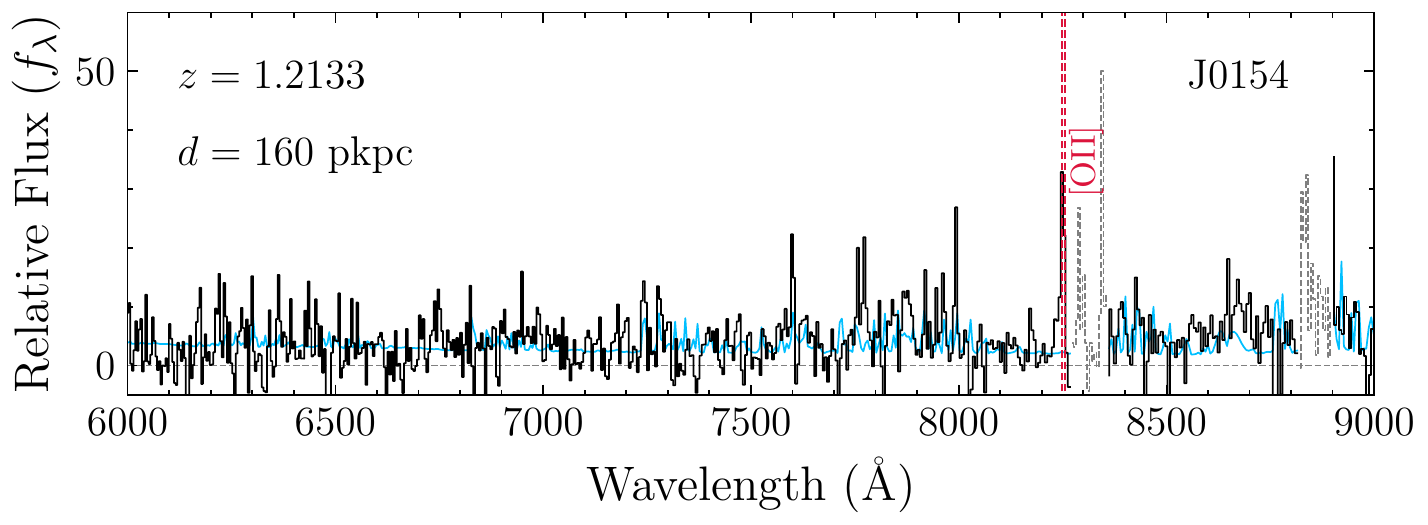}
\end{center}
\caption{Similar as Figure \ref{fig:g1} but for system ID: s20.
}
\label{fig:g20}
\end{figure}



\end{document}